\documentclass[10pt,a4paper]{book}

\usepackage{german}
\usepackage{amsmath}
\usepackage{amssymb}
\usepackage{epsfig}
\usepackage{floatfig}

\usepackage{fancyheadings}

\newcommand{\refformel}[1]{$ \left( \ref{#1} \right) $}
\newcommand{\graphik}[4]{ 
                        \begin{figure}[ht]
                            \vspace{0.5cm}
                            \begin{center}
                               \leavevmode
                               \epsfig{file=#1, height=#2}
                               \caption{{\small \it #3 \label{#4}}}
                            \end{center}
                        \end{figure}
                        }

\begin{document}

\initfloatingfigs

\newcommand{\clearemptydoublepage}{\newpage{\pagestyle{empty}\cleardoublepage}}
\newcommand{\verweistext}[1]{ \hspace{-0.6em} \cite{#1} \hspace{-0.6em} }

\pagestyle{fancyplain}
\lhead[\fancyplain{}{\rightmark}]{\fancyplain{}{}}
\rhead[\fancyplain{}{}]{\fancyplain{}{\rightmark}}
\cfoot[\fancyplain{}{}]{\fancyplain{}{}}
\lfoot[\fancyplain{\thepage}{\thepage}]{\fancyplain{}{}}
\rfoot[\fancyplain{}{}]{\fancyplain{\thepage}{\thepage}}

\setlength{\unitlength}{1cm}

\pagestyle{empty}
\begin{center}
{\huge{\bf{Melting of crystalline films with quenched random disorder}}} \\
\vspace{0.3cm}
{\large Diploma thesis}\\ 
\vspace{0.7cm} 
$\mbox{Peter Stahl}^*$ \\
{\it{Institut f"ur Theoretische Physik, Universit"at zu K"oln, Z"ulpicher Str. 77, 
}} \\
{\it{D-50937 K"oln, Germany}} 
\\
\vspace{1.5cm}

{\large \bf{Abstract}}
\end{center}

According to the Kosterlitz-Thouless-Theory two-dimensional solid films melt by 
the unbinding of dislocation pairs. A model including quenched random impurities
was already studied by Nelson [Phys. Rev. B 27 (1983) 2902] , who predicted a reentrance into the disordered
phase at low temperatures and weak disorder. New investigations of the 
physically related XY-model [e.g. T. Nattermann et al., J. Phys. (France) 5 (1995), 565] and a work of Cha and Fertig [Phys. Rev. Lett. 74 (1995) 4867] refuse this reentrant melting.
In this work we map the system onto a two-dimensional vector Coulomb gas and via a 
renormalization we derive flow equations both for the square and for the triangular 
lattice. An analysis of these flow equations shows a new behaviour in the low-
temperature range, where the reentrance into the non-crystalline phase with
short-range order is not found, but the crystalline phase with quasi-long-range
order is preserved below a critical disorder strength of $\bar{\sigma_c} = \frac{1}{16 \pi}$. 
Finally we estimate the influence of commensurate substrates and obtain phase diagrams,
which show that the melting by dislocation unbinding can only be expected, if the lattice constant of the crystalline 
film is a multiple of the lattice constant of the substrate potential.

\vspace{3.5cm}
$\mbox{}^*\mbox{now:}$
\begin{center}
{\it{Institute of Process Engineering, Swiss Federal Institute of Technology 
}} \\
{\it{(ETH Z"urich), CH-8092 Z"urich, Switzerland}} 
\\
\end{center}

\clearemptydoublepage

\pagestyle{fancyplain}
\lhead[\fancyplain{}{\rightmark}]{\fancyplain{}{}}
\rhead[\fancyplain{}{}]{\fancyplain{}{\rightmark}}
\cfoot[\fancyplain{}{}]{\fancyplain{}{}}
\lfoot[\fancyplain{\thepage}{\thepage}]{\fancyplain{}{}}
\rfoot[\fancyplain{}{}]{\fancyplain{\thepage}{\thepage}}

\pagenumbering{roman}

\tableofcontents
\clearemptydoublepage

\pagenumbering{arabic}

\chapter{Einleitung}
In der Festk"orperphysik versteht man unter einem dreidimensionalen Kristall eine (fast) unendliche Wiederholung identischer Struktureinheiten, die Translationssymmetrie aufweist. Der Anteil der Struktureinheiten, die die Oberfl"ache bilden soll verschwindend gering sein. In drei Dimensionen bilden alle Stoffe - sieht man von Helium und Quasikristallen wie ${\mbox{Al}}_{86}{\mbox{Mn}}_{14}$ ab - solche kristallinen Strukturen, die aber von eingefrorener Unordnung gest"ort werden k"onnen.
 
Daher geh"oren auch dreidimensionale Kristallstrukturen wie Metalle zum allt"aglichen Erfahrungsbereich und das Schmelzen der Kristalle zu Fl"ussigkeiten bei Temperaturerh"ohung kann unter geeignetem Druck beobachtet werden. Hierbei handelt es sich um einen Phasen"ubergang erster Ordnung, bei dem ein Sprung in der Entropie auftritt. Ausgel"ost wird er durch die energetische Anregung des Gitters. 

Obwohl sie auf Oberfl"achen h"aufig auftreten sind zweidimensionale kristalline Filme hingegen nicht so leicht zu beobachten, dennoch kann man das Schmelzen auch hier an einigen Systemen untersuchen. Dies sind unter anderen:
\begin{itemize}
\item Elektronen, die in einer einzigen Schicht etwa 100 {\AA} "uber der Oberfl"ache fl"ussigen Heliums aufgebracht werden. Dies geschieht mittels eines elektrischen Felds, welches senkrecht zur Oberfl"ache steht \verweistext{ccgga}. Bei gen"ugend geringer Dichte verhalten sie sich klassisch.
\item Polystyrolkugeln mit einem Durchmesser von etwa 0,3 $\mu$m bilden in L"osungen mit geeignetem pH-Wert Kolloide. Bringt man diese zwischen zwei Glasplatten mit einem Abstand von etwa 1 $\mu$m, formen diese Kugeln ein zweidimensionales Dreiecksgitter, nehmen also Kristallstruktur an \verweistext{mvw87}. 
\end{itemize}
 
Die Phasen"ubergangs-Temperatur ist dabei systemabh"angig. Sie wird von der Gitterkonstanten und den Lam\'e-Koeffizienten bestimmt. Dies f"uhrt dazu, da"s die erw"ahnte Elektronenschicht eine Schmelztemperatur von etwa 1 K hat, w"ahrend die Schmelztemperatur der Kolloide in der Gr"o"senordnung von $10^2$ K liegt.
  
Im Gegensatz zum dreidimensionalen Fall, gibt es bei freien zwei\-dimen\-sio\-nalen Kristallstrukturen f"ur $T>0$ aufgrund von Fluktuationen in den Moden der niederenergetischen phononischen Anregungen keine echte langreichweitige Ordnung, somit also auch keine beliebig langreichweitige Translationssymmetrie. Hier existiert eine quasi-langreich\-wei\-tige Ordnung, bei der die Korrelationsfunktion des Ordnungsparameters algebraisch abf"allt. Auch das Schmelzen kann in zwei Dimensionen einen anderen Charakter haben. Neben dem Phasen"ubergang erster Ordnung existiert die M"oglichkeit eines kontinuierlichen Kosterlitz-Thouless-Phasen"ubergangs, bei dem sich Defektpaare aufl"osen und freie topologische Defekte entstehen, was zur Aufl"osung der Kristallstruktur f"uhrt. Die oben genannten zweidimensionalen Systeme zeigen diesen kontinuierlichen Schmelzproze"s. Exakte Messungen sind hierbei nicht einfach, dennoch best"atigen die Ergebnisse diese im folgenden n"aher erl"auterte Theorie.
 
Dreidimensionale Kristalle sind gegen"uber topologischen Defekten wesentlicher stabiler als zweidimensionale, daher tritt in drei Dimensionen kein durch topologische Defekte induziertes Schmelzen auf \verweistext{mvw87}. 
 
Die Arbeit gliedert sich wie folgt: Nach einer einleitenden Vorstellung des Koster\-litz-Thouless-Phasen"ubergangs mit dem Hinweis auf verwandte Systeme, wird ein Modell f"ur zweidimensionale kristalline Filme mit Unordnung motiviert und in eine Coulombgas-Darstellung "uberf"uhrt. Die mit diesem Modell von Nelson \verweistext{nel83} gewonnenen Ergebnisse werden diskutiert. Es folgt anschlie"send eine Neuuntersuchung des Modells mittels eines dielektrischen Formalismus und mit Hilfe der Renormierungsgruppe. Nach Pr"asentation der Ergebnisse, die sowohl f"ur das Quadrat-, als auch f"ur das Dreiecksgitter keinen Wiedereintritt in die nicht-kristalline Phase bei tiefen Temperaturen zeigen, wird abschlie"send der Einflu"s eines Substrats untersucht.
 
\section{Der Kosterlitz-Thouless-Phasen"ubergang}
Es ist seit l"angerem bekannt, da"s topologische Defekte in zwei Dimensionen interessante Phasen"uberg"ange verursachen k"onnen. Einen theoretischen Ansatz zur Untersuchung dieser Art von Phasen"uber\-g"angen lieferten Kosterlitz und Thouless zu Beginn der 70er Jahre \verweistext{kt73, kos74}. Verschiedene zweidimensionale Systeme mit topologischer Unordnung, besitzen einen Phasen"ubergang, der mittels der Kosterlitz-Thouless-Theorie erkl"art werden kann:
\begin{itemize}
\item Das sogenannte XY-Modell ist ein zweidimensionales Spinsystem, dessen klassische Spins {\bf S}({\bf x})=S(cos$\theta$({\bf x}), sin$\theta$({\bf x})) (sie stellen hier auch den Ordnungsparameter dar) topologische Defekte in Form von Vortices ausbilden k"onnen. Die Spinphase $\theta$({\bf x}), die die Richtung des Ordnungsparameters beschreibt, "andert sich bei Umlauf um einen Vortexkern um 2k$\pi$, wobei k die Windungszahl des Vortex genannt wird. Somit ist ein Vortex ein Spinwirbel. Experimentell kann ein solches zweidimensionales Spinsystem auch durch dreidimensionale Systeme mit schwach gekoppelten Schichten realisiert werden. Der Phasen"ubergang entspricht hier den "Ubergang von der ferromagnetischen zur paramagnetischen Phase.
\item Ferner ist eine Anwendung der Theorie auf suprafluide Heliumfilme m"og\-lich, da auch hier Vortices gebildet werden k"onnen. Der Ordnungsparameter wird allerdings nicht durch einen zweidimensionaler Vektor, sondern eine "aquivalente komplexe Zahl dargestellt. Das suprafluide Helium geht mittels des Kosterlitz-Thouless-Phasen"ubergangs in den normal fl"ussigen Zustand "uber. 
\item Schlie"slich f"uhren topologische Defekte auch in den im folgenden behandelten zweidimensionalen kristallinen Filmen zu einem Phasen"ubergang von der kristallinen in die nicht-kristalline Phase. Hier liegen allerdings keine Vortices, sondern Versetzungen vor, die sich nicht alleine durch eine skalare Windungszahl beschreiben lassen, da sie auch eine Richtung haben.
\end{itemize} 
 
Anschaulich entspricht eine Versetzung im Quadratgitter dem Endpunkt einer zus"atzlich eingef"ugten Reihe von Gitterbausteinen. Dies ist in Abbildung 1.1a dargestellt. Der Burgers-Vektor, der entsprechend der Abbildung konstruiert wird, gibt Richtung und St"arke der Versetzung an; sind zwei Reihen eingef"ugt ist der Burgers-Vektor doppelt so gro"s, wie bei einer eingef"ugten Reihe. Wie in Abbildung 1.1b gezeigt, besteht eine Versetzung auf dem Dreicksgitter in der Kernregion aus einem Platz mit f"unf und einem Platz mit sieben Nachbarn. Die Versetzungen befinden sich auf den Gitterpunkten des dualen Gitters, welches beim Quadratgitter ebenfalls ein Quadratgitter, beim Dreiecksgitter ein hexagonales Gitter ist. Es handelt sich ausschlie"slich um sogenannte ``Stufen-Versetzungen'', da ``Schrauben-Versetzungen'' nur oberhalb von zwei Dimensionen auftreten k"onnen.
 
\graphik{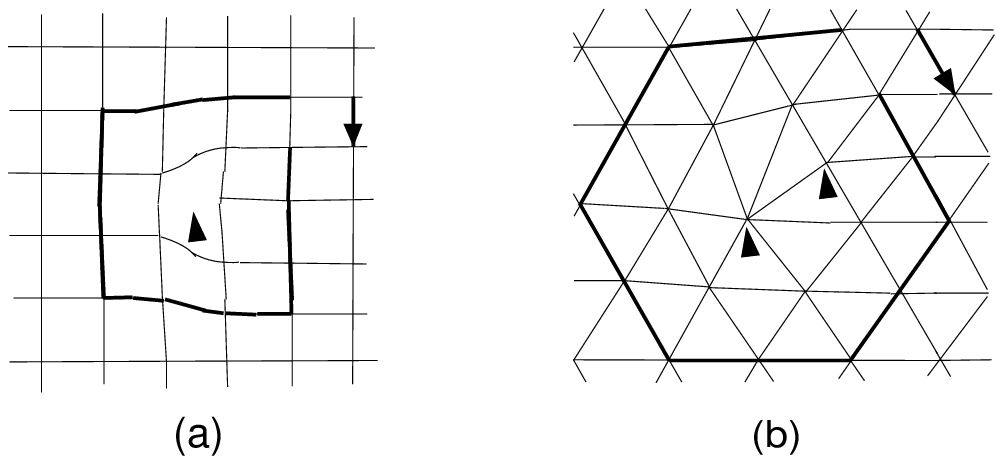}{5cm}{Elementare Versetzungen (a) im Quadrat- und (b) im Dreiecksgitter. Die durch dickere Linien eingezeichneten Wege sind wegen der Versetzungen nicht geschlossen. Die Burgers-Vektoren sind durch Pfeile dargestellt. Die Dreiecke markieren die Pl"atze die unregelm"a"sige Bonds aufweisen \verweistext{hal79}.}{101}
 
Die Energie einer einzelnen Versetzung mit Burgers-Vektor {\bf b} divergiert logarithmisch mit der Systemgr"o"se R, es gilt $E \propto {\bf b^2}\ln R$. Daher ist es bei tiefen Temperaturen die Wahrscheinlichkeit, ein gro"ses System mit einer einzelnen Versetzung vorzufinden, verschwindend gering. Betrachtet man den Burgers-Vektor als Vektorladung, so mu"s das System verschwindende Gesamtladung $\Sigma{\bf b_i} = 0$ aufweisen. Nur dann bleibt die Gesamtenergie auch f"ur beliebig gro"se Systeme endlich, da sich die obigen divergenten Beitr"age in diesem Fall mit Beitr"agen aus der Wechselwirkung der Versetzungen wegheben. Der Kos\-ter\-litz-Thouless-Pha\-sen\-"uber\-gang unterscheidet allgemein zwischen einer Tief\-tempe\-ratur- und einer Hoch\-tem\-pera\-tur\-phase. In der {\bf Tieftemperaturphase} sind gepaarte Versetzungen (Versetzungsdipole) neben den Phononen die elementaren Anregungen des Systems. Einen solchen Dipol f"ur das Quadratgitter zeigt Abbildung 1.2.
 
\graphik{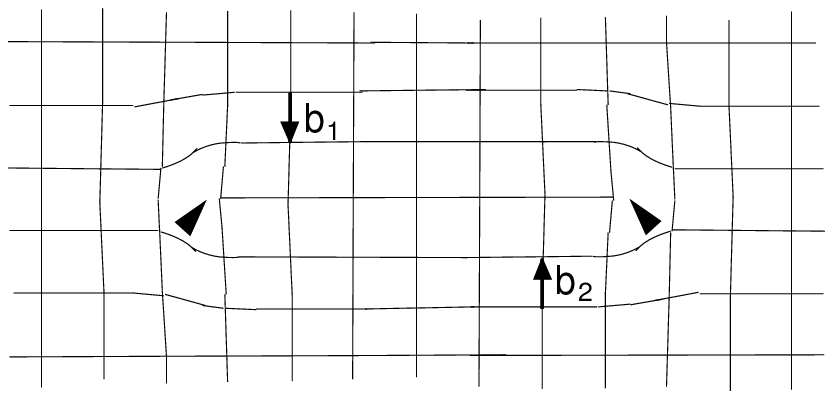}{5cm}{Versetzungsdipol auf dem Quadratgitter mit $\mbox{\bf b}_1 = - \mbox{\bf b}_2 = -\mbox{\bf e}_y$}{102}

Wie Mermin 1968 gezeigt hat \verweistext{mer68}, wird in zweidimensionalen Kristallen die langreichweitige Ordnung bei $T > 0$ durch Fluktuationen zerst"ort, die durch langwellige Phononen hervorgerufen werden. Es herrscht hier quasi-langreichweitige Ordnung vor, was hei"st, da"s die Translations-Kor\-rela\-tions\-funk\-tion des Translationsordnungsparameters $\rho_{\mbox{\bf \scriptsize G}}\left( \mbox{\bf r}\right)$ zwischen den Pl"atzen algebraisch abf"allt \verweistext{hal79}. Der Translationsordnungsparameter ist wie folgt definiert:
\begin{equation}
  \label{101a}
  \rho_{\mbox{\bf \scriptsize G}} \left( \mbox{\bf r} \right) = e^{i \mbox{\bf \scriptsize G} \mbox{\bf \scriptsize u(r)} },
\end{equation}
hierbei ist {\bf G} ein Vektor des reziproken Gitters und {\bf u(r)} ein Vektorfeld, da"s die Verschiebung der Kristallbausteine von ihrem Gleichgewichtsplatz mi"st. Die algebraisch abfallende Korrelationsfunktion nimmt folgende Form an:
\begin{equation}
  \label{102a}
  G( \mbox{\bf r} ) = \left<\rho_{\mbox{\bf \scriptsize G}} ( \mbox{\bf r} )^* \rho_{\mbox{\bf \scriptsize G}} ( \mbox{\bf 0})\right> \propto r^{- \eta_{\small G}}.
\end{equation}
$\left<...\right>$ steht f"ur eine thermische Mittelung. Der Exponent $\eta_{\small G}$ ist abh"angig von den Lam\'e-Koeffizienten $\lambda$, $\mu$, die Konstanten des Gitters sind 
\begin{equation}
  \label{103a}
  \eta_{\small G} = \frac{| {\mbox{\bf G}} |^2 T (3 \mu + \lambda)}{4 \pi \mu (2 \mu + \lambda)}.
\end{equation}
Separiert man die phononischen Systemanregungen von den Anregungen durch Versetzungen und transformiert den Anteil letzterer auf ein Coulombgas (Details werden in Kapitel 2 erl"autert), so erh"alt man f"ur die Energie eines Dipols
\begin{equation}
  \label{104a}
  E \propto J \left(\mbox{\bf b}_1 \mbox{\bf b}_2 \ln|{\mbox{\bf r}_1 - \mbox{\bf r}_2}| - \frac{\left(\mbox{\bf b}_1 \left(\mbox{\bf r}_1 - \mbox{\bf r}_2\right) \right) \left(\mbox{\bf b}_2 \left(\mbox{\bf r}_1 - \mbox{\bf r}_2\right) \right)} { |{\mbox{\bf r}_1 - \mbox{\bf r}_2}|^2} \right),
\end{equation}
J ist hierbei eine Kopplungskonstante, die von den Lam\'{e}-Koeffizienten des Gitters abh"angt: J = J($\lambda$, $\mu$).
 
Der Betrag der Burgers-Vektoren geht quadratisch in die Energie ein, es kann also in guter N"aherung angenommen werden, da"s nur Dipole mit Vektorbetr"agen $|\mbox{\bf b}| = a$ gebildet werden, da Dipole mit gr"o"seren Vektorbetr"agen demnach wesentlich unwahrscheinlicher sind.
Bei niedrigen Temperaturen ist nur eine geringe Anzahl an Versetzungsdipolen vorhanden, sie nimmt aber mit steigender Temperatur zu, wie auch die Dipolgr"o"se w"achst. Es kommt zu Abschirmungseffekten, denn f"ur Dipole mit weit auseinander liegenden Burgers-Vektoren wird die Kopplung durch dazwischen liegende kleinere Dipole reduziert. Dennoch bleibt in der Tieftemperaturphase die Kopplung auch im Limes $r \rightarrow \infty$ endlich ($r$: Dipolgr"o"se). 
 
In der {\bf Hochtemperaturphase} hingegen werden die Abschirmungseffekte so stark, da"s man f"ur $r \rightarrow \infty$ verschwindende Kopplung hat ($J \rightarrow 0$). Die Korrelationsfunktion f"allt hier exponentiell ab \verweistext{hal79}:
\begin{equation}
  \label{105a}
   G(r) \propto  e^{- \frac{r}{\xi}}.
\end{equation}
$\xi$ ist die Korrelationsl"ange. Oberhalb der "Ubergangstemperatur $T_m = J/16\pi$ l"osen sich die Versetzungspaare auf, in der Hoch\-temperatur\-phase liegen damit freie Versetzungen vor.
 
Der wesentliche Vorgang beim Kosterlitz-Thouless-Phasen"ubergang ist also das Dissoziieren der Versetzungsdipole. Es liegt keine kristalline Struktur mehr vor, da freie Versetzungen dazu f"uhren, da"s die Ordnung in der Hochtemperaturphase nicht mehr quasi-lang\-reich\-weitig, sondern kurzreichweitig ist. Die Korrelationsl"ange $\xi$ bei diesem Phasen"ubergang nimmt folgende Werte an:
\begin{eqnarray}
  \label{106a}
  \begin{array}{rclccc}
    \xi & \propto & \mbox{exp} \left[\left(\frac{b}{T - T_m}\right)^{\bar{\nu}}\right]    \hspace{3cm} & T & > & T_m \vspace{0.2cm} \\
    \xi & = & \infty & T & \le &T_m
  \end{array}
\end{eqnarray}
Hierbei ist $b$ eine materialabh"angige Konstante, der Exponent $\bar{\nu}$ nimmt auf dem Quadratgitter den Wert 0.5 und auf dem Dreiecksgitter den gen"aherten Wert 0.3696 an \verweistext{you79}.
 
Wie Kosterlitz in \verweistext{kos74} gezeigt hat, gilt f"ur den singul"aren Anteil der freien Energie
\begin{equation}
  \label{107a}
  F_{sing} \propto \xi^{-2}.
\end{equation} 
Mit \refformel{106a} folgt daraus, da"s sowohl die freie Energie als auch alle ihre Ableitungen bei $T \rightarrow T_m$ verschwindende singul"are Anteile haben und daher kontinuierlich sind. Somit ist der Kosterlitz-Thouless-Phasen"ubergang ein "Ubergang {\it unendlicher} Ordnung. Man kann die Phasen"ubergangstemperatur nur bestimmen, wenn man die Kopplung auf gro"sen L"angenskalen betrachtet, die bei $T_m$ verschwindet.

Eine besondere Eigenschaft des Phasen"ubergangs bei kristallinen Filmen ist, da"s in der Hochtemperaturphase zwar kurzreichweitige Ordnung bez"uglich der Translations-Kor\-rela\-tions\-funk\-tion vorliegt, aber eine Orientierungs-Kor\-rela\-tions\-funk\-tion mit einem Ordnungsparameter
\begin{eqnarray}
  \label{108a}
  \begin{array}{rclc}
  \psi(\mbox{\bf r}) & = & e^{4/\theta(\mbox{\bf \scriptsize r})}  \hspace{2cm}  & \mbox{f"ur Quadratgitter} \vspace{0.2cm} \\
  \psi(\mbox{\bf r}) & = & e^{6/\theta(\mbox{\bf \scriptsize r})}   & \mbox{f"ur Dreiecksgitter} 
  \end{array}
\end{eqnarray}
hier noch quasi-langreichweitige Ordnung aufweisen kann. $\theta(\mbox{\bf r})$mi"st die Bond-Orientierung relativ zu einer festen Referenzachse und ist eine Funktion, die alleine vom Verschiebungsfeld {\bf u(r)} abh"angt. Durch einen zweiten Kosterlitz-Thouless-"Ubergang bei $T_i > T_m$ erreicht man dann die isotrop fl"ussige Phase, in der sowohl bez"uglich Translation als auch bez"uglich Orientierung nur kurzreichweitige Ordnung vorliegt. Die Zwischenphase mit quasi-langreichweitiger Ordnung in der Orientierung, aber kurzreichweitige Ordnung bez"uglich der Translation wird im Fall des Quadratgitters ``tetratische'', im Fall des Dreiecksgitters ``hexatische'' Fl"ussigkristall-Phase genannt \verweistext{nh79}. Die hier auftretenden topologischen Defekte sind sogenannte Disklinationen: Auf einem geschlossenen Weg um eine Disklination in der hexatischen Phase "andert sich die Bond-Orientierung um ein Vielfaches von 60 Grad (in der tetratischen Phase 90 Grad). Unterhalb von $T_i$ liegen nur Disklinationspaare vor, die bei $T_i$ dissoziieren. Dabei entspricht ein Disklinationsdipol einer freien Versetzung, weswegen man auch Versetzungsdipole als eine Verbindung von vier Disklinationen betrachten kann. Abbildung 1.3 verdeutlicht die Phasenabfolge.
 
\graphik{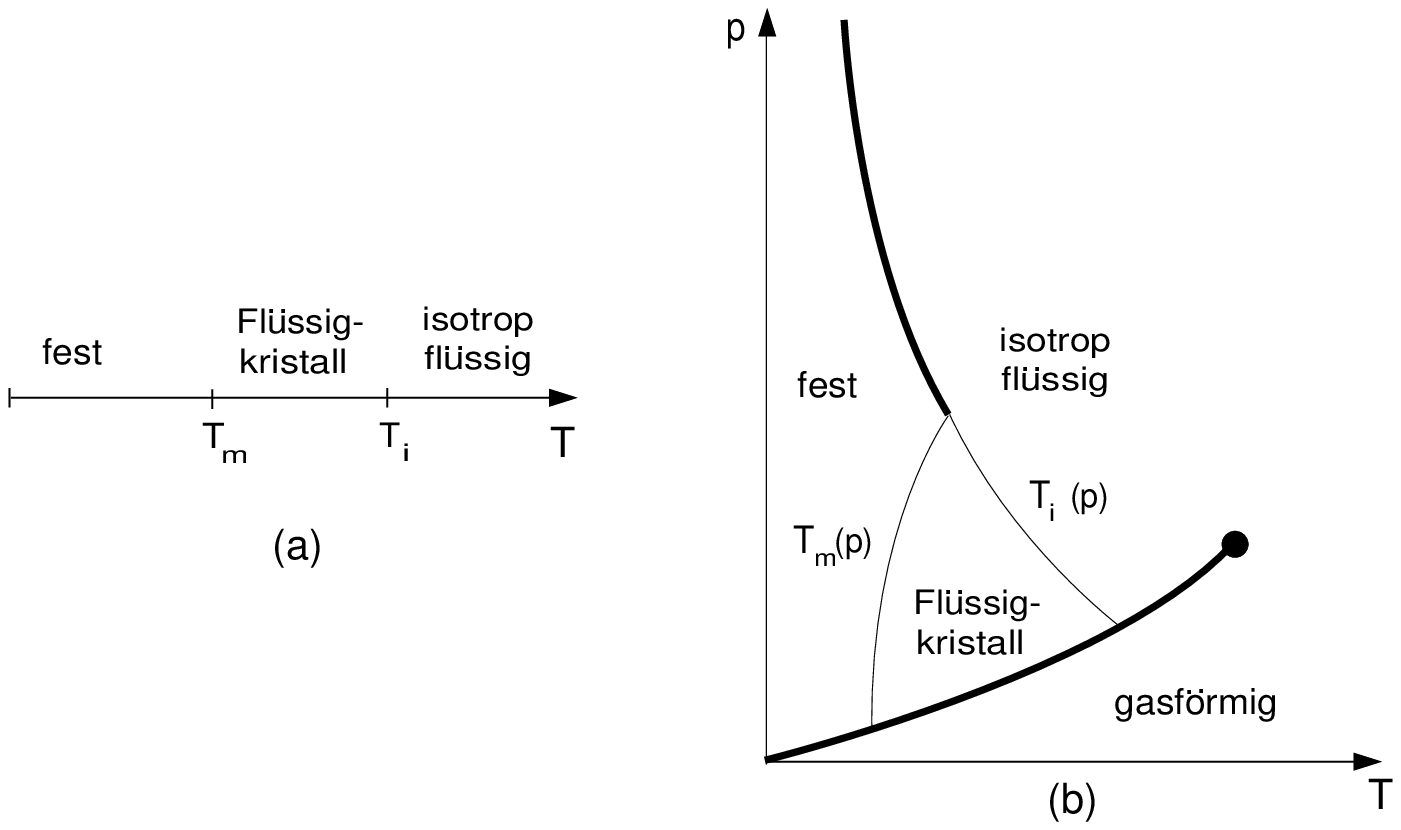}{6cm}{Phasendiagramm (a) des zweistufigen "Ubergangs von einem zweidimensionalen Kristall in eine isotrope Fl"ussigkeit, (b) zeigt ein spekulatives p-T-Phasendiagramm, in dem feste, fl"ussige, gasf"ormige und Fl"ussigkristall-Phase auftreten (nach \verweistext{nh79}). In (b) sind d"unne Linien Kosterlitz-Thouless-"Uberg"ange, massive Linien Phasen"uberg"ange erster Ordnung.}{103}
  
Der in Abbildung 1.3 eingezeichnete, rein spekulative Phasen"ubergang erster Ordnung zwischen fester und isotrop fl"ussiger Phase  ist im Rahmen der Kosterlitz-Thouless-Theorie dann m"oglich, wenn die Aufl"osung der Disklinationspaare rechnerisch vor der Aufl"osung der Versetzungspaare erfolgen soll. Dann fallen beide "Uberg"ange zusammen zu einem Phasen"ubergang erster Ordnung. 
 
Die folgende Tabelle veranschaulicht noch einmal die Phasenabfolge und das Verhalten der wesentlichen Gr"o"sen in den verschiedenen Phasen.
 
\begin{tabular}{|l||l|l|l|} \hline
& kristallin& tetratisch/hexatisch& isotrop fl"ussig \\ \hline \hline
& & & \\
\raisebox{2ex}{Versetzungen} & \raisebox{2ex}{gepaart} & \raisebox{2ex}{frei} & \raisebox{2ex}{frei} \\ \hline 
& & & \\
\raisebox{2ex}{Disklinationen} & \raisebox{2ex}{als Quartette} & \raisebox{2ex}{gepaart} & \raisebox{2ex}{frei} \\ \hline
\raisebox{-1ex}{Translations-} & \raisebox{-1ex}{quasi-lang-} &  &  \\
\raisebox{1ex}{Korrelation} & \raisebox{1ex}{reichweitig} & \raisebox{2ex}{kurzreichweitig} & \raisebox{2ex}{kurzreichweitig} \\ \hline
\raisebox{-1ex}{Orientierungs-} & & \raisebox{-1ex}{quasi-lang-} & \\ 
\raisebox{1ex}{Korrelation} & \raisebox{2ex}{langreichweitig} & \raisebox{1ex}{reichweitig} & \raisebox{2ex}{kurzreichweitig} \\ \hline
\end{tabular}

Abbildung 1.4 zeigt beispielhaft das beschriebene Schmelzen von der festen "uber die hexatische in die isotrop fl"ussige Phase am Beispiel eines Dreiecksgitters.
 
{\bf A separate downloading of this figure is possible.}

Abbildung 1.4: {\it Schmelzen eines Dreiecksgitters von der festen Phase (A) "uber die hexatische Phase (B, C) in die isotrop fl"ussige Phase (D). Schwarz eingef"arbte Bereiche markieren Versetzungen und Disklinationen (aus \verweistext{kus94}).}
 
In Abbildung 1.4(A) befindet sich das System noch im festen Zustand. Man findet hier vier Versetungspaare und vier ungebundene Versetzungen, die aber nahe am Rand der Darstellung liegen, deren Partner sich also au"serhalb des Darstellungsbereichs befinden. In (B) und (C) ist die Translations-Korrelation schon nicht mehr quasi-langreichweitig, w"ahrend dieses Verhalten f"ur die Orientierungs-Kor\-re\-la\-tion noch vorzufinden ist. Hier befindet sich das System in der hexatischen Phase, in der freie Versetzungen und gebundene Disklinationen vorliegen. In der isotrop fl"ussigen Phase (D) existiert keinerlei Ordnung mehr. 
 
Im folgenden soll aber nur auf den ersten Phasen"ubergang, also auf das Schmelzen der festen Phase eingegangen werden.
 
Die Frage, wann die Kosterlitz-Thouless-Theorie anwendbar ist und wann ein Phasen"ubergang erster Ordnung auftritt, der nicht auf einem Defektschmelzen beruht, ist noch nicht v"ollig gekl"art \verweistext{pok1}. Allerdings ist anzunehmen, da"s der im weiteren untersuchte Kosterlitz-Thouless-"Ubergang nur dann auftritt, wenn die hierf"ur notwendigen Bedingungen (beispielsweise eine kleine Versetzungskonzentration) erf"ullt sind. Starke Anharmonizit"aten wie eine starke Wechselwirkung mit einem Substrat k"onnen eine weitere Quelle f"ur einen Schmelz"ubergang erster Ordnung sein.  

Unter realen Bedingungen mu"s ein kristalliner Film immer auf ein Substrat aufgebracht werden. Gleichung \refformel{104a} gilt in obiger Form nur f"ur glatte Substrate. In Kapitel 5 wird versucht werden, den Substrateinflu"s abzusch"atzen. 

\section{Der Einflu"s von Unordnung}
Dem kristallinen Film kann zus"atzlich Unordnung in Form von eingefrorenen, im System nicht beweglichen, Verunreinigungen hinzugef"ugt werden. Realisierbar ist dies mit Fremdpartikeln, die Kristallpl"atze besetzen und so Verzerrungen der Kristallstruktur hervorrufen. Da die Verteilung der Verunreinigungen in realen Systemen als zuf"allig angenommen werden kann, ist auch die lokale Konzentration der Verunreinigungen zuf"allig um einen Mittelwert verteilt. Mi"st die Funktion $\delta c(\mbox{\bf r})$ die Abweichungen von diesem Mittelwert, so kann man annehmen, da"s eine gegebene Konfiguration $\delta c(\mbox{\bf r})$ mit der Wahrscheinlichkeit
\begin{equation}
  \label{110a}
  {\cal P}\left[\delta c(\mbox{\bf r}) \right] \propto \mbox{exp}\left(-\frac{1}{2\sigma} \int d^2 r \left[\delta c(\mbox{\bf r}) \right]^2 \right)
\end{equation}
auftritt \verweistext{nel83}. Somit sind die lokalen Abweichungen $\delta c(\mbox{\bf r})$ unkorreliert und gau"sverteilt mit der Varianz $\sigma$.
\begin{equation}
  \label{111a}
  \left[\delta c(\mbox{\bf r})\delta c(\mbox{\bf r'})\right]_D = \sigma \delta(\mbox{\bf r}-\mbox{\bf r'}).
\end{equation}
Hierbei bedeutet $[...]_D$ eine Mittelung "uber alle m"oglichen Unordnungskonfigurationen. Die St"arke der Unordnung ist gleichzusetzen mit der St"arke der Fluktuationen und wird durch die Varianz $\sigma$ ausgedr"uckt, die einer eingefrorenen Temperatur entspricht.

Das Schmelzen von solchen kristallinen Filmen mit eingefrorener Unordnung wurde bereits 1983 von Nelson untersucht \verweistext{nel83}. Abbildung 1.5 gibt das von ihm gefundene Phasendiagramm wieder. Auff"allig ist, da"s die quasi-langreichweitige Ordnung, die nur unterhalb einer kritischen Un\-ord\-nungs\-st"arke $\sigma_c$ existiert, f"ur niedrige Temperaturen $T<T_-(\sigma)$ wieder zerst"ort wird. Es findet demnach ein Wiedereintritt in die ungeordnete Fl"ussigkristall-Phase (hexatische, tetratische Phase) statt. 
 
\graphik{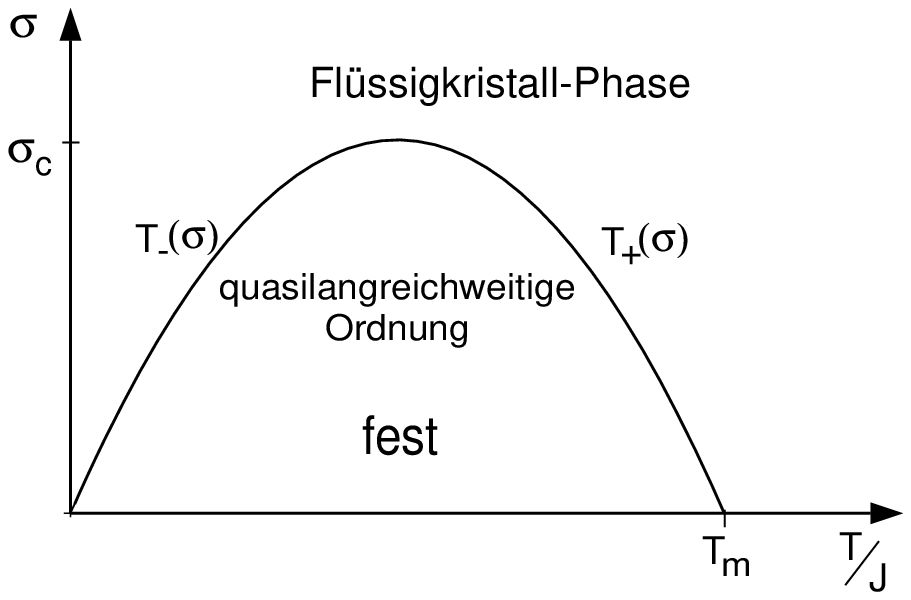}{4cm}{Phasendiagramm f"ur kristalline Filme mit Unordnung nach Nelson 1983.}{105}
 
Ein analoges Phasendiagramm haben Rubinstein, Shraiman und Nelson 1983 auch f"ur das physikalisch verwandte XY-Modell mit Phasenfrustration gefunden \verweistext{rsn83}. Simulationen von Josephson-Gittern zeigten allerdings keinen Wiedereintritt in die ungeordnete Phase \verweistext{fbl90, cd88}. Ozeki und Nishimori konnten schlie"slich zeigen, da"s ein solcher Wiedereintritt bei den vorliegenden Systemen unm"oglich ist \verweistext{on93}.
 
Neuere Arbeiten von Nattermann, Scheidl, Korshunov, Li und Tang \verweistext{nat95, kn96, tang96, sts97} finden dann auch einen anderen Verlauf der Phasengrenze f"ur das XY-Modell mit Phasenfrustration. Sie argumentieren, da"s der in \verweistext{rsn83} gefundene Wiedereintritt in die ungeordnete Phase nur ein Artefakt der N"aherung f"ur kleine Vortexfugazit"aten ist, die im Temperaturbereich $T<T^*$ falsch wird, da aufgrund des Unordnungspotentails hier hohe Vortexdichten auftreten k"onnen und somit die Wechselwirkung zwischen den Vortexdipolen wichtig wird. Eine Neubetrachtung liefert im Temperaturbereich $T<T^*$ eine der Temperaturachse parallele Phasengrenze bei $\sigma_c$, unterhalb der die quasi-langreichweitige Ordnung auch bei beliebig kleinen Temperaturen erhalten bleibt (vgl. Abbildung 1.6). Neuere Simulationen konnten dieses Ergebnis best"atigen \verweistext{mg97}.
 
\graphik{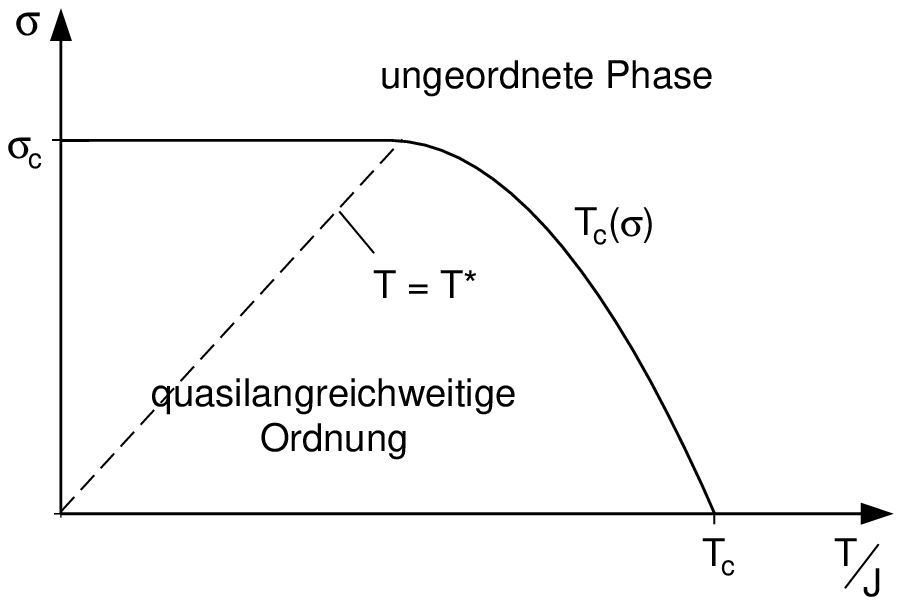}{4cm}{Modifiziertes Phasendiagramm f"ur das XY-Modell mit Phasenfrustration.}{106}
  
Aufbauend auf den in \verweistext{nat95, kn96, tang96, sts97} entwickelten Methoden sollen in dieser Arbeit modifizierte Flu"sgleichungen und Phasendiagramme auch f"ur kristalline Filme mit eingefrorener Unordnung gefunden werden.

\chapter{Das Modell}
\section{Herleitung des Modells}
In diesem Abschnitt wird ein Modell f"ur zweidimensionale kristalline Filme mit eingefrorenen Verunreinigungen hergeleitet. Mit Hilfe der Elasitizit"atstheorie, die die Mechanik fester K"orper beschreibt, erh"alt man eine Hamiltonfunktion f"ur das System. Hierbei werden die Verunreinigungen als lokale Quellen von Verzerrungen behandelt. Solche Verunreinigungen k"onnen beispielsweise Unebenheiten des Substrats, oder aber Fremdelemente im Gitter des kristallinen Films sein. Es handelt sich also um positionelle oder kompositionelle Unordnung. Abbildung 2.1 zeigt ein einzelnes gr"o"seres Atom in einem Gitter gebildet aus kleineren Partikeln. Wie man unschwer erkennt, resultieren Dehnungen und Kompressionen des Gitters. 
 
\graphik{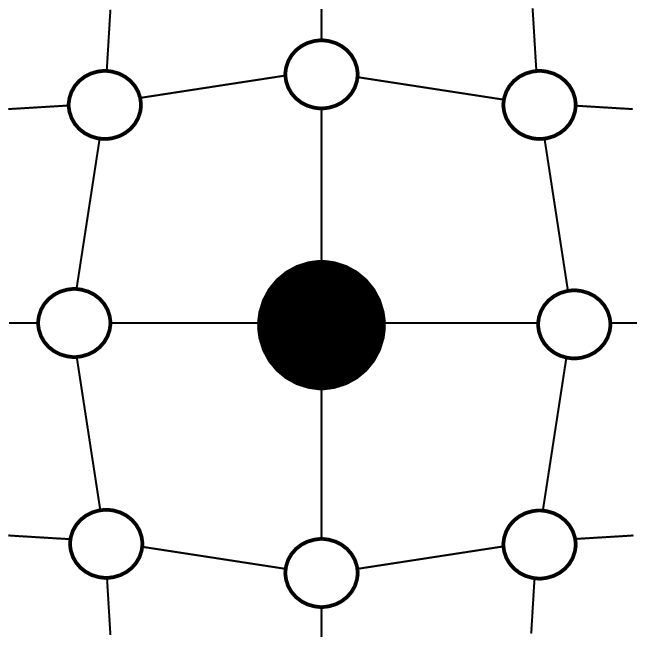}{7cm}{Quadratgitter mit eingebrachter Verunreinigung, die zu Verzerrungen f"uhrt \verweistext{nel83}.}{201}
 
Eine solche Verunreinigung soll eingefroren sein, was hei"st, da"s sie zwar durch langwellige Phononen angeregt werden kann, aber dennoch nicht in der Lage ist, mit anderen Gitterbausteinen den Platz zu wechseln.  Dies erfordet kleine Verunreinigungs-Diffusionskonstanten bei tiefen Temperaturen. Die Versetzungen hingegen sollen sich im thermischen Gleichgewicht befinden, welches sie mittels Gleiten durch den Kristall erreichen. Wir betrachten also Messungen nach Zeiten, die gr"o"ser sind als die Relaxationszeiten der Versetzungen, die aber wesentlich kleiner sind als die Relaxationszeiten der Verunreinigungen, deren Freiheitsgrade eingefroren sind und deren Konfiguration w"ahrend der Messung somit unver"andert bleibt. 
 
\subsection{Kristalliner Film ohne Unordnung}
Der zweidimensionale Kristall wird unter der Einwirkung innerer und "au"serer Kr"afte bis zu einem gewissen Grade deformiert. So erh"ohen auch Versetzungen die elastische Energie des Kristalls. In der "ublichen Kontinuumsn"aherung \verweistext{friedel, lanlif} wird die Deformation durch den Verschiebungsvektor {\bf u} dargestellt. Ist {\bf r} die Position eines Kristallpunktes ohne Deformation und {\bf r'} die Position mit Deformation, so gilt
\begin{equation}
  \label{201a}
  \mbox{\bf u} = \mbox{\bf r'}-\mbox{\bf r}.
\end{equation}
{\bf u(r)} ist damit das Verschiebungsfeld. Betrachtet man den Abstand zweier beliebiger, infinitesimal voneinander entfernten Punkte vor der Deformation d$l$ und nach der Deformation d$l'$, dann hat man mit Summenkonvention
\begin{eqnarray}
  \label{202a}
  \begin{array}{rcl}
    \mbox{d}{l'}^2 & = & \mbox{d}{x'}_i^2 = (\mbox{d}x_i + \mbox{d}u_i)^2 \vspace{0.2cm} \\
    \mbox{} & = & \mbox{d}l^2 + 2u_{ik} \mbox{d}x_i \mbox{d}x_k 
  \end{array}
\end{eqnarray}
dabei ist
\begin{equation}
  \label{203a}
    u_{ik}(\mbox{\bf r}) = \frac{1}{2} \left(\frac{\partial u_i({\bf r})}{\partial x_k} + \frac{\partial u_k({\bf r})}{\partial x_i} + \frac{\partial u_l({\bf r})}{\partial x_i} \frac{\partial u_l({\bf r})}{\partial x_k} \right).
\end{equation}
Der Tensor $u_{ik}$ ist der symmetrische {\it Verzerrungstensor}. Die auftretenden Deformationen sind stets klein in dem Sinne, da"s die "Anderung eines Abstandes im Kristall immer kleiner ist als der Abstand selbst. Daraus resultierend sind auch die Verschiebungen $u_i$ und deren Ableitungen immer klein, so da"s es gen"ugt im Verzerrungstensor nur Ableitungen erster Ordnung zu ber"ucksichtigen.
\begin{equation}
  \label{204a}
  u_{ik}(\mbox{\bf r}) = \frac{1}{2} \left(\frac{\partial u_i({\bf r})}{\partial x_k} + \frac{\partial u_k({\bf r})}{\partial x_i} \right).
\end{equation}
 
Da eine konstante Verschiebung keine "Anderung der Energie im Kristall hervorruft und da der Verzerrungtensor klein ist, kann man die Hamiltonfunktion nach dem Verzerrungstensor entwickeln. Asymmetrische Beitr"age niedriger Ordnung kann die Hamiltonfunktion nicht enthalten, wenn sie rotationsinvariant bleiben soll. Ferner kann die Entwicklung keine in $u_{ik}$ linearen Glieder enthalten, soll doch f"ur $u_{ik} = 0$ ein Minimum vorliegen. Somit ist die allgemeinste Form, die die Hamiltonfunktion in niedrigster Ordnung annehmen kann:
\begin{equation}
  \label{205a}
  H = \tilde{H} + \int d^2r K_{ijkl} u_{ij} u_{kl}
\end{equation}
Betrachtet man nur den von der Deformation abh"angigen {\it elastischen} Beitrag in $H$ und ber"ucksichtigt die in einem kontinuierlichen System vorliegenden Invarianzen, so bleibt
 \begin{equation}
  \label{206a}
  H_{el} = \frac{1}{2} \int d^2r \left( 2\mu u_{ik}^2(\mbox{\bf r}) + \lambda u_{ii}^2(\mbox{\bf r}) \right).
\end{equation}
mit den beiden unabh"angigen {\it Lam\'{e}-Koeffizienten} $\lambda$ und $\mu$.
Wie Landau und Lifshitz \verweistext{lanlif} gezeigt haben, gilt diese Kontinuumsform des Hamiltonians exakt auch f"ur das zweidimensionale Dreiecksgitter. F"ur das Quadratgitter findet man drei unabh"angige Koeffizienten, die obige Form der Hamiltonfunktion kann aber als N"aherung benutzt werden. 
 
Da die inneren Kr"afte einer Kristallfl"ache durch von au"sen, "uber den Fl"achenrand wirkende Kr"afte kompensiert werden, mu"s f"ur jede Komponente des Kraftvektors gelten
\begin{equation}
  \label{206b}
  F_i = \int_{\Omega}d^2r f_i(\mbox{\bf r}) = \int_{\partial \Omega} dx_j \sigma_{ij}(\mbox{\bf r}).
\end{equation}   
Nach den Integrationsregeln ist hierbei $f_i$ die Divergenz des symmetrischen {\it Spannungstensors} $\sigma_{ij}$
\begin{equation}
  \label{207a}
  f_i = \frac{\partial \sigma_{ik}}{\partial x_k}. 
\end{equation}
 
Betrachtet man nun eine kleine "Anderung des Verschiebungfelds {\bf u(r)} um den Wert $\delta${\bf u(r)}, so ist die hieraus resultierende Arbeit:
\begin{equation}
  \label{208a}
  \delta W = \int d^2r f_i \delta u_i
\end{equation}
Partielle Integration und Ausnutzung der Symmetrie des Tensors liefert
\begin{equation}
  \label{209a}
  \delta W = - \frac{1}{2} \int d^2r \sigma_{ik}(\mbox{\bf r}) \delta \left(\frac{\partial u_i({\bf r})}{\partial x_k} + \frac{\partial u_k({\bf r})}{\partial x_i} \right)
\end{equation}
oder
\begin{equation}
   \label{210a}
   \delta W = - \sigma_{ik}(\mbox{\bf r}) \delta u_{ik}(\mbox{\bf r}).
\end{equation}
 
Der Spannungstensor ist also gegeben durch
\begin{equation}
   \label{211a}
   \sigma_{ik} = \left( \frac{\partial E_{el}}{\partial u_{ik}} \right)_S,
\end{equation}
was ein weiteres Argument f"ur die Nicht-Existenz von linearen Gliedern des Verzerrungstensors in der Hamiltonfunktion liefert, k"onnten diese doch bei $u_{ik} = 0$ zu einer Spannung $\sigma_{ik} \neq 0$ f"uhren. Man erh"alt nun f"ur die Hamiltonfunktion  den folgenden Ausdruck \verweistext{kt73}:
\begin{equation}
   \label{212a}
   H_{el} = \frac{1}{2} \int d^2r \sigma_{ik}(\mbox{\bf r}) u_{ik}(\mbox{\bf r}),
\end{equation}
mit (wegen \refformel{206a})
\begin{equation}
   \label{213a}
   \sigma_{ik}(\mbox{\bf r}) = 2 \mu u_{ik}(\mbox{\bf r}) + \lambda \delta_{ik} u_{jj}(\mbox{\bf r}).
\end{equation}
 
Diese Hamiltonfunktion, die das System ohne den Einflu"s von Unordnung beschreibt, wurde bereits in mehreren Arbeiten ausf"uhrlich diskutiert \verweistext{you79, nh79, nel78}.

\subsection{Beschreibung der Unordnung}
Wie Abbildung 2.1 zeigt, sind die im Rahmen dieses Modells betrachteten Verunreinigungen fest lokalisierte Quellen von  Dehnungen oder Kompressionen des Gitters, die sph"arische Symmetrie bez"uglich Verunreinigungspositionen ${\bf r_I}$ aufweisen. Da das durch eine Verunreinigung erzeugte Verschiebungsfeld ${\bf u^{imp}(r)}$ ferner der Divergenzbedingung ${\bf \nabla u^{imp}(r)} \propto \delta({\bf r} - {\bf r_I})$ gen"ugen mu"s, hat es in Analogie zur Elektrodynamik in zwei Dimensionen die Form 
\begin{equation}
   \label{214a}
   {\bf u^{imp}(r)} = \gamma \frac{{\bf r}}{r^2} = \gamma {\bf \nabla} \ln |{\bf r}|.
\end{equation}
Dabei mi"st die Konstante $\gamma$ die ``St"arke'' des Defekts. Eine solche Verunreinigung bewirkt eine "Anderung $\Omega_0$ der Kristallfilmfl"ache, die der Defektst"arke proportional sein mu"s und eine positiven oder negativen Wert annehmen kann. Nach \verweistext{esh56} ist 
\begin{equation}
   \label{215a}
   \Omega_0 = 2 \pi \gamma \frac{2\mu + \lambda}{\mu + \lambda}.
\end{equation}  
 
Die hieraus resultierenden K"orperkr"afte lassen sich mit Hilfe von \refformel{207a} und \refformel{213a} berechnen.
\begin{eqnarray}
  \label{216a}
  \begin{array}{rcl}
    {\bf f(r)} & = & \mu {\bf \nabla}^2 {\bf u^{imp}(r)} + (\lambda + \mu) {\bf \nabla} ({\bf \nabla}{\bf u^{imp}(r)}) \vspace{0.2cm} \\
    \mbox{} & = & \Omega_0 {\bf \nabla} \delta({\bf r} - {\bf r_i}) (\mu + \lambda).
  \end{array}
\end{eqnarray}
 
Die Wechselwirkungsenergie einer Verunreinigung mit einem externen Verschiebungsfeld {\bf u(r)} erh"alt man hieraus mittels \refformel{208a}. Demnach nimmt die Hamiltonfunktion f"ur die Wechselwirkung zwischen der Unordnung und dem Verschiebungsfeld des Systems die folgende Form an \verweistext{nel83, kosv79} 
\begin{equation}
   \label{219a}
   H_{D} = - \Omega_0 (\mu + \lambda) \int d^2 r c({\bf r}) u_{kk}({\bf r}),
\end{equation}  
wobei $c({\bf r}) = \sum_i \delta({\bf r} - {\bf r_i})$ "uber alle Verunreinigungspositionen summiert. Auf gen"ugend gro"sen Wellenl"angen kann man die Diskretheit vernachl"assigen und c({\bf r}) als glatte Funktion behandeln. Die lokale Konzentration der Verunreinigungen ist nun zuf"allig um einen Mittelwert $c_0 = [c({\bf r})]_D$ verteilt. Es gilt
\begin{equation}
   \label{220a}
   c({\bf r}) = c_0 +  \delta c({\bf r}).
\end{equation}  
Das Verhalten von $\delta c({\bf r})$ ist bereits in \refformel{110a} und \refformel{111a} beschrieben worden. Der zu $c_0$ proportionale Term von $H_D$ verh"alt sich wie ein hydrostatischer Druck, der durch eine Dehnung des Gitters kompensiert werden kann \verweistext{nel83}. Somit ist nur der von  $\delta c({\bf r})$ abh"angige Anteil des Hamiltonians von Interesse.
 
Die Gesamthamiltonfunktion des Systems mit Unordnung lautet demnach
\begin{eqnarray}
  \label{221a}
  \begin{array}{rcl}
   H & = & H_{el} + H_D  \vspace{0.2cm} \\
   \mbox{} & = & \frac{1}{2} \int d^2 r [2\mu u_{ik}^{2}({\bf r}) + \lambda u_{jj}^{2}({\bf r})] - \Omega_0 (\mu + \lambda) \int d^2 r \delta c({\bf r}) u_{jj}({\bf r}).
  \end{array}
\end{eqnarray}
 
Ziel des n"achsten Abschnitts ist es, diese Hamiltonfunktion auf eine Coulomb\-gas-Beschreibung  abzubilden.

\section{Abbildung auf ein Coulombgas}
Das Verschiebungsfeld {\bf u(r)} kann in zwei Anteile zerlegt werden
\begin{equation}
   \label{230a}
   {\bf u(r)} = {\bf v(r)} + {\bf u^*(r)}.
\end{equation}
 
{\bf v(r)} stellt den Verschiebungsanteil, der durch Phononen hervorgerufen wird dar. Dieser Anteil bildet keine topologischen Defekte aus. F"ur alle geschlossenen Wege $\gamma$ im zweidimensionalen Film gilt
\begin{equation}
   \label{231a}
   \oint_{\gamma} {\bf dv(r)} = 0.
\end{equation}
 
Aus diesem Grund variiert der phononische Verzerrungstensor $v_{ik}({\bf r})$ analytisch, ohne Singularit"aten aufzuweisen.
 
Die Verschiebung aufgrund von Versetzungen ${\bf u^*(r)}$ hingegen erf"ullt f"ur geschlossene Wege $\gamma$, die eine Versetzung mit Burgers-Vektor {\bf b} einschlie"sen, entsprechend Abbildung 1.1 die Bedingung 
\begin{equation}
   \label{232a}
   \oint_{\gamma} {\bf du^*(r)} = {\bf b}.
\end{equation}
 
Ferner w"ahlen wir ${\bf u^*(r)}$ so, da"s ein System f"ur welches ${\bf u(r)} = {\bf u^*(r)}$ gilt, sich bei gegebener Versetzungskonfiguration in einem lokalen Energieminimum befindet. Somit d"urfen in einem solchen System keine inneren Kr"afte vorliegen. Wegen \refformel{207a} mu"s die Gleichgewichtsbedingung
\begin{equation}
   \label{232b}
   \frac{\partial \sigma_{ik}^*}{\partial x_k} = 0,
\end{equation}
"uberall, au"ser in den Versetzungskernen, erf"ullt sein. Hierbei ist
\begin{equation}
   \label{234a}
   \sigma_{ik}^*(\mbox{\bf r}) = 2 \mu u_{ik}^*(\mbox{\bf r}) + \lambda \delta_{ik} u_{jj}^*(\mbox{\bf r}).
\end{equation}
 
Setzt man diese Zerlegung in die Hamiltonfunktion ein, so erh"alt man einen phononischen und einen Versetzungsanteil, sowie einen Term der die Wechselwirkung zwischen Phononen und Versetzungen beschreibt
\begin{equation}
   \label{232c}
   H_{el} = H_{phon} + H_{Ver} + H_{WW},
\end{equation}
wobei
\begin{eqnarray}
  \label{232d}
  \begin{array}{lcl}
   H_{phon} & = & \frac{1}{2} \int d^2r \left( 2\mu v_{ik}^2({\bf r}) + \lambda v_{ii}^2({\bf r}) \right) \\
\\
   H_{Ver} & = & \frac{1}{2} \int d^2r \left( 2\mu {u_{ik}^*}^2({\bf r}) + \lambda {u_{ii}^*}^2({\bf r}) \right) \\
\\
   H_{WW} & = & \frac{1}{2} \int d^2r \left( 2\mu v_{ik}({\bf r}) u_{ik}^*({\bf r}) + \lambda v_{ii}({\bf r}) u_{jj}^*({\bf r})\right).
  \end{array}
\end{eqnarray}
 
Der Wechselwirkungsanteil nimmt nach explizitem Einsetzen des Tensors $v_{ik}$ folgende Form an
\begin{equation}
   \label{232e}
   H_{WW} = \frac{1}{2} \int d^2r \left(2 \mu \left(\frac{\partial v_i}{\partial x_k} + \frac{\partial v_k}{\partial x_i} \right) u_{ik}^* + 2 \lambda \frac{\partial v_i}{\partial x_i} u_{jj}^* \right).
\end{equation}
Dieser Ausdruck kann partiell integriert werden, wobei keine Randterme auftreten, da wir annehmen, da"s das Hinzuf"ugen der Versetzungen die Form des Systemrandes nicht ver"andert. Daher
\begin{equation}
\begin{split}
   \label{232f}
   H_{WW} & = -\frac{1}{2} \int d^2r \left[2 \mu \left(v_i \frac{\partial u_{ik}^*}{\partial x_k} + v_k \frac{\partial u_{ik}^*}{\partial x_i} \right) + \lambda \left(v_i \frac{\partial u_{jj}^*}{\partial x_i} + v_k \frac{\partial u_{jj}^*}{\partial x_k} \right) \right]\\
\\
   & = -\frac{1}{2} \int d^2r \left(v_i \frac{\partial \sigma_{ik}^*}{\partial x_k} + v_k \frac{\partial \sigma_{ki}^*}{\partial x_i} \right).
\end{split}
\end{equation}
Aufgrund von \refformel{232b} ergibt sich $H_{WW} = 0$, die Hamiltonfunktion zerf"allt also in einen rein phononischen Anteil und einen Anteil, der nur die Wechselwirkung der Versetzungen untereinander beschreibt. Da wir im folgenden am Verhalten der Versetzungen interessiert sind, betrachten wir nur $H_{Ver}$.

\subsection{Abbildung des Systems ohne Unordnung}
F"ur ein System ohne Unordnung nimmt die zu untersuchende Hamiltonfunktion nach \refformel{212a} folgende Form an
\begin{equation}
   \label{233a}
   H = \frac{1}{2} \int d^2r \sigma_{ik}^*(\mbox{\bf r}) u_{ik}^*(\mbox{\bf r}).
\end{equation}

Wegen \refformel{232b} gelten die folgenden beiden Gleichungen "uberall, au"ser in den Versetzungskernen
\begin{equation}
\begin{split}
   \label{235a}
   & \frac{\partial \sigma_{xx}^*}{\partial x} + \frac{\partial \sigma_{xy}^*}{\partial y} = 0 \\
   & \frac{\partial \sigma_{yx}^*}{\partial x} + \frac{\partial \sigma_{yy}^*}{\partial y} = 0
\end{split}
\end{equation}
Unter Ber"ucksichtigung der {\it Symmetrie} von $\sigma_{ik}^*$, ist die folgende Darstellung von $\sigma_{ik}^*$ die allgemeine Form der L"osung dieser beiden Gleichungen \verweistext{lanlif}:
\begin{equation}
   \label{236a}
   \sigma_{ik}^*({\bf r}) = \epsilon_{ij} \epsilon_{kl} \nabla_j \nabla_l \chi({\bf r})
\end{equation}
Die Funktion $\chi$ nennt man {\it Verzerrungsfunktion} und hat im Potential $\Phi$ ihr elektrodynamisches Analogon. Aufgrund der Beziehung zwischen $\sigma_{ik}^*$ und dem Verzerrungstensor $u_{ik}^*$ findet man
\begin{equation}
   \label{237a}
   u_{ik}^*({\bf r}) = \frac{1}{2\mu} \epsilon_{ij} \epsilon_{kl} \nabla_j \nabla_l \chi({\bf r}) - \frac{\lambda}{4 \mu (\lambda + \mu)} \nabla^2 \chi({\bf r}) \delta_{ik}.  
\end{equation}
 
F"uhrt man eine zweidimensionale vektorielle Versetzungsdichte ${\bf b(r)} = \\ \sum_{\alpha} {\bf b_{\alpha}} \delta({\bf r} - {\bf r_{\alpha}})$ ein, welche "uber alle Versetzungspositionen ${\bf r_{\alpha}}$ summiert, so l"a"st sich \refformel{232a} schreiben als
\begin{equation}
   \label{238a}
   \oint_{\gamma} du_{i}^*({\bf r}) = \oint_{\gamma} w_{ji}({\bf r}) dx_j = \int \epsilon_{kj} \nabla_k w_{ji}({\bf r}) d^2r = \sum_{\alpha} b_{\alpha i}.
\end{equation}
Hierbei ist $w_{ji} = \nabla_j u_{i}^*$, und es wird nur "uber die Burgers-Vektoren ${\bf b_{\alpha}}$ summiert, die von dem Weg $\gamma$ eingeschlossen werden. Dr"uckt man diese Gleichung mittels der Versetzungsdichte aus, so resultiert
\begin{equation}
   \label{239a}
   \epsilon_{kj} \nabla_k w_{ji}({\bf r}) = b_i({\bf r}).
\end{equation} 
 
Wendet man nun $\epsilon_{ij} \epsilon_{kl} \nabla_j \nabla_l$ auf beide Seiten der Gleichung \refformel{237a} an, so ergibt sich mit dem Kopplungskoeffizienten $J = 4 \mu (\mu + \lambda) a^2/(2 \mu + \lambda)$ \verweistext{friedel, chlu} 
\begin{eqnarray}
  \label{240a}
  \begin{array}{rcl}
   \frac{1}{J} {\bf \nabla}^4 \chi({\bf r}) & = & \frac{1}{2} \epsilon_{ij} \epsilon_{kl} \nabla_j \nabla_l (w_{ik} + w_{ki}) \vspace{0.2cm} \\
   \mbox{} & = & \frac{1}{2} \epsilon_{ij} \epsilon_{kl} \nabla_j \nabla_l (w_{ik} - w_{ki}) + \epsilon_{ij} \epsilon_{kl} \nabla_j \nabla_l w_{ki} \vspace{0.2cm} \\  
   \mbox{} & = & \epsilon_{ji} \nabla_j b_i({\bf r}) =: \eta({\bf r}).
  \end{array}
\end{eqnarray}
Der erste Term der zweiten Zeile entf"allt, da wir ein System ohne freie Disklinationen betrachten. Die oben definierte Funktion $\eta({\bf r})$ nennt sich {\it Quellenfunktion} und beschreibt die Verteilung der Versetzungen. Sie entspricht der elektrodynamischen Ladungsdichte $\rho$. Der Vergleich mit der Elektrodynamik, wo ${\bf \nabla}^2 \Phi({\bf r}) = - 4 \pi \rho({\bf r})$ gilt, w"ahrend hier ${\bf \nabla}^4 \chi({\bf r}) = J \eta({\bf r})$ ist, zeigt, da"s hier doppelt so oft abgeleitet wird. Dies ist notwendig, um den Vektorcharakter der betracheten Ladungen zu ber"ucksichtigen.  
 
Gel"ost wird ${\bf \nabla}^4 \chi({\bf r}) = J \eta({\bf r})$ f"ur $|{\bf r}| \gg a$ (die Gitterkonstante $a$ wird im folgenden bis einschlie"slich Kapitel 4 immer gleich 1 gesetzt, gefordert ist also $|{\bf r}| \gg 1$), mit Hilfe der Green-Funktion
\begin{equation}
   \label{241a}
   g({\bf r}) = \frac{1}{8 \pi} r^2 \left( \ln |{\bf r}| + C \right), 
\end{equation} 
die
\begin{equation}
   \label{242a}
   {\bf \nabla}^4 g({\bf r}) = \frac{1}{2 \pi}{\bf \nabla}^2 \left( \ln |{\bf r}| + C + 1 \right) = \delta({\bf r}) 
\end{equation}
erf"ullt. Denn setzt man f"ur die Verzerrungsfunktion
\begin{equation}
   \label{243a}
   \chi({\bf r}) = J \int d^2r' \eta({\bf r'}) g({\bf r} - {\bf r'})
\end{equation}
an, so gilt \refformel{240a}.
 
\refformel{233a} liefert mit \refformel{236a} und \refformel{237a} nach  zweimaliger partieller Integration folgenden Ausdruck f"ur die elastische Hamiltonfunktion \verweistext{kt73}
\begin{equation}
   \label{244a}
   H_{el} = \frac{1}{2} \int d^2r \eta({\bf r})\chi({\bf r}). 
\end{equation}
Die nicht ber"ucksichtigten Randterme liefern Beitr"age die proportional zu $\ln R$ sind, also mit der Systemgr"o"se divergieren. Sie entfallen allerdings, wenn das System Ladungsneutralit"at aufweist. Um die Energie endlich zu halten, mu"s daher
\begin{equation}
   \label{244b}
   \sum_{\alpha} {\bf b_{\alpha}} = 0.
\end{equation}

erf"ullt sein. 
 
Eine etwas aufwendigere Rechnung unter Verwendung der Ergebnisse f"ur $\chi$ und $\eta$ (vgl. Anhang A.1) f"uhrt schlie"slich zu der gesuchten Coulombgas-Darstellung der Hamiltonfunktion f"ur das Problem ohne Unordnung (${\bf r_{\alpha \beta}} := {\bf r_{\alpha}} - {\bf r_{\beta}}$) \verweistext{nh79, chlu}.
\begin{equation}
   \label{245a}
   H_{el} = - \frac{J}{8 \pi} \sum_{\alpha \neq \beta} \left({\bf b_{\alpha} b_{\beta}} \ln |{\bf r_{\alpha \beta}}| - \frac{({\bf b_{\alpha} r_{\alpha \beta}}) ({\bf b_{\beta} r_{\alpha \beta}})}{r_{\alpha \beta}^2} \right) + E_C \sum _{\alpha} |{\bf b_{\alpha}}|^2
\end{equation}
 
Auff"allig ist hierbei der von der "ublichen Elektrodynamik abweichende zweite Term, der sich aus dem Vektorcharakter der hier betrachteten Ladungen ergibt. Die Energie ist nicht nur vom Abstand der Ladungen voneinander, sondern auch von deren Stellung zueinander abh"angig. Dieser zweite Term nimmt beispielsweise f"ur ein Paar aus antiparallelen Burgers-Vektoren gleichen Betrags die Form $- \frac{J}{4\pi} b^2 \cos^2\theta$ an (vgl. Abbildung 2.2). Hier ist also bei gleichem Abstand ein Paar, bei dem die Burgers-Vektoren parallel zum Abstandsvektor stehen energetisch g"unstiger, als ein Paar, bei dem die Burgers-Vektoren senkrecht zum Abstandsvektor sind. 
 
\graphik{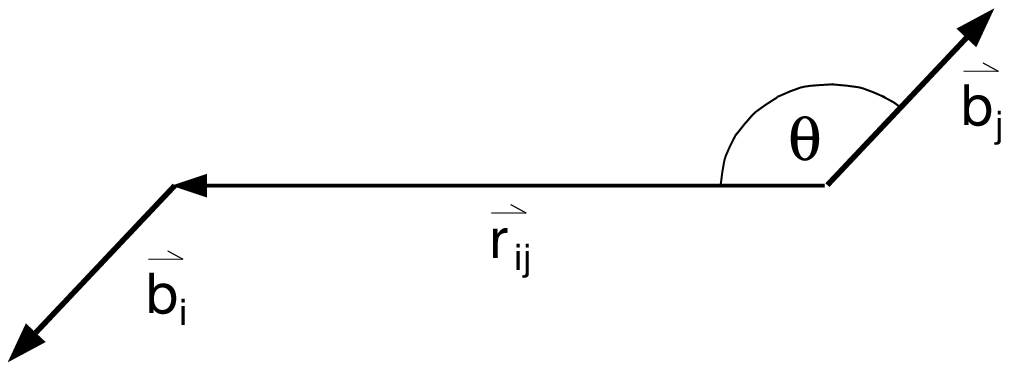}{4cm}{Schematische Darstellung eines Paars aus Burgers-Vektoren}{202}
 
Allerdings ist der Wert dieses zweiten Terms unabh"angig vom Betrag des Abstandes der Ladungen, er wird also bei sehr gro"sen Abst"anden klein gegen"uber dem ersten Term. Der dritte Term ist ein Ausdruck f"ur die Energie der Versetzungskerne (vgl. Anhang A.1) und zeigt auch keine Abstandsabh"angigkeit. 
 
\subsection{Abbildung des Unordnungsanteils}
Die Hamiltonfunktion f"ur die Wechselwirkung zwischen einer Verunreinigung und einem "au"seren Verschiebungsfeld {\bf u(r)} beschreibt Gleichung \refformel{219a}. F"ur das externe Feld setzen wir im folgenden Formel \refformel{237a} an. Physikalisch bedeutet das, da"s wir nur die Wechselwirkung zwischen der Unordnung und den Versetzungen ber"ucksichtigen, die direkte Unordnung-Unordnung-Wechsel\-wirkung aber vernachl"assigen. Gerechtfertigt ist dies, da diese Wechselwirkung von der Ordnung $\delta c({\bf r}) \delta c({\bf r'})$ ist und bei Unordnungsmittelung unter Benutzung von \refformel{111a} wegf"allt \verweistext{nel83}. Es gilt also
\begin{equation}
   \label{246a}
   H_D =  - \left(\frac{1}{2 \mu} - \frac{2 \lambda}{4 \mu (\lambda + \mu)} \right) (\mu + \lambda) \Omega_0 \int d^2r \delta c({\bf r}) \nabla^2 \chi({\bf r}).
\end{equation}

Bei Verwendung der Definition der Verzerrungsfunktion $\chi$ folgt sofort
\begin{equation}
   \label{247a}
   H_D = \frac{J}{2} \Omega_0 \int d^2r \delta c({\bf r}) \sum_{\alpha} \epsilon_{zij} b_{\alpha i} \nabla_j \frac{1}{2 \pi} \ln |{\bf r} - {\bf r_{\alpha}}|.
\end{equation}
 
$\epsilon_{zij}$ ist hier der "ubliche Tensor dritter Stufe, wobei die $z$-Richtung, die Richtung senkrecht zum Kristall ist. Summiert man \refformel{247a} mit der elastischen Hamiltonfunkton \refformel{245a}, ergibt sich abschlie"send die Hamiltonfunktion, die das vollst"andige Problem auf eine Coulombgas-Beschreibung abbildet \verweistext{nel83}
\begin{equation}
   \label{248a}
   \begin{split}
H = & - \frac{J}{8 \pi} \sum_{\alpha \neq \beta} \left({\bf b_{\alpha} b_{\beta}} \ln |{\bf r_{\alpha \beta}}| - \frac{({\bf b_{\alpha} r_{\alpha \beta}}) ({\bf b_{\beta} r_{\alpha \beta}})}{r_{\alpha \beta}^2} \right) + E_C \sum _{\alpha} |{\bf b_{\alpha}}|^2 \vspace{0.5cm} \\
 & + \frac{J}{4 \pi} \Omega_0 \int d^2r \delta c({\bf r}) \sum_{\alpha} \frac{{\bf e_z} \left[{\bf b_{\alpha}} \times ({\bf r} - {\bf r_{\alpha}}) \right]}{({\bf r} - {\bf r_{\alpha}})^2}.
   \end{split}
\end{equation}

Offenbar werden Versetzungen mit steigendem Betrag des Burgers-Vektors energetisch immer ung"unstiger. Daher ist anzunehmen, da"s bei tiefen Temperaturen lediglich solche Burgers-Vektoren auftreten, deren Betrag der Gitterkonstanten entspricht,  f"ur die also $|{\bf b_{\alpha}}| = 1$ gilt. Man sieht ferner, da"s sich antiparallele Burgers-Vektoren anziehen, w"ahrend parallele Burgers-Vektoren Energie gewinnen, wenn sie sich voneinander entfernen. Bei senkrecht zueinander stehenden Vektoren ist die Wechselwirkungsenergie unabh"angig vom Betrag ihres Abstands, f"allt doch der erste Term der Hamiltonfunktion in diesem Fall weg. Somit bilden sich bei gen"ugend tiefen Temperaturen und nicht zu starker Unordnung Dipolpaare, freie Versetzungen werden erst bei h"oheren Temperaturen oder starker Unordung energetisch m"oglich sein.  
 
Da bei tiefen Temperaturen und geringer Unordnung auch Ladungsneutralit"at ($\sum{\bf b_{\alpha}} = 0$) vorliegt, ist es m"oglich alle Versetzungen eindeutig zu Dipolen zu paaren. Dabei sind Dipole stets Versetzungspaare, deren Burgers-Vektoren antiparallel zueinander stehen. Wegen $|{\bf b_{\alpha}}| = 1$ sind Dipole somit auch ladungsneutral. Zur Bildung des ersten Dipols betrachtet man jede Versetzung und sucht die dazu n"achstliegende Versetzung mit antiparallelem Burgers-Vektor. Jenes der m"oglichen Paare mit der geringsten Separation bildet den ersten Dipol. Zur Bildung des zweiten Paares geht man unter Vernachl"assigung der beiden Versetzungen, die den ersten Dipol bilden, genauso vor. Nach gen"ugend vielen Schritten hat man so alle Versetzungen eindeutig gepaart.
 
Die Wechselwirkung zwischen zwei Versetzungen eines Dipols mit Separation $r$ wird durch Dipole mit kleiner Separation abgeschirmt. Dieser Effekt kann ber"ucksichtigt werden, in dem man alle Dipole mit einer Separation kleiner als $r$ nicht mehr als Ladungen, sondern nur noch als polarisierbares Medium auffa"st. Dieses Vorgehen f"uhrt zu einer Renormierung der physikalischen Gr"o"sen wie der Kopplungskonstanten, deren Wert dann von $r$ abh"angt.

\section{Bisherige Resultate}
\subsection{Wiedereintritt in die ungeordnete Phase}
Die im letzten Abschnitt hergeleitete Hamiltonfunktion f"ur das ungeordnete System \refformel{248a} wurde bereits 1983 von Nelson untersucht \verweistext{nel83}. Das wesentliche Resultat dieser Arbeit waren die folgenden Flu"sgleichungen, die mittels des Replikatricks und Renormierung, sowie mittels einer N"aherung f"ur kleine Werte der Fugazit"at $y$ hergeleitet wurden:
\begin{align}
  \label{249a}
  \frac{dJ}{dl} & = - \frac{3}{4} \pi \frac{J^2}{T} y^2 e^{\frac{JT - \bar{\sigma}J^2}{8 \pi T^2}} \left(2 I_0 \left(\frac{JT - \bar{\sigma}J^2}{8 \pi T^2}\right) - I_1 \left(\frac{JT - \bar{\sigma}J^2}{8 \pi T^2}\right) \right) \vspace{0.8cm} \\
  \label{250a}
  \frac{dy}{dl} & = \left(2 - \frac{JT - \bar{\sigma}J^2}{8 \pi T^2} \right) y + O(y^2) \vspace{0.8cm} \\
  \label{251a}
  \frac{d \sigma}{dl} & = 0
\end{align}
 
Es wird also eine Renormierung der Kopplungskonstante $J(l)$ und der Fugazit"at $y(l)$ gefunden, w"ahrend die Unordnungsst"arke $\sigma$ unrenormiert bleibt. Vom Gittertyp h"angt der Beitrag des zweiten Summanden in \refformel{250a} ab, der hier mit $O(y^2)$ bezeichnet ist.  Die sogenannte {\it Fugazit"at} $y$ besitzt folgenden Ausgangswert
\begin{equation}
   \label{252a}
   y_0 = y(r = 1) = e^{- \frac{E_C}{T}} = e^{-\frac{(\tilde{C} + 1) J}{8 \pi T}}.
\end{equation}
Auf gr"o"seren L"angenskalen $r := e^l$ ist die renormierte Fugazit"at $y = e^{-E_C(r)/T}$, wobei die renormierte Energie $E_C(r)$ eines Versetzungskerns seiner freien Energie $F_C = E_C - TS_C$ auf einer Fl"ache der Gr"o"se $r^2$ entspricht. Somit nimmt $y^2$ Werte an, die proportional sind zur Wahrscheinlichkeit Versetzungsdipole zu finden und proportional zu $r^4$ sind, der Zahl der durch Renormierung verlorengegangenen m"oglichen Dipolpostionen (vgl. hierzu Kapitel 4). Ferner wurde eine neudefinierte Unordnungsst"arke $\bar{\sigma} = \sigma \Omega_{0}^2$ und die "ublichen modifizierten Besselfunktionen $I_0$ und $I_1$ verwendet. F"ur $\sigma \rightarrow 0$ stimmen obige Flu"sgleichungen mit jenen Gleichungen "uberein, die f"ur Systeme ohne Unordnung gefunden wurden \verweistext{you79, nh79}. 
 
Eine numerische Integration der Flu"sgleichungen liefert die in Abbildung 2.3 gezeigte Darstellung des Hamiltonianflusses.
 
\graphik{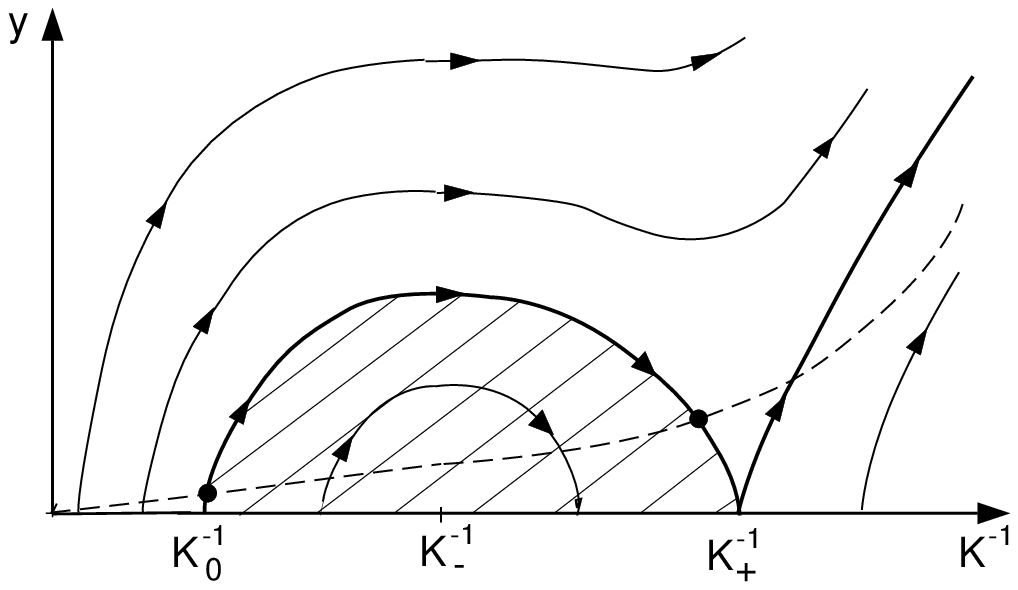}{6cm}{Hamiltonianflu"s im Falle $0 < \bar{\sigma} < \bar{\sigma}_c$. Die gestrichelte Linie stellt die Anfangsbedingung $y_0$ dar. Ferner ist $K := \frac{J}{T}$.}{203}
 
Die Anfangsbedingungen sind durch die $y_0$ darstellende gestrichelte Linie gekennzeichnet. Auf der Achse $y = 0$ befinden sich zwei ausgezeichnete Punkte $K_{-}^{-1}$ und $K_{+}^{-1}$ 
\begin{equation}
  \label{253a}
  K_{\pm}^{-1} = \frac{1}{32 \pi} \left[1 \pm (1 - 64 \pi \bar{\sigma})^{1/2} \right].
\end{equation}
Nur f"ur $K_{-}^{-1} < K^{-1} < K_{+}^{-1}$ geht der Flu"s zur $y=0$-Linie hin, in allen anderen F"allen flie"st er von dieser Linie weg. Die spezielle, dicker eingezeichnete, Linie, die die Fixlinie $y = 0$ bei $K_{0}^{-1}$ verl"a"st und bei $K_{+}^{-1}$ wieder in die Fixlinie einm"undet, separiert zwei Phasen. Man nennt sie daher auch {\it Separatrix}. Im schraffierten Bereich liegt die feste Phase, da hier der Flu"s immer bei $y=0$ und endlichen Werten von $K$ einm"undet. Wegen der oben erw"ahnten Proportionalit"at zwischen der Fugazit"at $y$ und der Dipolwahrscheinlichkeitsdichte gibt es auf gro"sen Skalen offenbar keine Dipole mehr, alle Versetzungen sind zu Dipolen kleinerer Separation gebunden. Im nicht-schraffierten Bereich strebt der Flu"s zu gro"sen $y$-Werten; man hat also selbst f"ur $r \rightarrow \infty$ eine hohe Dipolwahrscheinlichkeit. Ein Versetzungsdipol mit Separation $r \rightarrow \infty$ ist aber gleichbedeutend mit der Existenz freier Versetzungen, somit liegt hier die nicht-geordnete Phase vor.
 
Die feste Phase existiert offenbar nur f"ur Unordnungsst"arken, die kleiner sind als eine kritische St"arke $\bar{\sigma}_c$
\begin{equation}
  \label{254a}
  \bar{\sigma} \le \bar{\sigma}_c = \frac{1}{64 \pi},
\end{equation}
denn bei $\bar{\sigma}_c$ fallen $K_{-}^{-1}$ und $K_{+}^{-1}$ zusammen. Die realen Phasen"ubergangs-Temperaturen ergeben sich aus den Schnittpunkten der Anfangsbedingungs-Linie mit der Separatrix. Im Limes $E_C \rightarrow \infty$ f"allt die obere kritische Temperatur dann mit $K_{+}^{-1}$ zusammen. Deutlich wird, da"s auch im Falle sehr geringer Unordnung $\bar{\sigma} > 0$ bei gen"ugend tiefen Temperaturen ein {\it Wiedereintritt} in die ungeordnete Phase stattfindet. Das in Abbildung 1.5 dargestellte Phasendiagramm resultiert.
 
\subsection{Scheitern der N"aherung f"ur kleine Fugazit"aten}
Nelson hat in obiger Arbeit \verweistext{nel83} f"ur die Wahrscheinlichkeit, einen Versetzungsdipol vorzufinden, durchweg eine bosonische {\it Boltzmann - Verteilung} angesetzt. Ein Versetzungsdipol der Separation $r = |{\bf r_1} - {\bf r_2}|$ mit einem Winkel $\theta$ zwischen Abstandsvektor und Burgers-Vektoren (vgl. Abbildung 2.2) soll somit eine Wahrscheinlichkeit
\begin{equation}
  \label{255a}
  P(r, \theta) \propto \left[\exp \left(- \frac{E_1(r, \theta)}{T} \right) \right]_D
\end{equation}
haben. $[...]_D$ bedeutet wiederum eine Mittelung "uber alle Unordnungskonfigurationen, w"ahrend $E_1(r, \theta)$ die Energie eines solchen Versetzungsdipols mit $|{\bf b}| = 1$ ist. Nach \refformel{248a} gilt
\begin{equation}
   \label{256a}
   E_b(r, \theta) = \frac{J}{4 \pi} b^2 (\ln r - \cos^2 \theta) + 2 b^2 E_C + V({\bf r_1}) +  V({\bf r_2}),
\end{equation}
mit
\begin{equation}
   \label{257a}
   \begin{split}
   V({\bf r_{\alpha}}) & = \frac{J}{4 \pi} \Omega_0 \int d^2r \delta c({\bf r}) \frac{{\bf e_z} \left[{\bf b_{\alpha}} \times ({\bf r} - {\bf r_{\alpha}}) \right]}{({\bf r} - {\bf r_{\alpha}})^2} \vspace{0.8cm} \\
   & = \frac{J}{4 \pi} \Omega_0 \int d^2r \delta c({\bf r}) ({\bf e_z} \times {\bf b_{\alpha}}) {\bf \nabla} \ln |{\bf r} - {\bf r_{\alpha}}|.
   \end{split}
\end{equation}
Da die Energiekosten f"ur Versetzungen mit $\propto b^2$ ansteigen, w"ahrend ein m"oglicher Energiegewinn durch Wechselwirkung mit der Unordnung nur $\propto b$ sein kann, ist die Wahrscheinlichkeit Versetzungen mit einem Burgers-Vektor $|{\bf b}| > 1$ zu erzeugen verschwindend gering und wird im folgenden wie in \refformel{255a} vernachl"assigt.
 
Es ist zu beachten, da"s die letzten beiden Terme der Unordnungs-Wechsel\-wirkung in \refformel{256a} $V({\bf r_1}) +  V({\bf r_2})$ negative Werte annehmen k"onnen und sogar den gesamten Ausdruck $E_1(r, \theta)$ negativ machen k"onnen. Bei sehr tiefen Temperaturen dominieren solche Konfigurationen, hier ist also in den meisten F"allen $E_1(r, \theta) \gg T$ nicht mehr gew"ahrleistet. Folglich ergeben sich hier auch sehr gro"se Dipolwahrscheinlichkeiten $P(r, \theta) \gg 1$, was dazu f"uhrt, da"s Wechselwirkung zwischen den Versetzungspaaren ber"ucksichtigt werden m"ussen, man also kein verd"unntes System mehr vorliegen hat. Insbesondere wird die in \verweistext{nel83} durchgef"uhrte Entwicklung nach kleinen Potenzen der Fugazit"at unhaltbar, ist $y$ doch, wie bereits erw"ahnt, proportional zur Dipolwahrscheinlichkeit.
 
Da"s diese Problematik zu einem unphysikalischen Verhalten f"uhrt, sollen die folgenden Absch"atzungen zeigen. Entsprechend \refformel{250a} gilt
\begin{equation}
  \label{258a}
  \frac{dy}{dl} = \left(2 - \frac{J}{8 \pi T} + \pmb{\frac{J^2 \bar{\sigma}}{8 \pi T^2}} \right) y + O(y^2).
\end{equation}
Die $\frac{\bar{\sigma}}{T}$-Proportionalit"at des fett gedruckten Terms ist f"ur gro"se Temperaturen plausibel, wird doch der Einflu"s des Unordnungspotentials durch thermische Fluktuationen verringert. Im Fall $\bar{\sigma} > 0$ und $ T \rightarrow 0$ jedoch f"uhrt dieser Term zu einem unbegrenzt starken Anwachsen von $\frac{dy}{dl}$ und damit, wie in Abschnitt 2.3.1 beschrieben, zu einem Wiedereintritt in die ungeordnete Phase. Der $O(y^2)$-Term, der f"ur das quadratische Gitter null ist, nimmt im Fall des Dreiecksgitters lediglich kleine positive Werte an und kann daher bei dieser Betrachtung "ubergangen werden.

Versetzungsdipole mit einer Gr"o"se zwischen $r$ und $r+dr$ tragen mit $dF_V$ zur freien Energie pro Einheitsfl"ache bei. F"ur $dr \rightarrow 0$ kann dieser Beitrag nach Kosterlitz \verweistext{kos74} mittels 
\begin{equation}
   \label{258b}
   dF_V \approx - T \, r dr \frac{1}{4} \int_0^{2 \pi} d \theta \, [\ln Z]_D
\end{equation}
angen"ahert werden (vgl. auch Scheidl \verweistext{sts97} und Tang \verweistext{tang96}). Hierbei ist $Z$ die lokale Zustandssumme. Sie z"ahlt bei gegebenen Versetzungspositionen, deren Separation im Intervall $[r, r+dr]$ liegt, die m"oglichen Dipole. Im Falle des Quadratgitters sind vier Einstellm"oglichkeiten der Dipole zu ber"ucksichtigen, beim Dreiecksgitter gibt es sechs M"oglichkeiten.
 
Eine ausf"uhrliche Diskussion der Einstellm"oglichkeiten und orientierten Dipolwahrscheinlichkeiten $P(r, \theta)$ erfolgt in Abschnitt 3.2 (vgl. insbesondere Abbildung 3.2). Nimmt man f"ur das Quadratgitter analog zu  Nelson kleine Fugazit"aten und damit kleine Dipolwahrscheinlichkeiten an, so findet man wegen $\ln(1+x) \approx x$ f"ur $x \ll 1$:
\begin{equation}
   \label{258d}
   \begin{split}
   dF_V = & - \frac{T}{4} \int_0^{2 \pi} d \theta r dr [P(r, \theta) + P(r, \theta + \frac{\pi}{2})+ P(r, \theta + \pi) + P(r, \theta + \frac{3\pi}{2}) ] \\
        = & - T \, r dr \int_0^{2 \pi} d \theta \, P(r, \theta)
\end{split}
\end{equation}
Setzt man eine Boltzmann-verteilte Wahrscheinlichkeit $P(r, \theta) \propto r^{-\frac{J}{4 \pi T} \left(1 - \frac{J \bar{\sigma}}{T} \right)} \\
e^{\frac{J}{8 \pi T} \left(1 - \frac{J \bar{\sigma}}{T} \right) cos^2 \theta}$ an, so folgt
\begin{equation}
   \label{258e}
   dF_V \propto - \pi T \, r dr \, r^{-\frac{J}{4 \pi T} \left(1 - \frac{J \bar{\sigma}}{T} \right)} e^{\frac{J}{8 \pi T} \left(1 - \frac{J \bar{\sigma}}{T} \right)} I_0 \left(\frac{J}{8 \pi T} \left(1 - \frac{J \bar{\sigma}}{T} \right) \right).
\end{equation}
Unter Benutzung der von Nelson verwendeten Relation f"ur die Fugazit"at $y^2(r) = r^4 \, r^{-\frac{J}{4 \pi T} \left(1 - \frac{J \bar{\sigma}}{T} \right)}$ ergibt sich f"ur das Skalenverhalten der freien Energie
\begin{equation}
   \label{258f}
   \frac{dF_V}{dl} \propto - \pi T \, \frac{y^2}{r^2} e^{\frac{J}{8 \pi T} \left(1 - \frac{J \bar{\sigma}}{T} \right)} I_0 \left(\frac{J}{8 \pi T} \left(1 - \frac{J \bar{\sigma}}{T} \right) \right).
\end{equation}
Im Fall des Dreieckgitters findet man in niedrigster Ordnung, abgesehen von einem anderem konstanten Vorfaktor, dasselbe Resultat. Formal wurde dieses Ergebnis bereits von Kosterlitz \verweistext{kos74} f"ur das XY-Modell ohne Unordnung und von Tang \verweistext{tang96} und Scheidl \verweistext{sts97} f"ur den Bereich hoher Temperaturen des XY-Modells mit Unordnung gefunden, ihre Resultate unterscheiden sich lediglich durch Faktoren. 
 
Mittels der Beziehung $\frac{d S_V}{d l} = - \frac{\partial}{\partial T} \frac{d F_V}{d l}$ ist es nun m"oglich das Skalenverhalten der Entropie zu untersuchen. Hierbei ist es nicht n"otig die beiden hinteren Faktoren aus Gleichung \refformel{258f} zu betrachten, da sie auf gro"sen L"angenskalen $\ln r \gg \bar{\sigma}$ und $\ln r \gg \frac{E_C}{J}$ (nur hier gelten auch die verwendeten Relationen f"ur z.B. die Wahrscheinlichkeit) nicht zum physikalischen Verhalten beitragen. Somit w"urde eine Ber"ucksichtigung dieser Faktoren die Rechnung un"ubersichtlicher machen, aber das Ergebnis {\it nicht} beeinflussen.

Wir berechnen also lediglich:
\begin{equation}
   \label{258g}
   \begin{split}
   \frac{\partial}{\partial T} (Ty^2) = & y^2 + T \frac{\partial}{\partial T} y^2 \\
   = & y^2 + \frac{J}{4 \pi T^2} \, l \, \left( T - 2 \bar{\sigma}J \right) \, y^2
\end{split}
\end{equation}
Hieraus folgt, da"s die Entropie im Falle $T < T^* = 2 J \bar{\sigma}$ und f"ur $\bar{\sigma} \leq \bar{\sigma}_c$ unbeschr"ankt abnimmt. Vernachl"assigt wurde bisher die bei $l > 0$ vorhandene implizite Temperaturabh"angigkeit der abgeschirmten Gr"o"se $J$. Wie Scheidl \verweistext{sts97} bereits ausgearbeitet hat, k"onnen diese impliziten Abh"angigkeiten ber"ucksichtigt werden, indem man die Gr"o"se $\frac{\partial}{\partial T} \frac{d}{d l} (Ty^2)$ berechnet. Sind sowohl $\frac{\partial}{\partial T} (Ty^2)$ als auch $\frac{\partial}{\partial T} \frac{d}{d l} (Ty^2)$ positiv, so ist der Entropieflu"s auf jeden Fall positiv, da die $T$-Abh"angigkeit von $J$ eine Konsequenz der Abschirmung ist. Sind beide Ausdr"ucke negativ, liegt auch ein negativer Entropieflu"s vor.
 
Anstelle von $\frac{\partial}{\partial T} \frac{d}{d l} (Ty^2)$ kann man ebenso 
\begin{equation}
   \label{258h}
   \frac{\partial}{\partial T} \frac{d}{d l} \ln (Ty^2) = \frac{J}{4 \pi T} \left(1 - \frac{J \bar{\sigma}}{T} \right) - \frac{J^2 \bar{\sigma}}{16 \pi T^3} + O(y^2)
\end{equation}
berechnen. F"ur $ T \ge T^*$ ist der Entropieflu"s somit tats"achlich nicht negativ. Die unbeschr"ankte Abnahme der Entropie im Fall $T < T^* = 2 J \bar{\sigma}$ best"atigt sich allerdings.  Auf gro"sen L"angenskalen f"uhrt dies zu einer {\it negativen} Entropie. Dieses Verhalten ist offenbar eine Konsequenz der unkorrekten N"aherung f"ur kleine Fugazit"aten. Es ist daher notwendig den gesamten Bereich, in dem $T < T^*(\bar{\sigma})$ gilt neu zu untersuchen. Da $T^*(\bar{\sigma})$ die Phasengrenze gerade bei $\bar{\sigma} = \bar{\sigma}_c$ schneidet (vgl. Abbildung 1.6) f"allt der gesamte Bereich des Wiedereintritts in die ungeordnete Phase darunter.
 
\section{Absch"atzung der Phasengrenze}
In diesem Abschnitt soll die Phasengrenze im neu zu untersuchenden Bereich $T < T^*(\bar{\sigma})$ abgesch"atzt werden.
 
Da ein Phasen"ubergang in die ungeordnete Phase, in der Versetzungen nicht mehr zu Dipolen gebunden sind, mit der M"oglichkeit der Entstehung freier Versetzungen durch thermische Fluktuationen gleichzusetzen ist, untersuchen wir zun"achst das System bei $T = 0$ und schwacher Unordnung darauf, ob das Unordnungspotential hier die spontane Entstehung von Versetzungen erm"oglichen kann. Die Energie einer einzelnen Versetzung in einem System der Fl"ache $R \times R$ ohne Unordnung ist, wie man aus \refformel{233a} berechnen kann (vgl. \verweistext{hal79, chlu})
\begin{equation}
  \label{261a}
  E_{el} = \frac{J}{8 \pi} \ln R.
\end{equation}
 
Dies ist zu vergleichen mit dem m"oglichen Energiegewinn $E_D$ durch die Wechselwirkung $V({\bf r})$ zwischen der freien Versetzung und der Unordnung. Diese Wechselwirkung ist gau"sverteilt, da auch die Funktion $\delta c({\bf r})$ eine solche Verteilung aufweist. Die Varianz $\omega(R)$ der Wechselwirkung l"a"st sich mit Formel \refformel{257a} berechnen (vgl. Anhang B)
\begin{equation}
   \label{262a}
   \omega(R) = \left[V({\bf r_{\alpha}})^2 \right]_D = \frac{J^2}{16 \pi} \bar{\sigma} \ln R.
\end{equation}
Der {\it typische} Energiegewinn durch diese Wechselwirkung ist also
\begin{equation}
   \label{263a}
   E_D^{(typ)} = - J \left(\frac{\bar{\sigma}}{16 \pi} \ln R \right)^{1/2},
\end{equation}
und die freie Energie der einzelnen Versetzung ist damit:
\begin{equation}
   \label{264a}
   F = \frac{J}{8 \pi} \ln R - 2T \ln R - J \left(\frac{\bar{\sigma}}{16 \pi} \ln R \right)^{1/2}.
\end{equation}
Der zweite Term resultiert aus der Entropie, die $S = \ln(R^2)$ betr"agt, da eine einzelne Versetzung $R^2$ Pl"atze im System einnehmen kann. Vergleicht man bei $T \rightarrow 0$ den ersten mit dem letzten Term, so zeigt sich, da"s bei schwacher Unordnung $\bar{\sigma} \ll 1$ die Entstehung freier Versetzungen nicht zu erwarten ist. M"ochte man aber rigoros ausschlie"sen, da"s das Unordnungspotential die Entstehung freier Versetzung induzieren kann, so mu"s man nicht den typischen, sondern den maximal m"oglichen Energiegewinn $E_D^{(max)}$ der Wechselwirkung untersuchen. Dieser ist dann geben, wenn die Versetzung ihre energetisch g"unstigste Position im System einnimmt. 
 
Hierzu folgen wir der Methode von Korshunov und Nattermann \verweistext{nk96} und vernachl"assigen die Korrelationen zwischen den Werten des Unordnungspotentials an den verschiedenen Positionen. Wir betrachten also $R^2$ unabh"angige Variable $V({\bf r})$ die jeweils dieselbe Gau"sverteilung
\begin{equation}
   \label{264b}
   p(V) = \frac{1}{\sqrt{2 \pi \omega }} \exp \left(- \frac{V^2}{2 \omega} \right)
\end{equation}
aufweisen. Die Wahrscheinlichkeit genau den Wert $E_D^{(max)}$ als Maximalwert zu finden ist gegeben durch die Wahrscheinlichkeit, ihn an einem Platz vorzufinden, multipliziert mit der Wahrscheinlichkeit, da"s alle anderen Pl"atze eine geringere Energie aufweisen und multipliziert mit der Anzahl der Pl"atze $R^2$. Die Verteilung von $E_D^{(max)}$ ist also
\begin{equation} 
   \begin{split}
   \label{265a}
   P(E_D^{(max)}) & = p(E_D^{(max)}) \left[\int_{-\infty}^{E_{Dmax}} dV p(V) \right]^{R^2 - 1} R^2 \vspace{0.8cm} \\
   & = \frac{d}{dE_D^{(max)}}\left[\int_{-\infty}^{E_{Dmax}} dV p(V) \right]^{R^2}.
   \end{split}
\end{equation}
V"ollig analog zu \verweistext{nk96} l"a"st sich das Maximum dieser Verteilung berechnen, das wegen der geringen Breite dieser nicht-gau"sischen Verteilung mit dem wahrscheinlichsten Wert der Verteilung gleich gesetzt werden kann. Man findet
\begin{equation}
   \label{266a}
   \overline{E_D^{(max)}} \approx \sqrt{2 \omega \ln R^2}.
\end{equation}
Einsetzen von $\omega(R)$ liefert f"ur die gemittelte freie Energie einer Versetzung, die sich am energetisch g"unstigsten Platz des Systems befindet
\begin{equation}
   \label{267a}
    F = \frac{J}{8 \pi} \left(1 - \sqrt{16 \pi \bar{\sigma}}\right) \ln R.
\end{equation}
Auch diese Gleichung zeigt, da"s f"ur $T \rightarrow 0$ und kleine Unordnungsst"arke immer noch eine im Falle $R \rightarrow \infty$ unendliche Energie aufgebracht werden mu"s, um eine freie Versetzung zu erzeugen. Erst bei Unordnungsst"arken $\bar{\sigma} > \frac{1}{16 \pi}$ kann eine freie Versetzung die Energie des Systems absenken und daher sind ab diesem Wert ungebundene Versetzungen auch zu erwarten. 
 
Zus"atzlich zu obigen Betrachtungen kann untersucht werden, bis zu welcher Unordungsst"arke es bei $T = 0$ energetisch g"unstig ist, gebundene Versetzungsdipole zu erzeugen. Auch die Wechselwirkung eines solchen Paares mit der Unordnung ist gau"sverteilt. Die Varianz dieser Verteilung ist wegen \refformel{256a} f"ur einen Dipol mit der Separation $r = |{\bf r_{\alpha}} - {\bf r_{\beta}}|$ gegeben durch $\Delta^2 (r, \theta) = \left[(V({\bf r_{\alpha}}) + V({\bf r_{\beta}}))^2 \right]_D$. Beachtet man die Translationsinvarianz im System, so erh"alt man nach einiger Rechnung (vgl. Anhang B) f"ur Separationen $r \gg 1$
\begin{equation}
   \begin{split}
   \label{268a}
   \Delta^2(r, \theta) & = \left[(V({\bf r}) + V({\bf 0}))^2 \right]_D = 2 \left[V({\bf 0})V({\bf 0}) + V({\bf r})V({\bf 0}) \right]_D \vspace{0.8cm} \\
   & = \frac{J^2 \bar{\sigma}}{8 \pi} (\ln r - \cos^2 \theta),
   \end{split}
\end{equation}
hierbei ist $\theta$ wiederum der Winkel zwischen den Burgers-Vektoren und dem Abstandsvektor. 
 
Da die elastische Energie eines Versetzungsdipols durch
\begin{equation}
   \label{269a}
   E_{el} = \frac{J}{4 \pi} \ln r - \frac{J}{4 \pi} cos^2 \theta + 2 E_C
\end{equation}
gegeben ist, kann die Erzeugnung eines solchen Dipols bei $T = 0$ nur dann energetisch g"unstig sein, wenn die Wechselwirkung mit der Unordnung die Energiekosten durch $E_{el}$ aufhebt. Aufgrund der Definition von $E_C$ (vgl. Anhang A.1), kann $E_{el}$ selbst nicht negativ werden. Entsprechend Abbildung 2.4 mu"s also der Betrag dieser Wechselwirkung $V$ im schraffierten Bereich liegen.
 
\graphik{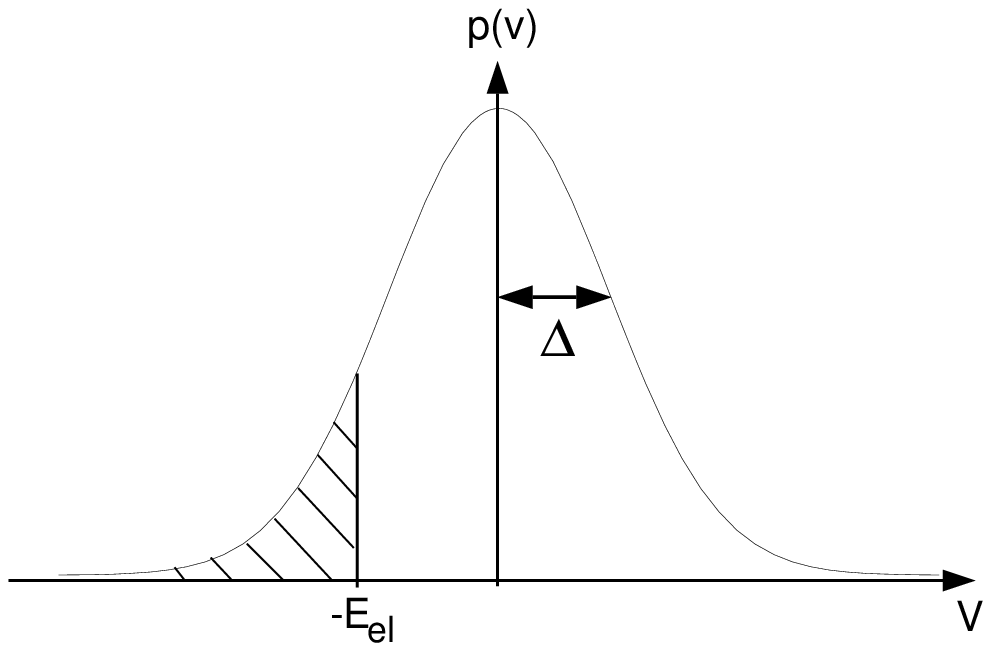}{6cm}{Gau"sverteilung der Unordnungswechselwirkung eines Versetzungsdipols}{204}
 
Daher ist die Wahrscheinlichkeit einen Versetzungsdipol mit Burgersvektor {\bf b} bei ${\bf r_{\alpha}}$ und mit {\bf -b} bei ${\bf r_{\beta}}$ zu finden
\begin{equation}
   \begin{split}
   \label{270a}
   P(r, \theta) & = \int_{- \infty}^{-E_{el}} \frac{dV}{\sqrt{2 \pi} \Delta} e^{- \frac{V^2}{2 \Delta^2}}\vspace{0.8cm} \\
   & \approx \frac{1}{\sqrt{2 \pi} \Delta} \frac{\Delta^2}{E_{el}} e^{- \frac{E_{el}^2}{2 \Delta^2}}\vspace{0.8cm} \\
   & = \sqrt{\frac{\Delta^2}{2 \pi E_{el}^2}}e^{- \frac{E_{el}^2}{2 \Delta^2}},
   \end{split}
\end{equation}
wobei $\ln r \gg \bar{\sigma}$ angenommen wurde. Setzt man hier die Ausdr"ucke f"ur $\Delta^2$ und $E_{el}$ ein, so resultiert, da man f"ur gro"se $r$ die Energie der Versetzungskerne vernachl"assigen kann ($J \ln r \gg E_C$, $\ln r \gg 1$),
\begin{equation}
  \label{271a}
  P(r, \theta) \approx \sqrt{\frac{\bar{\sigma}}{\ln r - \cos^2 \theta}} r^{- \frac{1}{4 \pi \bar{\sigma}}} e^{\frac{1}{4 \pi \bar{\sigma}} \cos^2 \theta}.
\end{equation}
 
Die Anzahl der m"oglichen Pl"atze f"ur einen Versetzungsdipol mit gro"ser Separation $r$ skaliert $\propto r^4$. Ber"ucksichtigt man nur die wesentlichen Abh"angigkeiten in $r$, so zeigt sich, da"s die Zahl der Pl"atze $P$, f"ur die es energetisch vorteilhaft ist, einen Versetzungsdipol zu erzeugen, wie folgt skaliert
\begin{equation}
   \label{272a}
    P \propto r^4 r^{- \frac{1}{4 \pi \bar{\sigma}}}.
\end{equation}
Auch hier stellt man fest, da"s im Falle $\bar{\sigma} > \frac{1}{16 \pi}$ die Zahl der Dipole mit der L"angenskala $r$ ansteigt, auf beliebig gro"sen Skalen also auch beliebig viele Dipole erzeugt werden. Wie in Abschnitt 2.3.1 erkl"art, befindet man sich somit in der ungeordneten Phase. F"ur $\bar{\sigma} < \frac{1}{16 \pi}$ hingegen findet man das in der festen Phase zu erwartende Verhalten. Diese Berechnung wie auch die vorherige Absch"atzung zeigen, da"s die kritische Unordnungsst"arke, oberhalb der keine feste Phase mit quasi-langreichweitiger Translationsordnung mehr bestehen kann, bei
\begin{equation} 
   \label{273a}
   \bar{\sigma}_c = \frac{1}{16 \pi}
\end{equation}
liegt. Dies steht im Widerspruch zu der von Nelson \verweistext{nel83} gefundenen kritischen Unordnungsst"arke von $\frac{1}{64 \pi}$. Dieser Unterschied resultiert aus einem Rechenfehler in \verweistext{nel83}, auch im folgenden werden f"ur $T > T^*$ die von Nelson gefundenen Flu"sgleichungen mit einer um den Faktor 4 st"arkeren Unordnung reproduziert. Der Wert $\bar{\sigma}_c = \frac{1}{16 \pi}$ ist ebenso von Cha und Fertig \verweistext{cf95} gefunden worden. Dem entsprechend ergibt sich f"ur den Wert von $T^*$ jetzt
\begin{equation}
   \label{274a}
   T^* = \frac{J \bar{\sigma}}{2}.
\end{equation}
 
Aufgrund vorangegangener Absch"atzungen ist im gesamten Bereich $T < T^*$ eine Phasengrenze zu erwarten, die bei $\bar{\sigma}_c$ parallel zur T-Achse verl"auft (vgl. Abbildung 1.6).  Dieser Verlauf w"urde im Gegensatz zu \verweistext{nel83} nicht den Ergebnissen von Ozeki und Nishimori \verweistext{on93} widersprechen, die einen Wiedereintritt in die ungeordnete Phase ausschlie"sen konnten und zeigten, da"s die Phasengrenze bei tiefen $T$ parallel zur Temperatur-Achse verlaufen mu"s, falls eine Phase mit quasi-langreichweitiger Ordnung "uberhaupt existiert.

\chapter{Dielektrischer Formalismus}
 
In Abschnitt 2.2.1 wurde der Zusammenhang zwischen der zweidimensionalen Elektrodynamik und dem vorliegenden Problem dargestellt. Wie in der Elektrodynamik, kann man bei der Renormierung das Verhalten auf gr"o"seren L"angenskalen $r$  mittels einer {\it dielektrischen Konstanten} $\varepsilon(r)$ beschreiben, die die Polarisierbarkeit von Versetzungpaaren kleinerer Separation ber"ucksichtigt. Da die Wahrscheinlichkeit, gebundene Versetzungen zu finden, in der geordneten Phasen f"ur $r \rightarrow \infty$ gegen null geht, ist hier ein endlicher Wert f"ur $\varepsilon(r)$ zu erwarten, w"ahrend in der ungeordneten Phase eine Divergenz der dielektrischen Konstante auf gro"sen L"angenskalen auftritt. 
 
Fordert man im noch nicht renormierten Gitter die G"ultigkeit der "ublichen elektrodynamischen Beziehung
\begin{equation}
   \label{301a}
   \nabla^4 \chi({\bf r}) = \frac{4 \pi \eta({\bf r})}{\varepsilon_0},
\end{equation}
so gilt offenbar $\varepsilon_0 = \varepsilon(1) = 4 \pi/J$. Mit Hilfe dieser Relation kann man eine renormierte Kopplungskonstante $J(r)$ f"ur beliebige L"angenskalen definieren
\begin{equation}
   \label{302a}
   J(r) := \frac{4 \pi}{\varepsilon(r)}.
\end{equation}
 
Eine Multipolentwicklung \verweistext{kt73} der Verzerrungsfunktion $\chi({\bf r})$ gibt Aufschlu"s "uber die Struktur der dielektrischen Konstanten.
 
\section{Multipolentwicklung}
Mit der am Ende von Abschnitt 2.2 geschilderten Vorgehensweise kann man alle  Versetzungen im System bei tiefen Temperaturen und kleiner Un\-ord\-nungs\-st"arke eindeutig zu Dipolen paaren.  
 
F"ur eine beliebige L"ange $\zeta$ k"onnen nun die Versetzungen in zwei Gruppen unterteilt werden. Die erste Gruppe besteht aus den Versetzungen, die zu Dipolen mit einer Separation $r < \zeta$ gebunden sind, die zweite aus denen, die gr"o"sere Dipole bilden. Indiziert man die erste Gruppe mit  $\Gamma$ und bezeichnet die Versetzungen, aus denen der Dipol $\Gamma$ besteht mit ${\bf r_{\Gamma_1}}$ und ${\bf r_{\Gamma_2}}$, so liegt dessen Dipolzentrum bei
\begin{equation}
   \label{302b}
   {\bf r_{\Gamma}} := \frac{1}{2} \left({\bf r_{\Gamma_1}} + {\bf r_{\Gamma_2}} \right).
\end{equation}
Die Quellenfunktion $\eta({\bf r}) = \epsilon_{ij} \nabla_i \sum_{\alpha} b_{\alpha, j} \delta ({\bf r} - {\bf r_{\alpha}})$ unterteilt sich somit wie folgt:
\begin{equation}
   \label{302c}
   \eta({\bf r}) = \eta_{ex}({\bf r}) + \sum_{\Gamma} \eta_{\Gamma}({\bf r})
\end{equation}
wobei
\begin{equation}
\begin{split}
   \label{302d}
   \eta_{ex}({\bf r}) & = \epsilon_{ij} \nabla_i \sum_{\bar{\alpha}} b_{\bar{\alpha}, j} \, \delta ({\bf r} - {\bf r_{\bar{\alpha}}}) \vspace{0.8cm} \\
   \eta_{\Gamma}({\bf r}) & = \epsilon_{ij} \nabla_i \left( b_{{\Gamma_1},j} \,\delta ({\bf r} - {\bf r_{\Gamma_1}}) + b_{{\Gamma_2},j} \, \delta ({\bf r} - {\bf r_{\Gamma_2}}) \right) \\
   & = \epsilon_{ij} \nabla_i \, b_{{\Gamma_1},j} \left(\delta ({\bf r} - {\bf r_{\Gamma_1}}) - \delta ({\bf r} - {\bf r_{\Gamma_2}}) \right).
\end{split}
\end{equation}
Hierbei ist $\sum_{\bar{\alpha}}$ eine Summation "uber alle Versetzungen, die nicht zu Dipolen mit einer Gr"o"se $r < \zeta$ gebunden sind, also der zweiten Gruppe angeh"oren. Zu beachten ist, da"s die Anzahl der Versetzungen, die in $ \eta_{ex}({\bf r})$ eingehen, von dem Wert von $\zeta$ abh"angt. Falls $\zeta$ so gro"s gew"ahlt ist, da"s alle Versetzungen Dipole mit $r < \zeta$ bilden, so ist $ \eta_{ex}({\bf r}) = 0$. Nach \refformel{243a} kann man die Verzerrungsfunktion nun ebenso wie die Quellenfunktion aufteilen $\chi({\bf r}) = \chi_{ex}({\bf r}) + \sum_{\Gamma} \chi_{\Gamma}({\bf r})$. Der Beitrag des Dipols $\Gamma$ zur Gesamtverzerrungsfunktion ist 
\begin{equation}
   \label{303a}
   \chi_{\Gamma}({\bf r}) = J \int d^2r' \eta_{\Gamma}({\bf r'}) g({\bf r} - {\bf r'}).
\end{equation}

Im folgenden wird der Beitrag des Dipols $\Gamma$ zur Verzerrungsfunktion an einem gen"ugend weit entfernten Ort ${\bf r}$ mittels einer Multipolentwicklung abgesch"atzt \verweistext{kt73}. Diese wird f"ur jeden Dipol durchgef"uhrt und die Beitr"age der Entwicklungen schlie"slich summiert. Wie im Anhang A.2 gezeigt wird entspricht dieses Vorgehen dem der Elektrodynamik. 
 
Zun"achst ist es zweckm"a"sig, zu einer Darstellung mit inneren Koordinaten ${\pmb \rho^{(\Gamma)}}$ des Dipols $\Gamma$ "uberzugehen. 
\begin{equation}
   \label{303b}
   {\pmb \rho^{(\Gamma)}} := {\bf r'} - {\bf r_{\Gamma}}
\end{equation}
Abbildung 3.1 verdeutlicht die vorliegende Situation. In inneren Koordinaten liegen die Versetzungen des Dipols $\Gamma$ bei ${\pmb \rho^{(\Gamma)}_1} = {\bf r_{\Gamma_1}} - {\bf r_{\Gamma}}$ und ${\pmb \rho^{(\Gamma)}_2} = {\bf r_{\Gamma_2}} - {\bf r_{\Gamma}}$.
 
\graphik{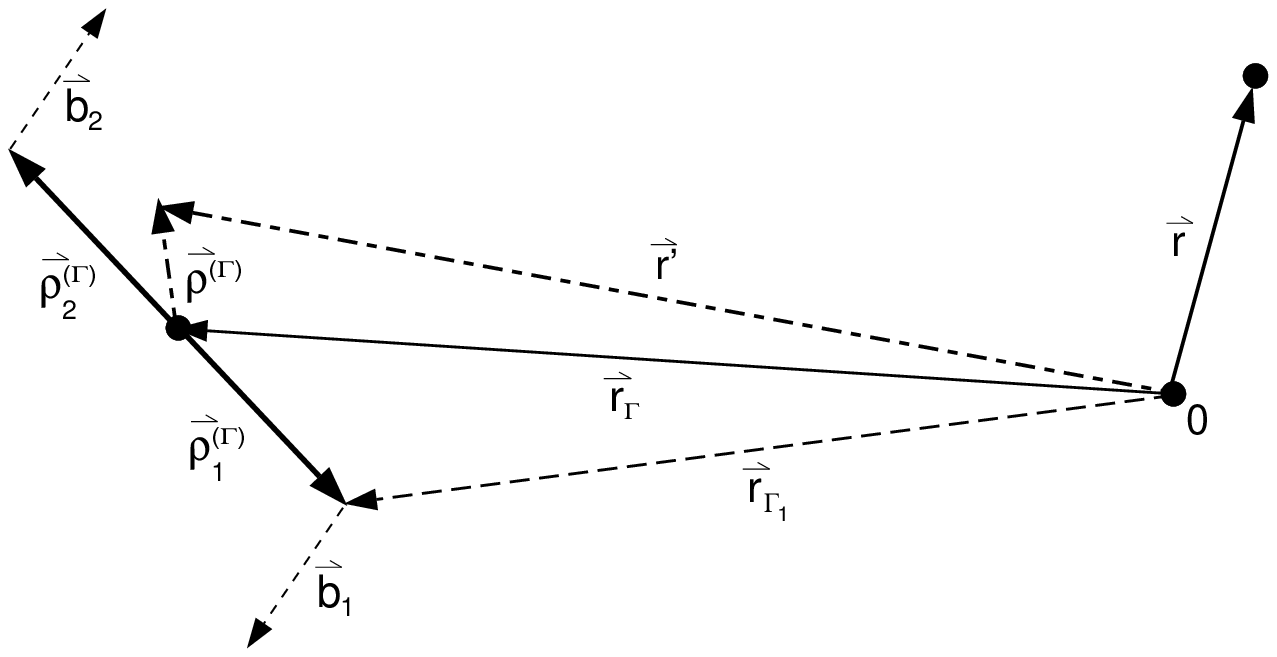}{6cm}{Ein Dipol mit Zentrum bei ${\bf r_{\Gamma}}$ liefert einen Beitrag zur Verzerrungsfunktion am Punkt ${\bf r}$. Die innere Koordinate ${\pmb \rho^{(\Gamma)}}$ wird vom Zentrum aus gemessen.}{301}
 
In inneren Koordinaten wird aus \refformel{303a}
\begin{equation}
   \label{303c}
   \chi_{\Gamma}({\bf r}) = J \int d^2 \rho^{(\Gamma)} \, \eta_{\Gamma} ({\pmb \rho^{(\Gamma)}} + {\bf r_{\Gamma}}) \, g({\bf r} - {\bf r_{\Gamma}} - {\pmb \rho^{(\Gamma)}}).
\end{equation}
Definiert man schlie"slich die Quellenfunktion mit $\tilde{\eta}_{\Gamma}({\pmb \rho^{(\Gamma)}}) := \eta_{\Gamma}({\pmb \rho^{(\Gamma)}} + {\bf r_{\Gamma}})$ um, so erh"alt man
\begin{equation}
   \label{304a}
   \chi_{\Gamma}({\bf r}) = J \int d^2 \rho^{(\Gamma)} \, \tilde{\eta}_{\Gamma}({\pmb \rho^{(\Gamma)}}) \, g({\bf r} - {\bf r_{\Gamma}} - {\pmb \rho^{(\Gamma)}}).
\end{equation}
Da ${\pmb \rho^{(\Gamma)}}$ eine interne Koordinate des Dipols ist, gilt in gen"ugend gro"sem Abstand vom Dipol $|{\pmb \rho^{(\Gamma)}}| \ll |{\bf r} - {\bf r_{\Gamma}}|$. Somit  kann obige Beziehung nach ${\pmb \rho^{(\Gamma)}}$ entwickelt werden.
\begin{equation}
\begin{split}
   \label{305a}
   \chi_{\Gamma}({\bf r}) = & \, J \int d^2 \rho^{(\Gamma)} \, \tilde{\eta}_{\Gamma}({\pmb \rho^{(\Gamma)}}) \, g({\bf r} - {\bf r_{\Gamma}}) \\
   & + \frac{1}{2} J \int d^2 \rho^{(\Gamma)} \, \tilde{\eta}_{\Gamma}({\pmb \rho^{(\Gamma)}}) \, {\rho^{(\Gamma)}}_i \, {\rho^{(\Gamma)}}_j \, \frac{\partial^2 g({\bf r} - {\bf r_{\Gamma}})}{\partial x_{\Gamma i} \partial x_{\Gamma j}}
\end{split}
\end{equation}
Hierbei handelt es sich um eine "ubliche Multipolentwicklung in zwei Dimensionen. Zu beachten ist, da"s man die Ableitungen der Green-Funktion nach Komponenten von ${\pmb \rho^{(\Gamma)}}$ durch Ableitungen nach ${\bf r_{\Gamma}}$ ersetzen kann. Ein Term erster Ordnung der Form
\begin{equation}
   \label{306a}
   J \int d^2 \rho^{(\Gamma)} \, \tilde{\eta}_{\Gamma}({\pmb \rho^{(\Gamma)}}) \, {\rho^{(\Gamma)}}_i \, \frac{\partial g({\bf r} - {\bf r_{\Gamma}})}{\partial x_{\Gamma i}}
\end{equation}
tritt hier nicht auf. Einsetzen der Quellenfunktion f"ur einen Dipol zeigt sofort, da"s er null ist. Wie bereits in Abschnitt 2.2.1 angemerkt, mu"s hier verglichen mit der Elektrodynamik doppelt so oft abgeleitet werden. Aus diesem Grund hat der Dipolterm obiger Multipolentwicklung die Form des klassischen Quadrupolterms.

Der Gesamtbeitrag aller Dipole mit $r < \zeta$ zur Verzerrungsfunktion am Ort {\bf r} ist damit
\begin{equation}
\begin{split}
   \label{306b}
   \chi^{Dip}({\bf r}) & = \sum_{\Gamma} \chi_{\Gamma}({\bf r}) \\
   & = J \sum_{\Gamma} (\int d^2 \rho^{(\Gamma)} \, \tilde{\eta}_{\Gamma}({\pmb \rho^{(\Gamma)}}) \, g({\bf r} - {\bf r_{\Gamma}}) \\
   & \quad + \frac{J}{2} \sum_{\Gamma} \int d^2 \rho^{(\Gamma)} \, \tilde{\eta}_{\Gamma}({\pmb \rho^{(\Gamma)}}) \, {\rho^{(\Gamma)}}_i \, {\rho^{(\Gamma)}}_j \, \frac{\partial^2 g({\bf r} - {\bf r_{\Gamma}})}{\partial x_{\Gamma i} \partial x_{\Gamma j}}  \\
   & = J \int d^2r' \eta^{Dip} ({\bf r'}) g({\bf r} - {\bf r'}) + J \int d^2r' C_{ij}^{Dip}({\bf r'}) \frac{\partial^2 g({\bf r} - {\bf r'})}{\partial {x'}_i \partial {x'}_j},
\end{split}
\end{equation}
wobei
\begin{align}
   \label{306c}
   \eta^{Dip} ({\bf r'}) & = \sum_{\Gamma} \int d^2 \rho^{(\Gamma)} \, \tilde{\eta}_{\Gamma}({\pmb \rho^{(\Gamma)}}) \, \delta({\bf r'} - {\bf r_{\Gamma}}), \\
   \label{306d}
   C_{ij}^{Dip}({\bf r'}) & =  \sum_{\Gamma} \int d^2 \rho^{(\Gamma)} \tilde{\eta}_{\Gamma}({\pmb \rho^{(\Gamma)}}) \, {\rho^{(\Gamma)}}_i \, {\rho^{(\Gamma)}}_j \, \delta({\bf r'} - {\bf r_{\Gamma}}).
\end{align}
  
M"ochte man nun eine makroskopische Verzerrungsfunktion erhalten, so ist "uber Fl"achen $\Delta A \gg \zeta^2$ zu mitteln. Danach treten Dipole mit einer Separation $r < \zeta$ nur noch als polarisierbares Dielektrikum auf. Da eine Mittelung einer Gr"o"se $B({\bf r})$ "uber eine Fl"ache $\Delta A$ allgemein die  Form $\langle B({\bf r}) \rangle = \frac{1}{\Delta A} \int_{\Delta A} d^2 \xi \, B({\bf r} + {\pmb \xi})$ besitzt, finden wir exemplarisch f"ur den ersten Term von \refformel{306b}
\begin{equation}
   \begin{split}
   \label{306e}
   \langle \chi_1^{Dip} ({\bf r}) \rangle & = \frac{J}{\Delta A} \int_{\Delta A} d^2 \xi \int d^2r' \eta^{Dip} ({\bf r'}) g({\bf r} + {\pmb \xi} - {\bf r'}) \\
\\
   & = \frac{J}{\Delta A} \int_{\Delta A} d^2 \xi \int d^2r''  \eta^{Dip} ({\bf r''} + {\pmb \xi}) g({\bf r} - {\bf r''}) \\ 
\\
   & = J \int d^2 r'' \langle \eta^{Dip} ({\bf r''}) \rangle g({\bf r} - {\bf r''}).
\end{split}
\end{equation}
Die hier auftretende Mittelung von \refformel{306c} ist
\begin{equation}
   \label{306f}
   \langle \eta^{Dip} ({\bf r'}) \rangle = \frac{1}{\Delta A} \sum_{\Gamma} \int d^2 \rho^{(\Gamma)} \, \tilde{\eta}_{\Gamma}({\pmb \rho^{(\Gamma)}}) \, \int_{\Delta A} d^2 \xi \, \delta({\bf r'} + {\pmb \xi} - {\bf r_{\Gamma}}).
\end{equation}
Die $\delta$-Funktion gibt f"ur einen Dipol $\Gamma$ nur dann einen Beitrag, wenn dessen Zentrum ${\bf r_{\Gamma}}$ innerhalb der Fl"ache $\Delta A$ um ${\bf r'}$ liegt. Da $\Delta A \gg \zeta$ gilt, ist die Zahl der Dipole, die vollst"andig in $\Delta A$ liegen wesentlich gr"o"ser als die Zahl der Dipole, die nur zur H"alfte, das hei"st mit nur einer Versetzung in dieser Fl"ache liegen.  
 
Die bisher betrachtete Mittelung pro Fl"acheneinheit ist geeignet f"ur makroskopische Felder. Die Gr"o"se $\eta^{Dip} ({\bf r'})$ l"a"st sich aber auch pro Dipol mitteln. Ist $\Delta N$ die mittlere Zahl von Dipolen mit $r < \zeta$ in einer Fl"ache der Gr"o"se $\Delta A$, so hat eine solche Mittelung hat die Form
\begin{equation}
   \label{306g}
   \langle B({\bf r}) \rangle_{\Delta N} = \frac{1}{\Delta N} \int_{\Delta A} d^2 \xi \, B({\bf r} + {\pmb \xi}).
\end{equation}
Mit der makroskopische Dipoldichte $n^{\zeta}({\bf r'})$ bei ${\bf r'}$ erh"alt man
\begin{equation}
   \label{306h}
   \langle \eta^{Dip} ({\bf r'}) \rangle = \langle \eta^{Dip} ({\bf r'}) \rangle_{\Delta N}  \, n^{\zeta}({\bf r'}).
\end{equation}
 
Schreibt man die makroskop. Verzerrungsfunktion $\chi ({\bf r}) = \chi_{ex} ({\bf r}) + [ \langle \chi ({\bf r}) \rangle ]_D$ in der Form
\begin{equation}
   \label{307a}
   \chi ({\bf r}) \approx J \int d^2r' \eta^{\zeta} ({\bf r'}) g({\bf r} - {\bf r'}) + J \int d^2r' C_{ij}^{\zeta}({\bf r'}) \frac{\partial^2 g({\bf r} - {\bf r'})}{\partial {x'}_i \partial {x'}_j},
\end{equation}
so ergibt sich f"ur die makroskopische Quellenfunktion nach einer zus"atzlichen Mittelung "uber die Unordnung $[...]_D$ entsprechend \refformel{302c}, \refformel{306c} und \refformel{306h} 
\begin{equation}
   \label{308a}
   \eta^{\zeta}({\bf r}) = \left[\left< \sum_{\Gamma} \int d^2r' \tilde{\eta}_{\Gamma}({\bf r'}) \delta({\bf r} - {\bf r_{\Gamma}}) \right>_{\Delta N} n^{\zeta}({\bf r}) \right]_D + \eta_{ex} ({\bf r}).
\end{equation}
 
Geht man f"ur den zweiten \refformel{306b} analog zum ersten Term vor, erh"alt man den Tensor $C_{ij}$ auf makroskopischem Niveau, der der elektrischen Polarisation entspricht
\begin{equation}
   \label{309a}
   C_{ij}^{\zeta} ({\bf r}) = \frac{1}{2} \left[ \left< \sum_{\Gamma} \int d^2r' \tilde{\eta}_{\Gamma} ({\bf r'}) {x'}_i {x'}_j \delta({\bf r} - {\bf r_{\Gamma}}) \right>_{\Delta N} n^{\zeta}({\bf r}) \right]_D.
\end{equation}
 
Partielle Integration und Ableiten liefern aus \refformel{307a}
\begin{equation}
\begin{split}
    \label{310a}
    \nabla^4 \chi ({\bf r}) & = J \eta^{\zeta}({\bf r}) + J \int d^2r' \frac{\partial^2}{\partial {x'}_i \partial {x'}_j} C_{ij}^{\zeta} ({\bf r'}) \delta ({\bf r} - {\bf r'}) \vspace{0.8cm} \\
    & = J \eta^{\zeta}({\bf r}) + J \frac{\partial^2}{\partial x_i \partial x_j} C_{ij}^{\zeta} ({\bf r}).
\end{split}
\end{equation}
 
In der Elektrodynamik ist die Polarisation eine Funktion des elektrischen Felds und damit eine Funktion der Ableitung des elektrostatischen Potentials. Analog dazu ist der Tensor $C_{ij}$ hier eine Funktion der zweifachen Ableitung der Verzerrungsfunktion
\begin{equation}
   \label{311a}
   C_{ij}^{\zeta} ({\bf r}) = \sum_{kl} a_{ijkl}(\zeta) \frac{\partial^2 \chi ({\bf r})}{\partial x_k \partial x_l} + \sum_{klmn} b_{ijklmn}(\zeta) \frac{\partial^2 \chi ({\bf r})}{\partial x_k \partial x_l} \frac{\partial^2 \chi ({\bf r})}{\partial x_m \partial x_n} + ....
\end{equation}
Ber"ucksichtigt man nur den ersten Term und setzt ein isotropes Medium oder ein Dreiecksgitter an, so gilt in symmetrisierter Form
\begin{equation}
   \label{312a}
   C_{ij}^{\zeta} ({\bf r}) = \frac{1}{2} \, C_1({\zeta}) \left(\frac{\partial^2 \chi ({\bf r})}{\partial x_i \partial x_j} + \frac{\partial^2 \chi ({\bf r})}{\partial x_j \partial x_i} \right) + C_2({\zeta}) \delta_{ij} \frac{\partial^2 \chi ({\bf r})}{\partial^2 x_k}.
\end{equation}
Diese spezielle Form von $a_{ijkl}({\zeta})$ beruht auf der Tatsache, da"s dieser Tensor -  wie schon der Tensor $K_{ijkl}$ aus \refformel{205a} -  den Gitterinvarianzen gen"ugen mu"s. Eingesetzt mit \refformel{310a} ergibt sich jetzt
\begin{equation}
\begin{split}
   \label{313a}
   \varepsilon({\zeta}) \, \nabla^4 \chi ({\bf r}) & = 4 \pi \eta^{\zeta} ({\bf r}) \vspace{0.8cm} \\
   & = \frac{4 \pi}{J} \nabla^4 \chi ({\bf r}) - 4 \pi (C_1({\zeta}) + C_2({\zeta})) \, \nabla^4 \chi ({\bf r}).
\end{split}
\end{equation}
Wir haben nun einen Ausdruck f"ur die dielektrische Konstante gefunden, n"amlich
\begin{equation}
   \label{314a}
   \varepsilon(r) = \varepsilon_0 - 4 \pi (C_1(r) + C_2(r)).
\end{equation}
 
Ein Vergleich mit der "ublichen Formulierung der Elektrodynamik zeigt, da"s $- C_1 - C_2$ somit einer elektrischen Suszeptibilit"at $\kappa$ entspricht, die sich durch eine Dipolwahrscheinlichkeit $P(r, \theta)$ und eine Dipolpolarisierbarkeit $\alpha(r, \theta)$ f"ur Dipole mit Separation $r$ und Winkel $\theta$ wiefolgt ausdr"ucken l"a"st 
\begin{equation}
   \label{315a}
   \kappa(r) = -C_1 - C_2 = \int_0^r d^2r' [P(r', \theta) \, \alpha(r', \theta)]_D.
\end{equation}  
Ziel der folgenden Abschnitte ist es die relevanten Gr"o"sen zu bestimmen, so da"s mittels eines expliziten Ausdruckes f"ur die dielektrische Konstante $\varepsilon(r)$ der Renormierungsflu"s der Kopplungskonstanten $J(r)$ hergeleitet werden kann. 
 
\section{Dipol-Wahrscheinlichkeit}
Im folgenden soll die Wahrscheinlichkeit $P(r, \theta)$ bestimmt werden, einen Versetzungsdipol mit Burgersvektoren {\bf b} bei ${\bf r_{\alpha}}$ und {\bf -b} bei ${\bf r_{\beta}}$ zu finden. Der Winkel ${\theta}$ ist wiederum der Winkel zwischen Abstandsvektor und Burgers-Vektoren, ferner ist $r = |{\bf r_{\alpha}} - {\bf r_{\beta}}|$ die Dipolgr"o"se.
 
F"ur das weitere Vorgehen beschr"anken wir uns auf die Untersuchung des Quadratgitters. Die meisten Argumente k"onnen unver"andert f"ur das Dreiecksgitter "ubernommen werden, notwendige Modifikationen werden in Kapitel 4 untersucht. Offenbar existieren vier m"ogliche Orientierungen f"ur einen Dipol mit Burgers-Vektoren bei ${\bf r_{\alpha}}$ und ${\bf r_{\beta}}$, die in Abbildung 3.2 dargestellt sind.
 
\graphik{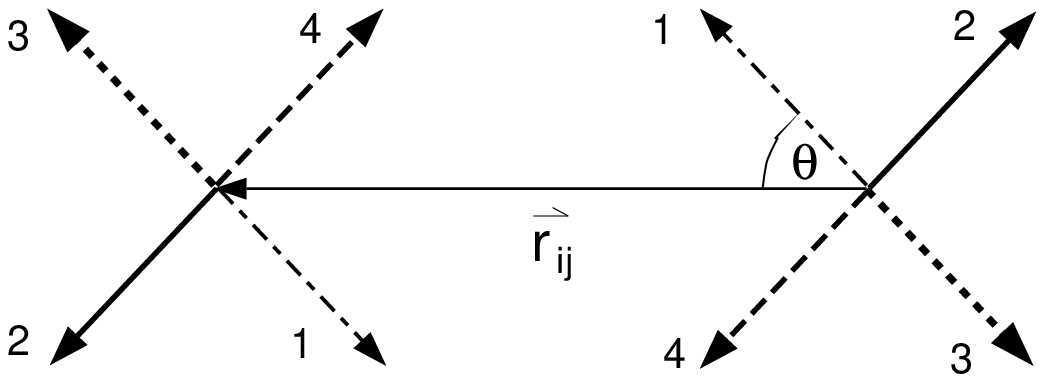}{4cm}{Schematische Darstellung der m"oglichen Orientierungen eines Dipols auf dem Quadratgitter}{302}
 
Die mit 1 und 3 gekennzeichneten Einstellungen haben beide die elastische Energie
\begin{equation}
   \label{316a}
   E = \frac{J}{4 \pi} \ln r - \frac{J}{4 \pi} \cos^2 \theta + 2 E_C,
\end{equation}
und einen Beitrag aus der Wechselwirkung mit der Unordnung von $V$, beziehungsweise von $-V$. F"ur die Einstellungen 2 und 4 ist 
\begin{equation}
   \label{317a}
   \tilde{E} = E \left(\theta + \frac{\pi}{2} \right) = \frac{J}{4 \pi} \ln r - \frac{J}{4 \pi} \sin^2 \theta + 2 E_C,
\end{equation}
w"ahrend die Unordnungs-Wechselwirkung hier mit $\tilde{V}$ oder $-\tilde{V}$ angesetzt wird. Ber"ucksichtigt man die M"oglichkeit, da"s kein Dipolpaar gebildet wird (Gewicht = 1, da keine Energie), so lautet die lokale Zustandssummme
\begin{equation}
   \label{317b}
   Z = 1 + w_+ + w_- + \tilde{w}_+ + \tilde{w}_-,
\end{equation} 
wobei die Gewichte durch
\begin{equation}
   \label{318a}
   w_{\pm} = e^{-\beta(E \pm V)} \quad \text{und} \quad \tilde{w}_{\pm} = e^{-\beta(\tilde{E} \pm \tilde{V})}
\end{equation}
gegeben sind. Somit ist die normierte Wahrscheinlichkeit ein {\it orientiertes} Paar mit dem Vektor $\theta$ zwischen Burgers-Vektoren und Abstandsvektor zu finden (das sind die Einstellungen 1 und 3 aus Abbildung 3.2)
\begin{equation}
   \label{319a}
   P(r, \theta) = \frac{1}{2} \left[\frac{w_+ + w_-}{1 + w_+ + w_- + \tilde{w}_+ + \tilde{w}_-} \right]_{D}.
\end{equation}
 
Da die Berechnung dieses Ausdrucks recht aufwendig ist, vergleichen wir ihn mit der Wahrscheinlichkeit $p(r, \theta)$, die gegeben w"are, g"abe es nur die Orientierungsm"oglichkeiten 1 und 3. Dann w"are die Wahrscheinlichkeit $p$ 
\begin{equation}
   \label{320a}
   p(r, \theta) = \frac{1}{2} \left[\frac{w_+ + w_-}{1 + w_+ + w_-} \right]_{D}.
\end{equation}
Offenbar ergibt sich direkt aus \refformel{319a} und \refformel{320a} 
\begin{equation}
   \label{321a}
   P(r, \theta) \le p(r, \theta).
\end{equation}
Da die Gesamtwahrscheinlichkeit Dipole mit Abstand $r$ zu finden durch zus"atzliche Einstellm"oglichkeiten auf keinen Fall absinken kann, mu"s auch die winkelgemittelte Dipolwahrscheinlichkeit $P(r)$ mit vier erlaubten Burgers-Vektor\-rich\-tungen gr"o"ser sein, als die winkelgemittelte Wahrscheinlichkeit $p(r)$ f"ur nur zwei erlaubte Vektorrichtungen. Es ist
\begin{align}
   \label{321b}
   P(r) & = \int_0^{2 \pi} \frac{d \theta}{2 \pi} [2P(r, \theta) + 2P(r, \theta + \pi / 2)] = 4 \int_0^{2 \pi} \frac{d \theta}{2 \pi} P(r, \theta) \\
   \label{321c}
   p(r) & = 2 \int_0^{2 \pi} \frac{d \theta}{2 \pi} p(r, \theta).
\end{align}
Mit \refformel{321a} folgt hieraus die Ungleichung
\begin{equation}
   \label{321d}
   \frac{1}{2} \int_0^{2 \pi} \frac{d \theta}{2 \pi} p(r, \theta) \le \int_0^{2 \pi} \frac{d \theta}{2 \pi} P(r, \theta) \le \int_0^{2 \pi} \frac{d \theta}{2 \pi} p(r, \theta).
\end{equation}
Da aber im folgenden Vorfaktoren nicht wesentlich sind (auch in \verweistext{nat95} wurden nur die Hauptabh"angigkeiten betrachtet), k"onnen wir somit als N"aherung
\begin{equation}
   \label{322a}
   P(r) \propto p(r) = 2 \int_0^{2 \pi} \frac{d \theta}{2 \pi} p(r, \theta)
\end{equation}
ansetzen.
 
\subsection{Wahrscheinlichkeit bei $T = 0$}
Setzt man die Gewichte ein, so liefert \refformel {322a}
\begin{equation}
\begin{split}
   \label{323a}
   p(r, \theta) & = \frac{1}{2} \: 2 \int_0^{\infty} \frac{dV}{\sqrt{2 \pi \Delta^2}} \, e^{-\frac{V^2}{2 \Delta^2}} \, \frac{e^{-\beta(E+V)} + e^{-\beta(E-V)}}{1 + e^{-\beta(E+V)} + e^{-\beta(E-V)}} \vspace{1.5cm} \\
\\
   & = \int_0^{\infty} \frac{dV}{\sqrt{2 \pi \Delta^2}} \, e^{-\frac{V^2}{2 \Delta^2}} \, \frac{e^{-\beta(E-V)}}{1 + e^{-\beta(E-V)}}.
\end{split}
\end{equation}
Die Summanden $w_+$ k"onnen, da sie einen postiven Beitrag aus der Unordnungswechselwirkung haben, bei tiefen Temperaturen allgemein ver\-nach\-l"as\-sigt werden. F"ur $T=0$ gilt insbesondere
\begin{equation}
   \label{324a}
   \frac{e^{-\beta(E-V)}}{1 + e^{-\beta(E-V)}} =
   \begin{cases}
   \: 0 & \quad \text{falls} \; E-V > 0 \\
   \: 1 & \quad \text{falls} \; E-V < 0
   \end{cases}
\end{equation}
Somit hat man f"ur gro"se $r$ ($\ln r \gg \bar{\sigma}$ und $\ln r \gg E_C/J$)
\begin{equation}
   \label{325a}
   \begin{split}
   p(r, \theta) & = \int_{E}^{\infty} \frac{dV}{\sqrt{2 \pi \Delta^2}} \, e^{- \frac{V^2}{2 \Delta^2}}\vspace{0.8cm} \\
\\
   & \approx \sqrt{\frac{\bar{\sigma}}{\ln r - \cos^2 \theta}} r^{- \frac{1}{4 \pi \bar{\sigma}}} e^{\frac{1}{4 \pi \bar{\sigma}} \cos^2 \theta},
\end{split}
\end{equation}
wie in \refformel{271a} aus Abschnitt 2.4 bereits hergeleitet worden ist. Da wir Flu"sgleichungen aber nicht nur bei $T=0$ berechnen m"ochten, ist eine Untersuchung auch bei h"oheren Temperaturen notwendig.
 
\subsection{Wahrscheinlichkeit bei $T<T^*$}
F"ur $T>0$ gilt \refformel{324a} nicht. Hier liefert auch der Fall $E-V > 0$ einen Beitrag zur Integration
\begin{equation}
  \label{326a}
  \begin{split}
  p(r, \theta) & = \int_0^{E} \frac{dV}{\sqrt{2 \pi \Delta^2}} \, e^{-\frac{V^2}{2 \Delta^2}} \, e^{-\beta(E-V)} + \int_E^{\infty} \frac{dV}{\sqrt{2 \pi \Delta^2}} \, e^{-\frac{V^2}{2 \Delta^2}} \\
\\
  & = \tilde{p}(r, \theta) + p_{T=0}(r, \theta),
   \end{split}
\end{equation}
der zu berechnen ist. Mittels Substitution erh"alt man f"ur diesen Anteil 
\begin{equation}
   \label{327a}
   \tilde{p}(r, \theta) = \frac{1}{\sqrt{\pi}} \, e^{-\beta E + \frac{1}{2} \beta^2 \Delta^2} \, \int_{\frac{\beta \Delta^2 - E}{\sqrt{2} \Delta}}^{\frac{\beta \Delta}{\sqrt{2}}} e^{-t^2} dt.
\end{equation}
 
Allgemein erf"ullt im Falle gro"ser Werte $x$ die Funktion $\Phi(x) := \frac{2}{\sqrt{\pi}} \int_0^x e^{-t^2} dt$ die Relation
\begin{equation}
   \label{328a}
   1 - \Phi(x) = \frac{1}{\sqrt{\pi}x} \, e^{-x^2} \, \left(1 - \frac{1}{2 x^2} + ... \right),
\end{equation}
die sich im Fall $T<T^*$ anwenden l"a"st, da $\beta$ hinreichend gro"s ist: 
\begin{equation}
\label{329a}
\begin{split}
   \tilde{p}(r, \theta) & = \frac{1}{2} \left[\Phi\left(\frac{\beta \Delta}{\sqrt{2}}\right) - \Phi\left(\frac{\beta \Delta^2 - E}{\sqrt{2} \Delta} \right) \right] \, e^{\frac{1}{2} \beta^2 \Delta^2 - \beta E}\\
\\
   & \approx \frac{1}{2 \sqrt{\pi}} \left[ \frac{\sqrt{2} \Delta}{\beta \Delta^2 - E} e^{-\frac{E^2}{2 \Delta^2}} - \frac{\sqrt{2} \Delta}{\beta \Delta^2} e^{-\beta E} \right]\\
\\
   & = \frac{E}{\beta \Delta^2 -E} \, p_{T=0}(r, \theta),
\end{split}
\end{equation}
dabei wird der zweite Term der zweiten Zeile aufgrund der niedrigen Temperatur vernachl"assigt. Mit \refformel{325a} ergibt sich f"ur
\begin{equation}
\label{330a}
\begin{split}
   p(r, \theta) & = \frac{\beta \Delta^2}{\beta \Delta^2 - E} p_{T=0}(r, \theta) \\
\\
   & \propto  r^{- \frac{1}{4 \pi \bar{\sigma}}} e^{\frac{1}{4 \pi \bar{\sigma}} \cos^2 \theta}.
\end{split}
\end{equation}
Diese Formel ist offenbar eine Verallgemeinerung der Aussage f"ur $T=0$. Es zeigt sich, da"s im gesamten neu zu betrachtenden Bereich $T<T^*$ die f"ur verschwindende Temperatur gefundenen Abh"angigkeiten dominieren. Die Dipolwahrscheinlichkeit $P(r)$ ist in diesem Bereich somit
\begin{equation}
   \label{330b}
   \begin{split}
   P(r) & \propto 2 \int_0^{2 \pi} \frac{d \theta}{2 \pi} r^{- \frac{1}{4 \pi \bar{\sigma}}} e^{\frac{1}{4 \pi \bar{\sigma}} \cos^2 \theta} \\
\\
   & \propto 2 r^{- \frac{1}{4 \pi \bar{\sigma}}} e^{\frac{1}{8 \pi \bar{\sigma}}} I_0 \left(\frac{1}{8 \pi \bar{\sigma}} \right),
\end{split}
\end{equation}  
denn es gilt $e^{x \cos^2 \theta} = e^{\frac{x}{2}} e^{\frac{x}{2} \cos 2 \theta}$, und die modifizierte Besselfunktion ist durch $I_0(x) = \int_0^{2 \pi} \frac{d \theta}{2 \pi} e^{x \cos 2 \theta}$ definiert.

\subsection{Wahrscheinlichkeit bei $T \ge T^*$}
Analog zum obigen Fall gilt auch im hohen Temperaturbereich \refformel{327a}. Hier strebt allerdings $\frac{\beta \Delta}{\sqrt{2}} \rightarrow 0$, w"ahrend die untere Integrationsgrenze f"ur gro"se L"angenskalen unbeschr"ankt ins Negative w"achst. Wegen $\int_{-\infty}^0 e^{-x^2} dx = \frac{\sqrt{\pi}}{2}$ gilt im diesem Fall
\begin{equation}
   \label{331a}
   \tilde{p}(r, \theta) \propto e^{\frac{1}{2} \beta^2 \Delta^2 - \beta E}.
\end{equation}
Da dieser Term den Beitrag von $p_{T=0}$ bei hohen Temperaturen "uberwiegt, kann letzterer vernachl"assigt werden. Setzt man die bekannten Werte f"ur $E$ und $\Delta$ (aus \refformel{268a}) ein, so resultiert abschlie"send f"ur $p(r, \theta)$
\begin{equation}
   \label{332a}
   p(r, \theta) \propto
   \begin{cases}
   \: r^{- \frac{1}{4 \pi \bar{\sigma}}} e^{\frac{1}{4 \pi \bar{\sigma}} \cos^2 \theta} & \quad \text{f"ur} \; T<T^* \\
&\\
   \: r^{-\frac{J}{4 \pi T} \left(1-\frac{J \bar{\sigma}}{4 T} \right)} e^{\frac{J}{4 \pi T} \left(1-\frac{J \bar{\sigma}}{4 T} \right) \cos^2 \theta} & \quad \text{f"ur} \; T \ge T^*
   \end{cases}
\end{equation}
wobei die Bedingungen $\ln r \gg \bar{\sigma}$ und $\ln r \gg \frac{E_C}{J}$ erf"ullt sein m"ussen.
 
F"ur die winkelgemittelte Wahrscheinlichkeit $P(r)$ einen Dipol mit Separation $r$ zu finden, erh"alt man daher
\begin{equation}
   \label{332b}
   P(r) \propto
   \begin{cases}
   \: 2 r^{- \frac{1}{4 \pi \bar{\sigma}}} e^{\frac{1}{8 \pi \bar{\sigma}}} I_0 \left(\frac{1}{8 \pi \bar{\sigma}} \right) & \quad \text{f"ur} \; T<T^* \\
&\\
   \: 2 r^{-\frac{J}{4 \pi T} \left(1-\frac{J \bar{\sigma}}{4 T} \right)} e^{\frac{J}{8 \pi T} \left(1-\frac{J \bar{\sigma}}{4 T} \right)} I_0 \left(\frac{J}{8 \pi T} \left(1-\frac{J \bar{\sigma}}{4 T} \right) \right) & \quad \text{f"ur} \; T \ge T^*
   \end{cases}
\end{equation}
 
Beide Proportionalit"aten der Wahrscheinlichkeitsdichte liefern f"ur $T = T^*= \frac{J \bar{\sigma}}{2}$ dasselbe Resultat, sie sind somit konsistent. Durch den Ansatz \refformel{319a} f"ur eine {\it normierte} Wahrscheinlichkeit ergibt sich bei hohen Temperaturen zwar eine Boltzmann-verteilte Wahrscheinlichkeit, die \verweistext{nel83} im gesamten Temperaturbereich angesetzt hat. Bei niedrigen Temperaturen $T < T^*$, wo die Boltzmann-Verteilung zu Widerspr"uchen in Nelsons Arbeit f"uhrt, liefert obiger Ansatz eine {\it Fermi-verteilte} Wahrscheinlichkeit. Aufgrund der Ber"ucksichtigung des fermionischen Charakters der Versetzungsdipole treten bei tiefen Temperaturen nun im Rahmen des G"ultigkeitsbereichs obiger Formeln keine sehr gro"sen Dipolwahrscheinlichkeiten $P(r) \gg 1$ mehr auf. Daher ist es gerechtfertig im folgenden dielektrischen Formalismus die Versetzungspaare als {\it nicht-wechselwirkende Fermionen} zu behandeln. 
 
\section{Polarisierbarkeit}
 
Mit obiger Wahrscheinlichkeitsdichte ist es m"oglich, die elektrische Suszeptibilit"at zu berechnen. Dazu wird der {\it Dipoltensor} $Q_{ij}(r, \theta)$ eines Dipols mit Abstand $r$ und Winkel $\theta$ ben"otigt, der nach \refformel{309a} die Form
\begin{equation}
\label{334a}
\begin{split}
   Q_{ij}(r, \theta) & = \frac{1}{2} \int d^2r \, \eta_{r, \theta}({\bf r}) \, x_i x_j \\
\\
& = \frac{1}{2} \epsilon_{ik} \epsilon_{jl} (b_k \epsilon_{lm} x_m + b_l \epsilon_{km} x_m)
\end{split}
\end{equation}
annimmt. Hierbei ist $\eta_{r, \theta}({\bf r})$ die Quellenfunktion des Dipols. Die Wechselwirkung dieses Dipols mit einem von au"sen angelegten homogenen Feld entspricht formal der Quadrupolwechselwirkung der zweidimensionalen Elektrodynamik, wobei aufgrund der Definition von $Q_{ij}$ kein Vorfaktor auftritt. Diese Wechselwirkung ist 
\begin{equation}
   \label{335a}
   W = \sum_{ij} Q_{ij} E_{ij}.
\end{equation}
Der Tensor $E_{ij}$ symbolisiert die zweifache Ableitung der Verzerrungsfunktion $E_{ij} := \nabla_i \nabla_j \chi$. Die Verzerrung, die $\chi$ darstellt, wird durch eine von au"sen an das System angelegte Konfiguration hervorgerufen. Analog zu Elektrodynamik, wo man Konfigurationen betrachtet, die zu einem homogenen, konstanten E-Feld f"uhren, sollen die Konfigurationen hier so gew"ahlt sein, da"s der Tensor $E_{ij}$ konstant ist. 
 
Auch hier gen"ugt es, sich bei der Untersuchung auf den Fall mit zwei erlaubten Burgers-Vektorrichtungen zu beschr"anken, da die hier untersuchte Polarisierbarkeitsdichte und damit die Suszeptibilit"at durch die Einf"uhrung zus"atzlicher Ausrichtungen nicht abnehmen darf. Es gilt somit eine zu \refformel{321d} analoge Relation f"ur diese Gr"o"sen.
 
Die {\it Polarisationsdichte} $q_{ij}(r, \theta)$ f"ur einen Versetzungsdipol mit Burgers-Vektor {\bf b} bei ${\bf r_{\alpha}}$ und mit {\bf -b} bei ${\bf r_{\beta}}$ ist somit
\begin{equation}
   \label{336a}
   q_{ij}(r, \theta) = \left[ \frac{w_+ - w_-}{1 + w_+ + w_-} Q_{ij}(r, \theta) \right]_D,
\end{equation}
mit
\begin{equation}
   \label{337a}
   w_{\pm} = \exp \left[ -\beta(E \pm V \pm W) \right].
\end{equation}
 
Mittels dieser Polarisationsdichte l"a"st sich formal eine Gr"o"se $A_{ijkl}$ definieren, aus deren Komponenten dann mit \refformel{312a} die Polarisierbarkeitsdichte der Versetzungspaare bei Abstand $r$ ermittelt werden kann. 
\begin{equation}
   \label{338a}
   A_{ijkl}(r) := 2 \, \frac{1}{2} \, \int_0^{2 \pi} \frac{d \theta}{2 \pi} \left[ \frac{\partial q_{ij}(r, \theta)}{\partial E_{kl}} \right]_D \bigg\arrowvert_{E=0}
\end{equation}
 
Diese Definition entspricht dem Vorgehen in der Elektrodynamik, wo ein ebenso angesetzter Tensor die Berechnung der Polarisierbarkeit erlaubt. Diese Analogie wird ausf"uhrlich in Anhang C.1 gezeigt.
 
Der Vorfaktor $\frac{1}{2}$ ist notwendig aufgrund der Tatsache, da"s in \refformel{336a} eine Orientierung nur modulo 180 Grad ausgezeichnet ist, der Faktor 2, da es auf dem Quadratgitter neben den beiden in \refformel{336a} ber"ucksichtigten Burgers-Vektor - Orientierungen mit Winkel $\theta$ noch zwei weitere mit Winkel $\theta + \frac{\pi}{2}$ gibt, die denselben Beitrag zur Polarisationsdichte liefern.

\subsection{Polarisierbarkeit ohne Unordnung}
Zun"achst werten wir \refformel{338a} f"ur den Fall ohne Unordnung aus. Hier erh"alt man nach der Ableitung
\begin{equation}
\label{339a}
\begin{split}
    A_{ijkl}(r) & = \int_0^{2 \pi} \frac{d \theta}{2 \pi} \left[ \frac{ \beta Q_{ij} Q_{kl} [(w_+ + w_-) (1 + w_+ + w_-) -  (w_+ - w_-)^2]}{(1 + w_+ + w_-)^2} \right] \bigg\arrowvert_{E=0} \\
\\
& = - \int_0^{2 \pi} \frac{d \theta}{2 \pi} \beta Q_{ij} Q_{kl} \left[ \frac{w_+ + w_-}{1 + w_+ + w_-} \right]  \bigg\arrowvert_{E=0} \\
\\
& = - \int_0^{2 \pi} \frac{d \theta}{2 \pi} \beta Q_{ij} Q_{kl} \, 2 \, p(r, \theta),
\end{split}
\end{equation}
dabei wurde benutzt, da"s ohne Unordnung und ohne "au"seres Feld $w_+ - w_- = 0$ gilt. Wie in Anhang C.2 hergeleitet wird, gilt folgende n"utzliche Relation
 
\begin{equation}
\label{340a}
\begin{split}
   & \int_0^{2 \pi} \frac{d \theta}{2 \pi} \, Q_{ij} Q_{kl} \, e^{x \cos^2 \theta}\\
\\
   & = \frac{1}{4} \epsilon_{ir} \epsilon_{js} \epsilon_{kt} \epsilon_{lu} \, e^{\frac{x}{2}} \, \left[Q_1(r) (\delta_{rt} \delta_{su} + \delta_{ru} \delta_{ts}) + Q_2(r)  \delta_{rs} \delta_{tu} \right],
\end{split}
\end{equation}
wobei
\begin{equation}
\label{341a}
\begin{split}  
   Q_1(r) & = \frac{1}{2} r^2 \, I_0\left(\frac{x}{2} \right)\\
   Q_2(r) & = -\frac{1}{2} r^2 \, I_1\left(\frac{x}{2} \right).
\end{split}
\end{equation} 
   
$I_0$ und $I_1$ sind modifizierte Besselfunktionen, deren Definition durch
\begin{equation}
\label{342a}
\begin{split}
   I_0(x) & := \int_0^{2 \pi} \frac{d \theta}{2 \pi} e^{x \cos 2 \theta}\\
\\
   I_1(x) & := \int_0^{2 \pi} \frac{d \theta}{2 \pi} \cos 2 \theta \, e^{x \cos 2 \theta}
\end{split}
\end{equation}
gegeben ist. Definiert man nun die Gr"o"sen $A_1(r)$ und $A_2(r)$ wie folgt
\begin{equation}
   \label{343a}
   C_i(r) = \int_0^r d^2r' A_i(r') \quad \text{mit} \: i = (1,2) ,
\end{equation}
so resultiert aufgrund von \refformel{312a} und der speziellen Form der Dipol-Wahr\-schein\-lich\-keit ohne Unordnung ($p(r, \theta) = r^{-\frac{J}{4 \pi T}} e^{\frac{J}{4 \pi T} \cos^2 \theta}$), die die Benutzung von \refformel{340a} erm"oglicht:
\begin{equation}
\label{344a}
\begin{split}
   \frac{1}{2} \, A_1(r) & = A_{xyxy}(r) = A_{xyyx}(r) = A_{yxxy}(r) = A_{yxyx}(r)\\
\\
   & = - \frac{2}{T}\, \frac{1}{4} \, r^{-\frac{J}{4 \pi T}} \, e^{\frac{J}{8 \pi T}} \, \frac{r^2}{2} \, I_0 \left(\frac{J}{8 \pi T} \right) \\
\\
\\
   2 A_2(r) & = A_{xxxx}(r) + A_{xxyy}(r) - A_1(r) = ...\\
\\
   & = -\frac{r^{-\frac{J}{4 \pi T}}}{2T} \, e^{\frac{J}{8 \pi T}} \, r^2 \, \left(I_0 \left(\frac{J}{8 \pi T} \right) - I_1 \left(\frac{J}{8 \pi T} \right) - I_0 \left(\frac{J}{8 \pi T} \right) \right) \\
\\
   & = \frac{1}{2T} \, r^{-\frac{J}{4 \pi T}} \, e^{\frac{J}{8 \pi T}} \, r^2 \, I_1 \left(\frac{J}{8 \pi T} \right).
\end{split}
\end{equation}
 
Mit Gleichungen \refformel{314a} und  \refformel{315a} zeigt sich nun, da"s die Polarisierbarkeitsdichte $A(r)$ folgenden Wert annimmt
\begin{equation}
\label{345a}
\begin{split}
   A(r) & = - \left(A_1(r) + A_2(r) \right) \\
\\
   & = \frac{r^2}{2T} \, r^{-\frac{J}{4 \pi T}}  e^{\frac{J}{8 \pi T}} \left( I_0 \left(\frac{J}{8 \pi T} \right) -\frac{1}{2} I_1 \left(\frac{J}{8 \pi T} \right) \right).
\end{split}
\end{equation}
 
Diese Polarsierbarkeitsdichte stimmt mit den von Kosterlitz und Thouless \verweistext{kt73} angegebenen Werten f"ur $C_1$ und $C_2$ "uberein.
 
F"ur die {\it Polarisierbarkeit} $\alpha(r)$ eines Versetzungspaares folgt unter Beachtung der Relation \refformel{315a}
\begin{equation}
\label{346a}
\begin{split}
   \alpha(r) & = \frac{A(r)}{2\, \frac{1}{2} \int \frac{d \theta}{2 \pi} p(r, \theta)}\\
\\
   & = \frac{\int d \theta \, \frac{r^2}{2T} \left(1-\frac{1}{2} \cos 2 \theta \right) r^{-\frac{J}{4 \pi T}} e^{\frac{J}{4 \pi T} \cos^2 \theta}}{\int d \theta \, r^{-\frac{J}{4 \pi T}} e^{\frac{J}{4 \pi T} \cos^2 \theta}} \\
\\
   & = \frac{r^2}{2T} \left(1 - \frac{1}{2} <\cos 2 \theta>_{ang} \right).
\end{split}
\end{equation}
 
Ohne Unordnung ist dieser Wert f"ur die Polarisierbarkeit bereits von Young \verweistext{you79} auf eine etwas andere Weise hergeleitet worden. Das Symbol $<...>_{ang}$ bedeutet eine Mittelung "uber die Winkel. Der Wichtungsfaktor ist die Dipolwahrscheinlichkeit und damit vom Temperaturbereich abh"angig.
 
\subsection{Polarisierbarkeit mit Unordnung}
Dieses bis hierher entwickelte Prinzip zur Behandlung der Polarisierbarkeit ohne Unordnung kann durch eine Erweiterung auch auf F"alle mit Unordnung "ubertragen werden. Hierzu ist eine Unordnungsmittelung in Formel \refformel{339a} notwendig. Der Tensor $A_{ijkl}(r)$ wird modifiziert zu
\begin{equation}
\label{347a}
    A_{ijkl}(r) = \int_0^{2 \pi} \frac{d \theta}{2 \pi} \, \beta Q_{ij} Q_{kl} \underbrace{\left[ \frac{w_+ + w_- + 4 w_+ w_-}{(1 + w_+ + w_-)^2} \bigg\arrowvert_{E=0} \right]_D}_{:= \pi(r, \theta)}.
\end{equation}
Es ist also gegen"uber dem bisher Berechneten zus"atzlich der Ausdruck $\pi(r, \theta)$ auszuwerten. Beschr"ankt man sich dabei in Z"ahler und Nenner auf die f"uhrenden Terme, so folgt 
\begin{equation}
\label{348a}
\begin{split}
   \pi(r, \theta) & \approx 2 \, \int_0^{\infty} \frac{dV}{\sqrt{2 \pi \Delta^2}} \, e^{-\frac{V^2}{2 \Delta^2}} \frac{e^{-\beta (E-V)}}{1 + e^{-2\beta (E-V)}} \\
\\
   & \approx 2 \, \left[\int_0^{E} \frac{dV}{\sqrt{2 \pi \Delta^2}} \, e^{-\frac{V^2}{2 \Delta^2} -\beta(E-V)} + \int_E^{\infty} \frac{dV}{\sqrt{2 \pi \Delta^2}} \, e^{-\frac{V^2}{2 \Delta^2} +\beta(E-V)} \right].
\end{split}
\end{equation}
Unter Benutzung der Methoden und Schreibweise aus Abschnitt 3.2.2 l"a"st sich dies vereinfachen zu
\begin{equation}
\label{349a}
\begin{split}
   \pi(r, \theta) & = 2 \tilde{p}(r, \theta) + \frac{2}{\sqrt{\pi}} \, e^{\beta E + \frac{1}{2} \beta^2 \Delta^2} \, \int_{\frac{\beta \Delta^2 + E}{\sqrt{2} \Delta}}^{\infty} e^{-t^2} dt \\
\\
   & = 2 \tilde{p}(r, \theta) + e^{\beta E + \frac{1}{2} \beta^2 \Delta^2} \, \left[ \Phi \left(\frac{\beta \Delta^2 + E}{\sqrt{2} \Delta} \right) - \Phi \left(\infty \right) \right] \\
\\
   & = 2 \tilde{p}(r, \theta) + 2 \, \frac{E}{E + \beta \Delta^2} \, p_{T=0}(r, \theta).
\end{split}
\end{equation}
 
Im hohen Temperaturbereich $T > T^*$ gilt $\tilde{p} \gg p_{T=0}$, daher kann man den zweiten Term in \refformel{349a} vernachl"assigen. Ein identisches Vorgehen wie im vorherigen Abschnitt ist m"oglich, da auch hier die Form von $\tilde{p}$ die Anwendung von \refformel{340a} erlaubt. Wir erhalten f"ur die Polarisierbarkeitsdichte und die Polarisierbarkeit
\begin{align}
   \label{350a}
   A(r) & = \frac{r^2}{2T} \, r^{-\frac{4TJ - J \bar{\sigma}}{16 \pi T^2}}  e^{\frac{4TJ - J \bar{\sigma}}{32 \pi T^2}} \left[ I_0 \left(\frac{4TJ - J \bar{\sigma}}{32 \pi T^2} \right) -\frac{1}{2} I_1 \left(\frac{4TJ - J \bar{\sigma}}{32 \pi T^2} \right) \right]\\
   \label{351a}
   \alpha(r) & = \frac{r^2}{2T} \left(1 - \frac{1}{2} <\cos 2 \theta>_{ang>} \right).
\end{align}
 
F"ur Temperaturen unterhalb der Grenze $T^*$ gibt Relation \refformel{330a}
\begin{equation}
\label{352a}
\begin{split}
   \pi(r, \theta) & = 2 \, \left(\frac{E}{\beta \Delta^2 - E} + \frac{E}{E + \beta \Delta^2} \right) \, p_{T=0}(r, \theta) \\
\\
    & \approx 2 \, \frac{T}{T + T^*} \, p_{T=0}(r, \theta).
\end{split}
\end{equation}
Nach Anwendung des bekannten Verfahrens findet man auch hier die  gesuchten Gr"o"sen.
 
Insgesamt ergibt sich so f"ur die Polarisierbarkeitsdichte $A(r)$:
\begin{equation}
   \label{353a}
   A(r) =
   \begin{cases}
   \: \frac{r^2}{T + T^*} \, r^{- \frac{1}{4 \pi \bar{\sigma}}} e^{\frac{1}{8 \pi \bar{\sigma}}} \left[ I_0 \left(\frac{1}{8 \pi \bar{\sigma}} \right) -\frac{1}{2} I_1 \left(\frac{1}{8 \pi \bar{\sigma}} \right) \right] & \quad \text{f"ur} \; T<T^* \\
&\\
   \: \frac{r^2}{2T} \, r^{-\frac{4TJ - J \bar{\sigma}}{16 \pi T^2}}  e^{\frac{4TJ - J \bar{\sigma}}{32 \pi T^2}} \left[ I_0 \left(\frac{4TJ - J \bar{\sigma}}{32 \pi T^2} \right) -\frac{1}{2} I_1 \left(\frac{4TJ - J \bar{\sigma}}{32 \pi T^2} \right) \right] & \quad \text{f"ur} \; T \ge T^*
   \end{cases}
\end{equation}
 
Analog ist die Polarisierbarkeit 
\begin{equation}
   \label{354a}
   \alpha(r) =
   \begin{cases}
   \: \frac{r^2}{T+T^*} \left(1 - \frac{1}{2} <\cos 2 \theta>_{ang<} \right) & \quad \text{f"ur} \; T<T^* \\
&\\
   \: \frac{r^2}{2T} \left(1 - \frac{1}{2} <\cos 2 \theta>_{ang>} \right) & \quad \text{f"ur} \; T \ge T^*
   \end{cases}
\end{equation}
Die Notationen der Winkelmittelung sollen den Unterschied der zu benutzenden Gewichte verdeutlichen. Es sind
\begin{equation}
\label{355a}
\begin{split}
   <...>_{ang<} &: e^{\frac{1}{4 \pi \bar{\sigma}} \cos^2 \theta} \\
   <...>_{ang>} &: e^{\frac{J}{4 \pi T}(1 - \frac{J \bar{\sigma}}{4T}) \cos^2 \theta}.
\end{split}
\end{equation}
 
Diese Resultate sind konsistent bei $T = T^*$. Die angestrebten Flu"sgleichungen bei Renormierung k"onnen jetzt hergeleitet werden, indem man die gewonnen Ergebnisse in \refformel{314a} einsetzt und \refformel{302a} beachtet. Das n"achste Kapitel wird diese Flu"sgleichungen und ihre physikalischen Konsequenzen behandeln.

\chapter{Flu"sgleichungen und Phasendiagramme}
\section{Ergebnisse f"ur das Quadratgitter}
In Kapitel 3 sind alle notwendigen Relationen hergeleitet worden, um das Schmelzen kristalliner Filme mit quadratischem Gitter zu analysieren. Das Skalenverhalten der dielektrischen Konstante und damit der Kopplungskonstante bei Renormierung kann aus den Ergebnissen gewonnen werden. Dieses Skalenverhalten l"a"st direkte Schl"usse auf das gesuchte Phasendiagramm zu, welches die feste Phase mit quasi-langreichweitiger Ordnung bez"uglich Translation von einer nicht-kristallinen Phase trennt, die diese Ordnung nicht zeigt. 

\subsection{Herleitung der Flu"sgleichungen}
Zun"achst ist durch \refformel{314a} der Wert der renormierten dielektrischen Konstanten in Abh"angigkeit der L"angenskala gegeben. Mit der Beziehung zwischen Suszeptibilit"at zur Polarisierbarkeitsdichte aus \refformel{343a} gilt daher
\begin{equation}
   \label{401a}
   \varepsilon(r+dr) = \varepsilon(r) + 4 \pi A(r) \, 2 \pi r \, dr.
\end{equation}
Nun ist es m"oglich die "Anderung der Kopplungskonstanten $J(r)$ zu ermitteln, die resultiert, wenn man bei der Renormierung von einer L"angenskala $r$ zu einer Skala $r + dr$ "ubergeht, das hei"st den Abschirmeffekt derjenigen Versetzungspaaren zu ber"ucksichtigen, deren Separation zwischen $r$ und $r + dr$ liegt. \refformel{302a} gibt mit Hilfe einer Entwicklung 
\begin{equation}
\label{402a}
\begin{split}
   J(r + dr) & = \frac{4 \pi}{\varepsilon(r+dr)} = \frac{4 \pi}{\varepsilon(r) + 8 \pi^2 \, A(r) \, r \, dr} \\
\\
   & \approx J(r) - J(r)^2 \, A(r) \, 2 \pi r \, dr + O(A^2).
\end{split}
\end{equation}
 
Hieraus folgt sofort die Flu"sgleichung der Kopplungskonstanten. Setzt man $l:= \ln r$, so ist
\begin{equation}
   \label{403a}
   \frac{dJ}{dl} = r \frac{dJ}{dr} = - J(r)^2 \, A(r) \, 2 \pi r^2 + O(A^2).
\end{equation}
 
Wir f"uhren nun den bereits in Kapitel 2 behandelten Begriff der {\it Fugazit"at} $y$ ein. Wegen \refformel{269a} ist der Ausgangswert dieser Fugazit"at im nicht-renormier\-ten System mit $r = 1$ gegeben durch
\begin{equation}
   \label{404a}
   y_0 = e^{-\frac{E_C}{T}} = e^{-\frac{(\tilde{C} + 1)J}{8 \pi T}},
\end{equation}
die Konstante $\tilde{C}$ betr"agt nach Anhang A etwa $\frac{\pi}{2}$. In renormierten Systemen wird die Fugazit"at als $y = e^{-F_C/T}$ aufgefa"st, wobei $F_C$ die freie Energie eines Versetzungskerns auf der entsprechenden Skala $r$ ist. Da die Zahl der durch Renormierung verlorengegangenen m"oglichen Dipolpl"atze pro Platzpaar $\propto r^4$ skaliert und die Wahrscheinlichkeit einen Versetzungsdipol zu finden $p(r, \theta)$ betr"agt, ergibt sich f"ur die renormierte Fugazit"at 
\begin{equation}
   \label{405a}
   y(r)^2 = \, <r^4 \, p(r, \theta)>.
\end{equation}
Hierbei bezeichnet $<...>$ eine n"aher zu bestimmende Art der Winkelmittelung. Ber"ucksichtigt man f"ur hohe Temperaturen $T>T^*$, den bisher vernachl"assigten Beitrag der Energie der Versetzungskerne $E_C$ zur Dipolwahrscheinlichkeit, so modifiziert sich \refformel{332a} zu
\begin{equation}
   \label{406a}
   p(r, \theta) \propto r^{-\frac{J}{4 \pi T} \left(1-\frac{J \bar{\sigma}}{4 T} \right)} e^{\frac{J}{4 \pi T} \left(1-\frac{J \bar{\sigma}}{4 T} \right) \cos^2 \theta} e^{-\frac{2 E_C}{T}}.
\end{equation}
Dies zeigt, da"s \refformel{405a} f"ur kleine $r$ der Forderung $y_0 = e^{-E_C/T}$ nur dann gen"ugen kann, wenn die Winkelmittelung die bereits f"ur die Polarisierbarkeit(vgl. \refformel{355a}) benutzte Form hat:
\begin{equation}
   \label{407a}
   y(r)^2 = \, <r^4 \, p(r, \theta)>_{ang}.
\end{equation}
Diese Definition der renormierten Fugazit"at stimmt mit den von Nelson, Halperin und Young genutzten Ans"atzen \verweistext{nel83, you79, nh79} "uberein. F"ur tiefe Temperaturen $T<T^*$ findet man $y_0 = e^{-E_C/T}$ allerdings nicht, wenn man die Dipolwahrscheinlichkeit entsprechend \refformel{332a} ansetzt, welche  aber auch nur f"ur gro"se Werte von $r$ korrekt ist. Eine genauere Untersuchung f"ur kleine L"angenskalen gibt auch hier das richtige Ergebnis.
 
Schlie"slich bleibt noch zu bemerken, da"s mittels dieser Definition die Fugazit"at im relevanten Bereich $\bar{\sigma} < \sigma$ und $T<T^*$ klein bleibt. Hier gilt auf gro"sen Skalen $|y| \ll 1$. Wir f"uhren jetzt noch eine n"utzliche Variable $\tau^*$ ein, die im folgenden die expliziten Fallunterscheidungen zwischen hohem und niedrigem Temperaturbereich "uberfl"ussig macht
\begin{equation}
   \label{408a}
   \tau^* = min(1, \frac{2T}{J \bar{\sigma}})
\end{equation}
 
Somit nimmt $\tau^*$ im Bereich $T>T^*$ den Wert 1 an, im Bereich $T<T^*$ den Wert $\tau^* = \frac{T}{T^*}$. Insbesondere ist diese Definition konsistent bei $T=T^*$. F"ur die renormierte Fugazit"at erh"alt man hiermit auf gro"sen Skalen
\begin{equation}
   \label{410a}
   y(l)^2 = r^4 \, r^{-\frac{J \tau^*}{4 \pi T} \left(1-\frac{J \bar{\sigma} \tau^*}{4 T} \right)} \Big\arrowvert_{r=e^l}.
\end{equation}
 
Die hier sowie im folgenden auftretenden Gr"o"sen sind immer renormiert. Mit dieser Fugazit"at und \refformel{403a} ergeben sich die gesuchten Renormierungs-Flu"s\-gleich\-ungen f"ur das quadratische zweidimensionale Gitter:
\begin{align}
\label{411a}
   \frac{dJ}{dl} = & - \frac{2 \pi J^2}{T + T(\tau^*)^{-1}} \, y^2 \, \exp \left({\frac{J \tau^*}{8 \pi T} \left(1 - \frac{J \bar{\sigma} \tau^*}{4 T} \right)} \right) \notag \\
   & \times \left[I_0 \left( \frac{J \tau^*}{8 \pi T} \left(1 - \frac{J \bar{\sigma} \tau^*}{4 T} \right) \right) - \frac{1}{2} I_1 \left( \frac{J \tau^*}{8 \pi T} \left(1 - \frac{J \bar{\sigma} \tau^*}{4 T} \right) \right) \right] \notag \\
\\
\label{412a}
   \frac{dy}{dl} = & \left(2 - \frac{J \tau^*}{8 \pi T} \left(1 - \frac{J \bar{\sigma} \tau^*}{4 T} \right) \right) \, y \notag \\
\\
\label{413a}
   \frac{d \bar{\sigma}}{dl} = & 0
\end{align}
 
Bei diesen Flu"sgleichungen wurden lediglich die ersten beiden Ordnungen in $y$ ber"ucksichtigt, was aufgrund obiger Ausf"uhrungen gerechtfertigt ist. Eine Renormierung der Unordnungsst"arke kann mit dem angewandten Formalismus nicht gefunden werden. Die von Tang \verweistext{tang96} und Scheidl \verweistext{sts97} gefundene Unordnungsrenormierung am XY-Modell m"usste bei Verwendung der von ihnen genutzten Formalismen auch bei diesem Problem auftreten.  
 
Die Flu"sgleichungen stimmen f"ur Systeme ohne Unordnung mit den Ergebnissen von Nelson, Halperin und Young "uberein \verweistext{you79, nh79}. Den Fall mit Unordnung hat Nelson \verweistext{nel83} lediglich am Dreiecksgitter untersucht, aber eine Anwendung der dort benutzten Methodik auf das Quadratgitter g"abe eine "Ubereinstimmung im gesamten Bereich $T>T^*$. Bei tiefen Temperaturen w"urde sich mit Nelsons Methodik auch f"ur das Quadratgitter ein Wiedereintritt in die nicht-kristalline Phase ergeben, den unsere Gleichungen nicht aufweisen.
 
\subsection{Flu"s und Phasendiagramm}
Wertet man die Flu"sgleichung der Fugazit"at, so ist folgendes Verhalten offensichtlich
\begin{equation}
   \label{414a}
   \frac{dy}{dl}
   \begin{cases}
   \: < 0 & \quad \text{falls} \; T<T^* \, \text{und} \, \bar{\sigma} < \frac{1}{16 \pi} \\
   & \quad \text{oder} \; T\ge T^* \, \text{und} \, \bar{\sigma} < \frac{4T}{J} - \frac{64 \pi T^2}{J^2}\\
\\
   \: > 0 & \quad \text{sonst}
   \end{cases}
\end{equation}   
Wie bereits in Abschnitt 2.3.1 erl"autert, kann in einem Bereich, in dem $y$ auf gro"sen Skalen immer weiter ansteigt, keine geordnete Phase vorliegen. Aufgrund der Proportionalit"at zwischen $y^2$ und der Dipolwahrscheinlichkeit, wird diese dann f"ur $r \rightarrow \infty$ sehr gro"s, was gleichbedeutend mit der Existenz freier Versetzungen ist. Eine Bedingung f"ur die feste Phase ist also $\frac{dy}{dl} < 0$ auf gro"sen L"angenskalen.
 
M"ochte man den durch \refformel{411a} bis \refformel{413a} erzeugten Flu"s analog zu Abbildung 2.3 veranschaulichen, so ist es zun"achst notwendig 
\begin{equation}
   \label{415a}
   \frac{dy}{dK^{-1}} = \frac{dy}{dl} \frac{dl}{dJ} \frac{dJ}{dK}\, (-K^2)
\end{equation}
mit $K:= \frac{J}{T}$ zu berechnen. Im Bereich $K^{-1} > \frac{\bar{\sigma}}{2}$ ist keine Ver"anderung des von Nelson gewonnen Flu"sbildes (vgl. Abbildung 2.3) zu erwarten, da die Flu"sgleichungen im Bereich hoher Temperatur unver"andert sind. Interessant hingegen ist das Verhalten im tiefen Temperaturbereich.
 
Numerische Integration von \refformel{415a} liefert f"ur einen festen Wert von $\bar{\sigma}$ den in Abbildung 4.1 dargestellten Renormierungsflu"s.

\graphik{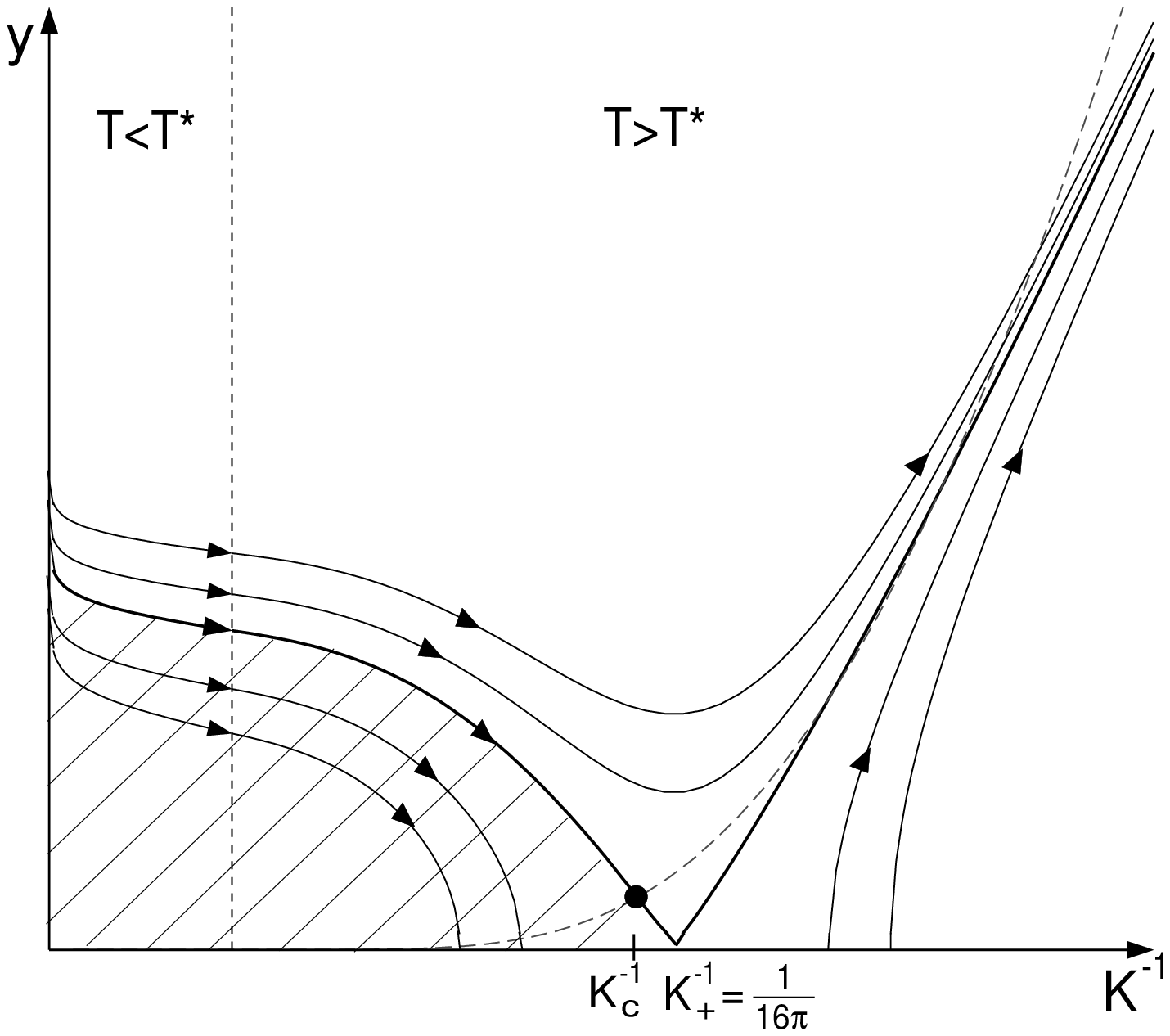}{7cm}{Renormierungsflu"s des Quadratgitters bei $\bar{\sigma} = \frac{1}{32 \pi}$}{401}
 
Der Startpunkt des Flusses befindet sich immer auf der gestrichelten Linie $y_0$. Die dick eingezeichnete Linie ist die Separatrix, die einen schraffierten Bereich abtrennt. Sie m"undet bei $K_+^{-1}$ in die Linie $y = 0$. Der Schnittpunkt der Separatrix mit der Linie $y_0$ liefert einen Wert $K_c^{-1}$. F"ur Anfangswerte mit $K^{-1} < K_c^{-1}$, also bei gen"ugend tiefer Temperatur, liegt der Startpunkt im schraffierten Bereich, wo der Flu"s immer bei $y = 0$ und einem endlichen Wert von $K$ endet. Hier befindet man sich somit in der festen Phase. Liegt der Startpunkt au"serhalb des schraffierten Bereichs, w"achst die Fugazit"at immer st"arker an, man befindet sich in der nicht-kristallinen Phase. 
 
Somit ist $T_c(\bar{\sigma}) =  K_c^{-1}J$, die reale Phasen"ubergangs-Temperatur, sie f"allt f"ur $E_C \rightarrow \infty$ mit $T_+$ zusammen. Analysiert man f"ur alle Werte von $\bar{\sigma}$ den Renormierungsflu"s, so ergibt sich das Phasendiagramm f"ur das System, dargestellt in Abbildung 4.2.
 
\graphik{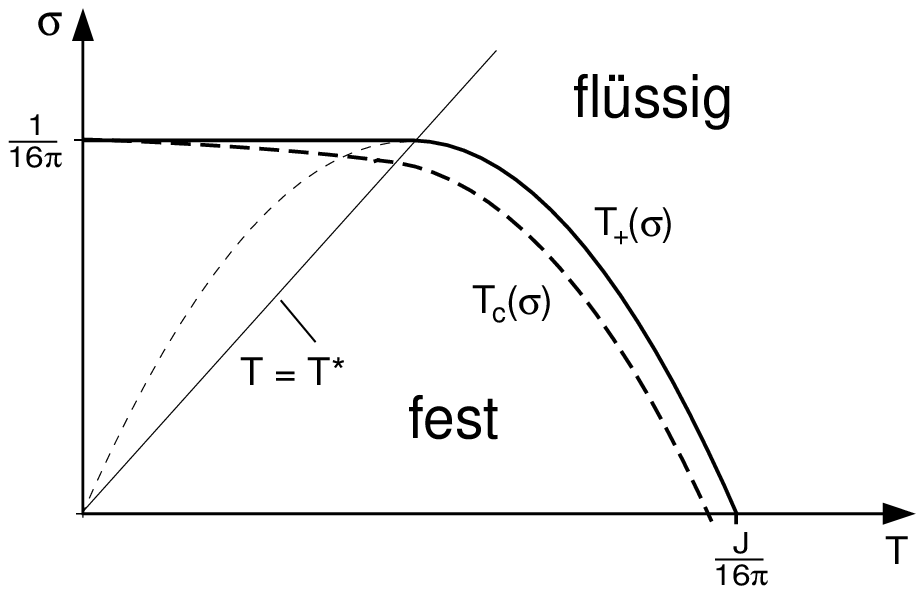}{6cm}{Phasendiagramm f"ur das Quadratgitter. Es zeigt die Abh"angigkeit zwischen kritischer Temperatur und Unordnungsst"arke}{402}
 
Die durchgezogene Linie stellt die renormierte Phasengrenze $T_+(\bar{\sigma})$ dar, w"ah\-rend die reale Phasengrenze $T_c(\bar{\sigma})$ durch die gestrichelte Linie wiedergegeben ist. Im Gegensatz zum Phasendiagramm von Nelson (d"un  gestrichelte Linie) findet man hier keinen Wiedereintritt in die ungeordnete Phase. Daher steht dieses Ergebnis auch nicht im Widerspruch zu der Arbeit von Ozeki und Nishimori \verweistext{on93}. 

\section{Ergebnisse f"ur das Dreiecksgitter}
Der wesentliche Unterschied zwischen Quadrat- und Dreiecksgitter, der dazu f"uhrt, da"s die Ergebnisse des Quadratgitters nicht einfach auf das Dreiecksgitter "ubertragbar sind, ist die Tatsache, da"s auf dem Dreiecksgitter auch drei Burgers-Vektoren so kombiniert werden k"onnen, da"s sie in der Summe ladungsneutral sind. Wir werden diesen Unterschied mit Hilfe des von Young \verweistext{you79} benutzten Formalismus behandeln.
 
\subsection{Modifikationen f"ur das Dreiecksgitter}
Eine solche ladungsneutrale Kombination von drei Burgers-Vektoren, im folgenden auch Tripol genannt, zeigt Abbildung 4.3. Der Abstand $r'$ sei der kleinste Abstand zweier Burgers-Vektoren. "Uberschreitet die Renormierung nun die L"ange $r'$, so werden die Vektoren 1 und 2 zu {\it einem} Burgers-Vektor zusammengefa"st. Auf gr"o"seren L"angenskalen liegt somit effektiv ein Dipol vor. Tripole k"onnen daher in die Berechnung einbezogen werden, in dem man erlaubt, da"s {\it einer} der beiden Burgers-Vektoren jedes Versetzungspaares mit Separation $r$ aus zwei Burgers-Vektoren mit einem Abstand $r' < r$ zusammengesezt ist.
 
Offenbar "andert sich hiermit die Dipolwahrscheinlichkeit, was Neuuntersuchungen erforderlich macht.
 
\graphik{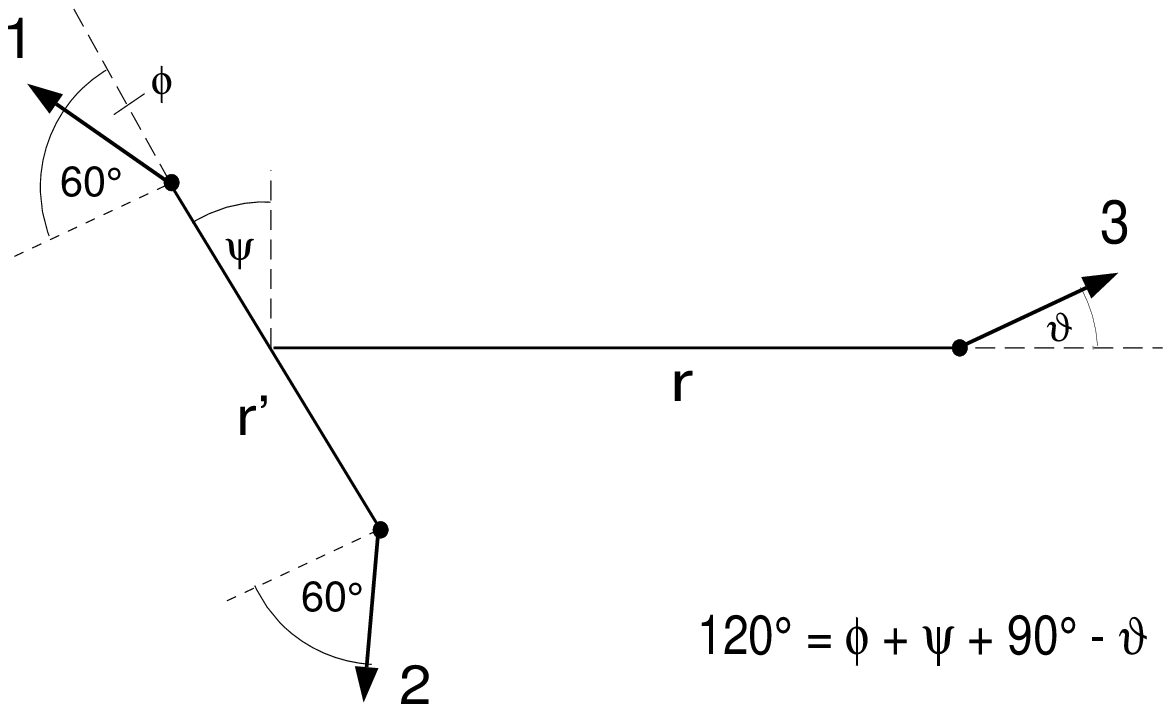}{6cm}{Darstellung eines Tripols. Fa"st man die Vektoren 1 und 2 zusammen, so erh"alt man effektiv einen Dipol mit Separation $r$ und Winkel $\theta$}{403}
 
Zun"achst ist es notwendig die elastische Energie ohne Unordnung dieser Konfiguration zu ermitteln. Wie auch f"ur Dipole auf dem Dreiecksgitter, hat ein Tripol bei festen Postionen sechs Orientierungsm"oglichkeiten f"ur die Burgers-Vektoren. Abbildung 4.3 zeigt eine davon. F"ur den Fall $r \gg r'$ findet man aufgrund der Relation \refformel{248a}
\begin{equation}
   \label{416a}
   \begin{split}
   E_{Tri} = & \: 2 \, \frac{J}{8 \pi} \ln r + \frac{J}{8 \pi} \ln r' + 3 E_C + \frac{J}{4 \pi} \cos \theta \, \cos (\theta + \frac{2 \pi}{3}) \\
   &  + \frac{J}{4 \pi} \cos \theta \, \cos (\theta - \frac{2 \pi}{3}) + \frac{J}{4 \pi} \cos \phi \, \cos (\phi + \frac{2 \pi}{3}). 
\end{split}
\end{equation}
Einfache Umformungen mit $\gamma := \psi - \theta$ ergeben
\begin{equation}
   \label{417a}
   \begin{split}
   E_{Tri} & = \frac{J}{4 \pi} \ln r + \frac{J}{8 \pi} \ln r' + 3 E_C - \frac{J}{4 \pi} \cos^2 \theta \\
   & \quad - \frac{J}{4 \pi} \cos^2 \gamma + \frac{J}{16 \pi} \\
\\
   & = E_{Dip} + \frac{J}{8 \pi} \ln r' - \frac{J}{4 \pi} \cos^2 \gamma + E_C,
\end{split}
\end{equation}
wobei $E_{Dip}$ die bereits in \refformel{269a} angegebene elastische Energie eines Versetzungsdipols mit Abstand $r$ ist. 
 
Au"ser der elastischen Energie mu"s auch die Varianz der Wechselwirkung des Tripols mit der Unordnung berechnet werden. Diese Varianz ist
\begin{equation}
   \label{418a}
   \Delta^2_{Tri} = \left[(V({\bf r_1}) + V({\bf r_2}) + V({\bf r_3}))^2 \right]_D 
\end{equation}
Bei Berechnung dieses Ausdrucks ist zu ber"ucksichtigen, da"s die Burgers-Vek\-toren hier nicht, wie bei Dipolen antiparallel zueinanderstehen, sondern der Winkel zwischen ihnen jeweils $\frac{2 \pi}{3}$ betr"agt. Ist $\Delta^2_{Dip} = \Delta^2(r, \theta)$ die Varianz der Unordnungswechselwirkung eines Dipols entsprechend \refformel{268a}, so gilt wegen der Burgers-Vektor-Stellung
\begin{equation}
   \label{419a}
   \Delta^2_{Tri} = \Delta^2_{Dip} + [V({\bf r_1})^2]_D + 2\, [V({\bf r_2})V({\bf r_1})]_D.
\end{equation}
Aufgrund der Berechnungen aus Anhang B erh"alt man
\begin{equation}
   \label{420a}
   \Delta^2_{Tri} = \Delta^2_{Dip} + \frac{J^2 \bar{\sigma}}{16 \pi} \ln r' + \frac{J^2 \bar{\sigma}}{8 \pi} \cos \phi \, \cos(\phi + \frac{2 \pi}{3}). 
\end{equation}
Formt man dies analog zu \refformel{417a} um, so resultiert abschlie"send
\begin{equation}
   \label{421a}
   \Delta^2_{Tri} = \Delta^2_{Dip} + \frac{J^2 \bar{\sigma}}{16 \pi} \ln r' - \frac{J^2 \bar{\sigma}}{8 \pi} \cos^2 \gamma + \frac{J^2 \bar{\sigma}}{32 \pi}. 
\end{equation}
 
Die modifizierte Dipolwahrscheinlichkeit setzt sich nun aus dem reinen Dipolanteil und dem Tripolanteil zusammen. Entsprechend \refformel{332a} gilt
\begin{equation}
   \label{422a}
   p(r, \theta) \propto
   \begin{cases}
   \:  \exp \left(-\frac{E^2_{Dip}}{2 \Delta^2_{Dip}} \right) + \exp \left(-\frac{E^2_{Tri}}{2 \Delta^2_{Tri}} \right)  & \quad \text{f"ur} \; T<T^* \\
&\\
   \: \exp \left(\frac{1}{2} \beta \Delta^2_{Dip} - \beta E_{Dip} \right) + \exp \left(\frac{1}{2} \beta \Delta^2_{Tri} - \beta E_{Tri} \right) & \quad \text{f"ur} \; T \ge T^*
   \end{cases}
\end{equation}
 
Setzt man alle bekannten Gr"o"sen ein und benutzt den Parameter $\tau^*$, so ist
\begin{equation}
   \label{423a}
   p(r, \theta) = r^{-\frac{J \tau^*}{4 \pi T} \left(1 - \frac{J \bar{\sigma} \tau^*}{4 T} \right)} \, e^{\frac{J \tau^*}{4 \pi T} \left(1 - \frac{J \bar{\sigma} \tau^*}{4 T} \right) \cos^2 \theta} \, \left(1 + e^{-\frac{J \tau^*}{16 \pi T} \left(1 - \frac{J \bar{\sigma} \tau^*}{4 T} \right)} \, q(r) \right).
\end{equation}

Hierbei gibt $q(r)$ die Zahl der M"oglichkeiten an, da"s einer der beiden Burgers-Vektoren (daher Faktor 2) des Dipols aus zwei Vektoren kleineren Abstands kombiniert ist:
\begin{equation}
   \label{424a}
   q(r) = 2 \, \int_0^{2 \pi} \int_1^r d \gamma \, dr' \, r' \,  {r'}^{-\frac{J \tau^*}{8 \pi T} \left(1 - \frac{J \bar{\sigma} \tau^*}{4 T} \right)} \, e^{\frac{J \tau^*}{4 \pi T} \left(1 - \frac{J \bar{\sigma} \tau^*}{4 T} \right) \cos^2 \gamma}
\end{equation}

Indem man erlaubt, da"s einer der Vektoren, die den schon zusammengesetzten Dipolvektor bilden wiederum aus zwei Vektoren noch geringerer Separation gebildet ist, kann dieser Betrachtung die von Young geforderte Selbstkonsistenz gegeben werden. Diese Art der Einf"uhrung von Quadrupolen etc. f"uhrt allerdings lediglich zu Korrekturen h"oherer Ordnung, die im folgenden nicht von Interesse sind und auf die daher verzichtet werden kann.
 
Soll wie f"ur das Quadratgitter die Relation
\begin{equation}
   \label{425a}
   y(r)^2 = \, <r^4 \, p(r, \theta)>_{ang}
\end{equation} 
gelten, so ist die Fugazit"at $y$ f"ur das Dreiecksgitter wie folgt zu definieren
\begin{equation}
   \label{426a}
   y(l)^2 = r^4 \, r^{-\frac{J \tau^*}{4 \pi T} \left(1 - \frac{J \bar{\sigma} \tau^*}{4 T} \right)} \, \left(1 + e^{-\frac{J \tau^*}{16 \pi T} \left(1 - \frac{J \bar{\sigma} \tau^*}{4 T} \right)} \, q(r) \right) \Big\arrowvert_{r=e^l}.
\end{equation}
 
Bei Untersuchung kleiner L"angenskalen gilt auch bei dieser Definition der geforderte Anfangswert
\begin{equation}
   \label{427a}
   y_0 = e^{-\frac{E_C}{T}}.
\end{equation} 
 
Somit sind effektive Dipolwahrscheinlichkeit und Fugazit"at dem Dreiecksgitter angepa"st worden. Die modifizierten Flu"sgleichungen folgen hieraus sofort.

\subsection{Flu"sgleichungen f"ur das Dreiecksgitter}
Wie schon im Falle des Quadratgitters entwickeln wir auch hier die Flu"sgleichungen bis zur zweiten Ordnung in $y$. Mit den Ab\-lei\-tungs\-regeln ergibt sich f"ur den Renormierungsflu"s der Fugazit"at
\begin{equation}
   \label{428a}
   \frac{dy}{dl} = \frac{r}{2y} \, \frac{dy^2}{dr}.
\end{equation}
 
Arbeitet man mit der modifizierten Fugazit"at, bleibt der Flu"s der Kopplungskonstante in den betrachteten Ordnungen formal unver"andert. Es erfolgt lediglich die "Anderung des Vorfaktors, da in \refformel{338a} der erste Faktor eine 3 statt einer 2 sein mu"s, befindet man sich doch jetzt auf dem Dreiecksgitter, wo es insgesamt sechs Orientierungsm"oglichkeiten f"ur einen Dipol mit festen Positionen gibt. Die Renormierungs-Flu"sgleichungen f"ur das Dreiecksgitter nehmen somit folgende Form an:
\begin{align}
\label{429a}
   \frac{dJ}{dl} = & - \frac{3 \pi J^2}{T + T(\tau^*)^{-1}} \, y^2 \, \exp \left({\frac{J \tau^*}{8 \pi T} \left(1 - \frac{J \bar{\sigma} \tau^*}{4 T} \right)} \right) \notag \\
   & \times \left[I_0 \left( \frac{J \tau^*}{8 \pi T} \left(1 - \frac{J \bar{\sigma} \tau^*}{4 T} \right) \right) - \frac{1}{2} I_1 \left( \frac{J \tau^*}{8 \pi T} \left(1 - \frac{J \bar{\sigma} \tau^*}{4 T} \right) \right) \right] \\
\notag \\
\label{430a}
   \frac{dy}{dl} = & \left(2 - \frac{J \tau^*}{8 \pi T} \left(1 - \frac{J \bar{\sigma} \tau^*}{4 T} \right) \right) \, y \notag\\
   & + 2 \pi \, \exp \left({\frac{J \tau^*}{16 \pi T} \left(1 - \frac{J \bar{\sigma} \tau^*}{4 T} \right)} \right) \, I_0 \left( \frac{J \tau^*}{8 \pi T} \left(1 - \frac{J \bar{\sigma} \tau^*}{4 T} \right) \right) \, y^2 \\
\notag \\
\label{431a}
   \frac{d \bar{\sigma}}{dl} = & 0
\end{align}
 
Wie bereits erw"ahnt ist mit dem verwendeten Formalismus keine Renormierung der Unordnungsst"arke festzustellen, hierzu gelten die schon f"ur das Quadratgitter gemachten Bemerkungen. Wie schon die Flu"sgleichungen des Quadratgitters, stimmen die des Dreiecksgitters mit den von Nelson, Halperin und Young gefundenen Gleichungen f"ur Systeme ohne Unordnung "uberein. Zus"atzlich ist die "Ubereinstimmung mit Nelsons Resultaten f"ur ungeordnete Systeme \verweistext{nel83} im Bereich $T>T^*$ zu erw"ahnen. Ebenso wie im Fall des Quadratgitters aber zeigen obige Flu"sgleichungen auch f"ur das Dreiecksgitter nicht den von Nelson vorhergesagten Wiedereintritt in die ungeordnete Phase.  
 
\subsection{Phasendiagramm f"ur das Dreiecksgitter}
Aufgrund des zweiten Summanden in der Flu"sgleichung der Fugazit"at kann $\frac{dy}{dl}$ hier nicht so einfach abgesch"atzt werden wie im Fall des Quadratgitters. Offenbar bleibt \refformel{414a} aber g"ultig f"ur $y \rightarrow 0$, so da"s $K_+^{-1} \: (K:= \frac{J}{T})$ nicht ver"andert wird und keine "Anderung der renormierten Phasengrenze $T_+(\bar{\sigma})$ eintritt.
 
Dennoch ist es notwendig den Flu"s genauer zu untersuchen, so da"s er "uber numerische Intergration $\frac {dy}{dK^{-1}}$ veranschaulicht werden mu"s. Geht man analog zum Quadratgitter vor, so erh"alt man den in Abbildung 4.4 dargestellten Renormierungsflu"s. Auch hier ergibt sich im Bereich $K^{-1} > \frac{\bar{\sigma}}{2}$ keine "Anderung gegen"uber Nelsons Ergebnissen. 

\graphik{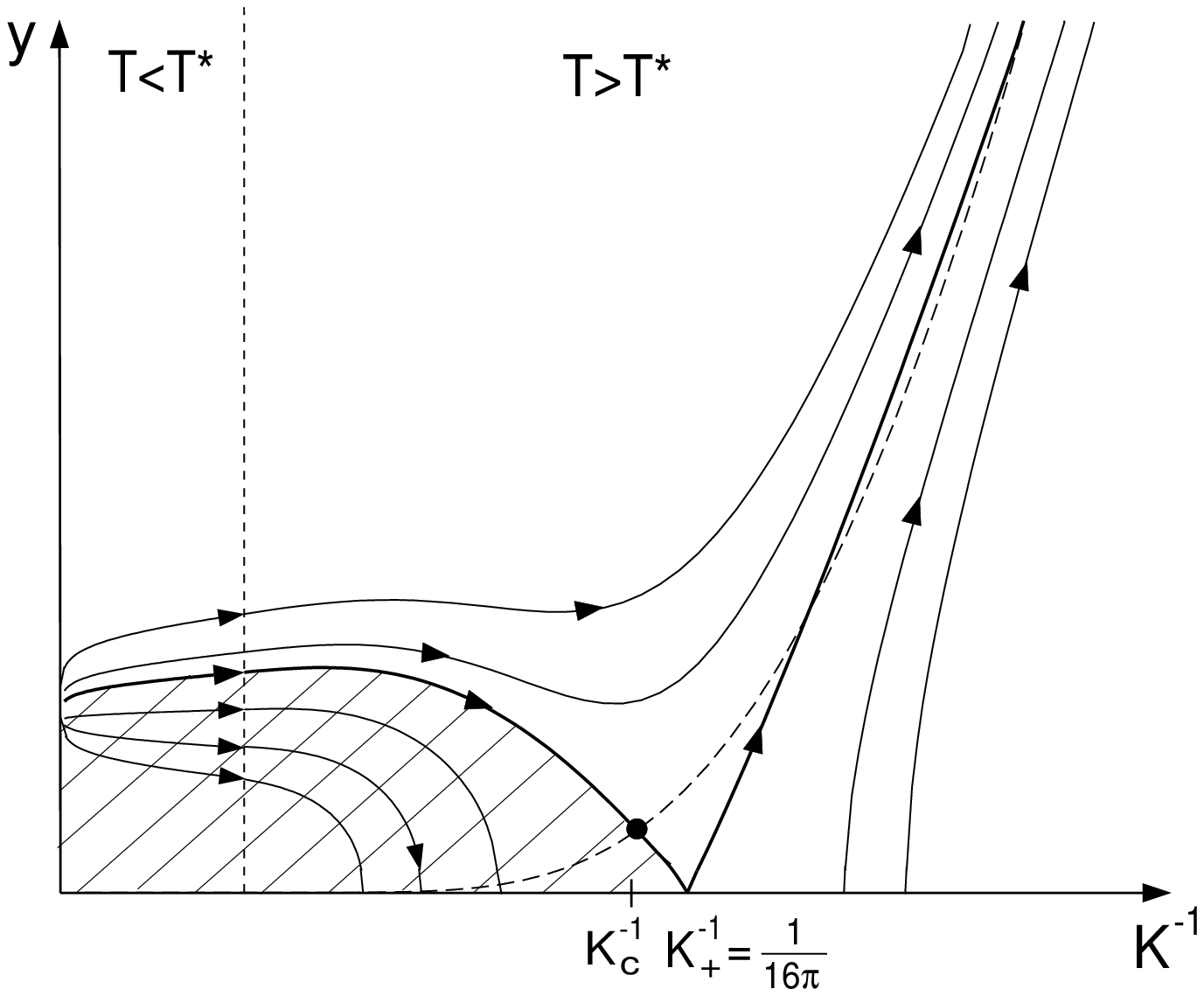}{7cm}{Renormierungsflu"s des Dreiecksgitters bei $\bar{\sigma} = \frac{1}{32 \pi}$}{404}
 
Auch hier befindet man sich nur f"ur Startpunkte im schraffierten, durch die dicker eingezeichnete Separatrix abgetrennten Bereich, in der kristallinen festen Phase. F"ur Startpunkte au"serhalb dieses Bereichs endet der Flu"s bei beliebig gro"sen Fugazit"aten, also in der nicht-kristallinen Phase. Da alle realen Startpunkte auf der gestrichelten Linie $y_0$ liegen, gibt der Schnittpunkt $K_c^{-1}$ die reale Phasen"ubergangs-Temperatur.
 
Ein auffallender Unterschied zum Renormierungsflu"s des Quadratgitters ist, da"s der Flu"s f"ur $K^{-1} \rightarrow 0$ nicht im Unendlichen, sondern bei einem festen, nur von $\bar{\sigma}$ abh"angigen Wert, n"amlich bei
\begin{equation}
   \label{432a}
   y(K^{-1} = 0) = \left(\frac{1}{8 \pi \bar{\sigma}} - 2 \right) \, e^{-\frac{1}{16 \pi \bar{\sigma}}} \, \frac{1}{2 \pi I_0 \left(\frac{1}{8 \pi \bar{\sigma}} \right)} 
\end{equation}
beginnt. Dies hat keine direkte physikalische Konsequenz, da der Wert f"ur $\bar{\sigma} < \frac{1}{16 \pi}$ immer gr"o"ser als null ist und daher kein Wiedereintritt in die ungeordnete Phase m"oglich ist. Dennoch stellt sich die Frage, ob dieses Verhalten physikalisch ist, oder nur durch zu grobe Absch"atzungen in Abschnitt 4.2.1 hervorgerufen wird. 
 
Analysiert man hier den den Renormierungsflu"s f"ur alle Werte von ${\bar{\sigma}}$, so erh"alt man das in Abbildung 4.5 wiedergegebene Phasendiagramm des Dreiecksgitters. 
 
\graphik{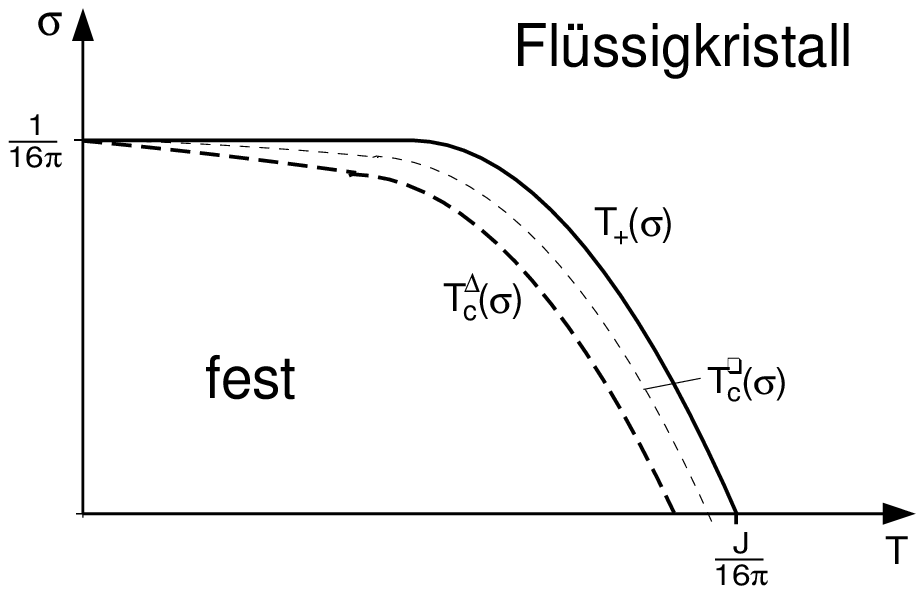}{6cm}{Phasendiagramm f"ur das Dreiecksgitter. Es zeigt die Abh"angigkeit zwischen kritischer Temperatur und Unordnungsst"arke}{405}
 
Wiederum stellt die durchgezogene Linie die renormierte Phasengrenze $T_+({\bar{\sigma}})$ dar, die gegen"uber dem Quadratgitter unver"andert bleibt. Lediglich die reale Phasengrenze $T_c^{\vartriangle}({\bar{\sigma}})$ liegt bei einer tieferen Temperatur als die beim Quadratgitter $T_c^{\square}({\bar{\sigma}})$. Ein wesentlicher Unterschied zwischen den Phasendiagrammen des Quadrat- und des Dreiecksgitters kann also nicht festgestellt werden. In beiden F"allen erfolgt kein Wiedereintritt in die ungeordnete Phase.
 
\section{Entropieflu"s}
Abschlie"send stellt sich die Frage, ob die gewonnen Ergebnisse zu einem nicht negativen und damit physikalischen Entropieflu"s f"uhren. Wie bereits in Abschnitt 2.3.2 angemerkt, kann der Beitrag zur freien Energiedichte $dF_V$, den Versetzungsdipole mit einer Gr"o"se zwischen  $r$ und $r+dr$ leisten, durch 
\begin{equation}
   \label{433a}
   dF_V \approx - T \, r dr \frac{1}{4} \int_0^{2 \pi} d \theta \, [\ln Z]_D
\end{equation}
angen"ahert werden. Hierbei ist $Z$ die lokale Zustandssumme, die f"ur das Quadratgitter durch \refformel{317b} gegeben ist. In diesem Fall ist somit
\begin{equation}
   \label{434a}
   dF_V \approx - T \, r dr \frac{1}{4} \int_0^{2 \pi} d \theta \, [\ln (1 + w_+ + w_- + \tilde{w}_+ + \tilde{w}_-)]_D,
\end{equation}
da vier m"ogliche Dipoleinstellungen existieren. Deren elastische Energien $E$ und $\tilde{E}$ (gegeben durch \refformel{316a} und \refformel{317a}) lassen sich nun in abstandsabh"angige Anteile $E_r$ und $\tilde{E}_r$, sowie in winkelabh"angige Anteile $E_w$ und $\tilde{E}_w$ aufteilen, so da"s $E = E_r + E_w$ und $\tilde{E} = \tilde{E}_r + \tilde{E}_w$ gilt. Auf gen"ugend gro"sen L"angenskalen $\ln r \gg \frac{E_C}{J}$ erh"alt man so
\begin{equation}
   \label{435a}
   E_r = \tilde{E}_r = \frac{J}{4 \pi} \ln r 
\end{equation}
und
\begin{equation}
   \label{436a}
   E_w = - \frac{J}{4 \pi} \cos^2 \theta , \quad \tilde{E}_w = - \frac{J}{4 \pi} \sin^2 \theta.
\end{equation}
 
Ableiten von \refformel{434a} nach $e^{- \beta E_r}$ liefert
\begin{equation}
   \label{437a}
   e^{- \beta E_r} \frac{\partial (dF_V)}{\partial (e^{- \beta E_r})} = - T \, r dr \frac{1}{4} \int_0^{2 \pi} d \theta \left[\frac{w_+ + w_- + \tilde{w}_+ + \tilde{w}_-}{1 + w_+ + w_- + \tilde{w}_+ + \tilde{w}_-} \right]_{D}.
\end{equation}
Mit den Berechnungen aus Abschnitt 3.2 erh"alt man sofort
\begin{equation}
   \label{438a}
   e^{- \beta E_r} \frac{\partial (dF_V)}{\partial (e^{- \beta E_r})} = - \frac{\pi}{2} T \, r dr \, P(r),
\end{equation}
wobei $P(r)$ die durch \refformel{332b} gegebene winkelgemittelte Wahrscheinlichkeit ist, einen Dipol mit Separation $r$ bei vorgegebenen Versetzungspositionen zu finden. F"ur das Dreiecksgitter gilt diese Relation auch, allerdings ist hier f"ur die Wahrscheinlichkeit $P(r)$ der sich aus \refformel{423a} ergebende Ausdruck einzusetzen.
 
Im Fall $T \ge T^* = \frac{J \bar{\sigma}}{2}$ folgt daraus 
\begin{equation}
   \label{439a}
   d F_V = - \frac{\pi}{2} T  \, r dr \, P(r),
\end{equation}
da die $P(r)$ hier linear in $e^{- \beta E_r}$ sind. F"ur den Flu"s der freien Energiedichte ergibt sich
\begin{equation}
   \label{440a}
   \begin{split}
   \frac{dF_V}{dl} = & - \frac{\pi}{2} T r^2 P(r)\\
\\
   = & - \pi T \frac{y^2}{r^2} e^{\frac{J}{8 \pi T} \left(1 - \frac{J \bar{\sigma}}{4T} \right)} I_0 \left(\frac{J}{8 \pi T} \left(1 - \frac{J \bar{\sigma}}{4T} \right) \right).
\end{split}
\end{equation}
Dieses Resultat entspricht, abgesehen von Vorfaktoren, den von Tang \verweistext{tang96} und Scheidl \verweistext{sts97} gefundenen Ergebnissen f"ur das ungeordnete XY-Modell im Bereich $T > T^*$.
 
F"ur die Fugazit"at $y$ in \refformel{440a} ist der zum jeweiligen Gittertyp passende Ausdruck anzusetzen. Offenbar entspricht Gleichung \refformel{440a} der in Abschnitt 2.3.2 gewonnenen Gleichung \refformel{258f}. Dort konnte bereits gezeigt werden, da"s der hieraus mittels $\frac{d S_V}{d l} = - \frac{\partial}{\partial T} \frac{d F_V}{d l}$ folgende Flu"s der Entropie im Fall $T \ge T^*$ positiv ist. Beim "Ubertragen der dort benutzten Argumente auf \refformel{440a} ist lediglich $\bar{\sigma}$ durch $\frac{\bar{\sigma}}{4}$ zu ersetzen. F"ur $T \ge T^*$ liegt somit kein unphysikalisches Verhalten vor.
 
Der interessantere Fall ist $T < T^*$. Hier wurde der Entropieflu"s bei dem von Nelson vorgeschlagenen Verhalten negativ und hier kommen auch die in dieser Arbeit entwickelten Modifikationen zum Tragen. F"ur das Quadratgitter folgt in diesem Temperaturbereich aus \refformel{438a} unter Benutzung von \refformel{332b}
\begin{equation}
   \label{441a}
   \frac{\partial (dF_V)}{\partial (e^{- \beta E_r})} = - \pi T \, r dr \frac{1}{e^{- \beta E_r}} \int_0^{2 \pi} \frac{d \theta}{2 \pi} e^{-\frac{(E_r + E_w)^2}{2 \Delta^2}},
\end{equation}
wobei wie in Anhang B hergeleitet $\Delta^2 = \frac{J^2 \bar{\sigma}}{8 \pi}(\ln r - \cos^2 \theta)$ gilt. Integration liefert als wesentlichen Term auf gro"sen L"angenskalen 
\begin{equation}
   \label{442a}
   dF_V = - \pi \, r dr  \int_0^{2 \pi} \frac{d \theta}{2 \pi} \frac{\Delta^2}{E_r + E_w} e^{-\frac{(E_r + E_w)^2}{2 \Delta^2}}.
\end{equation}
Setzt man hier die expliziten Ausdr"ucke f"ur $E_r$, $E_w$ und $\Delta$ ein, so findet man auf gen"ugend gro"sen L"angenskalen ($\ln r \gg \frac{E_C}{J}$ und $ \ln r \gg \bar{\sigma}$)
\begin{equation}
   \label{443a}
   dF_V = - \frac{\pi}{2} \pi T^*  \, r dr \, P(r),
\end{equation}
und f"ur den Flu"s der freien Energiedichte im Bereich $T < T^*$
\begin{equation}
   \label{444a}
   \begin{split}
   \frac{dF_V}{dl} = & - \frac{\pi}{2} T^* r^2 P(r)\\
\\
   = & - \pi T^* \frac{y^2}{r^2} e^{\frac{1}{8 \pi \bar{\sigma}}} I_0 \left(\frac{1}{8 \pi \bar{\sigma}} \right) .
\end{split}
\end{equation}
Auch dieses Ergebnis stimmt mit dem von Tang und Scheidl gefundenen Verhalten des ungeordneten XY-Modells bis auf Vorfaktoren "uberein. F"ur das Dreiecksgitter ergibt sich bei Verwendung der passenden Wahrscheinlichkeit $P(r)$ und der passenden Fugazit"at $y$ die gleiche Beziehung.
 
Zur Berechnung des Entropieflusses mittels $\frac{d S_V}{d l} = - \frac{\partial}{\partial T} \frac{d F_V}{d l}$ beschr"anken wir uns in \refformel{444a} auf die wesentlichen Faktoren, die das Verhalten auf gro"sen Skalen bestimmen. Wir betrachten daher nur $\frac{dF_V}{dl} \propto - T^* y^2$. Die vernachl"assigten Faktoren haben keinen Einflu"s auf den Entropieflu"s.
 
Wegen
\begin{equation}
   \label{445a}
   \frac{\partial}{\partial T} (T^* y^2) = 0,
\end{equation}
was sowohl beim Quadrat-, als auch beim Dreiecksgitter gilt, liegt kein Entropieflu"s vor. Zwar wurden hierbei die impliziten Temperaturabh"angigkeiten vernachl"assigt, da allerdings auch 
\begin{equation}
   \label{446a}
   \frac{\partial}{\partial T} \frac{d}{dl} (T^* y^2) = 0 + O(y^4)
\end{equation}
f"ur beide Gittertypen gefunden wird, liefern auch diese impliziten Abh"angigkeiten (vgl. hierzu auch Abschnitt 2.3.2 und \verweistext{sts97}) keinen Entropieflu"s f"ur $T < T^*$. Dieses Verhalten, welches Tang \verweistext{tang96} ebenso f"ur den Tieftemperaturbereich des ungeordneten XY-Modells findet, erkl"art er mit einem Einfrieren der Defektpaare bei $T = T^*$. Festzuhalten bleibt, da"s anders als bei Nelson \verweistext{nel83} der Entropieflu"s im Bereich $T < T^*$ nicht unphysikalisch wird.

\chapter{Einflu"s eines Substrats}
 
Bisher haben wir lediglich einen freien zweidimensionalen Film betrachtet. Unter realen Bedingungen liegt ein solcher Film allerdings immer auf einem Substrat auf, mit dem der Film auch wechselwirkt. Die in den vorangegangenen Kapiteln gefundenen Ergebnisse sind in diesem Fall nur dann anwendbar, wenn man die Wechselwirkung mit dem Substrat auf gro"sen L"angenskalen vernachl"assigen kann, die Ordnung im kristallinen Film also nicht gest"ort wird. 
 
F"ur ein glattes Substrat ist dies erf"ullt, da die Wechselwirkung zwischen Substrat und kristallinem Film konstant ist. In Experimenten ist ein Film aber meist auf Substrate aufgebracht, die, wie der Film selbst, eine kristalline Struktur aufweisen, und deren Gitterkonstante die gleiche Gr"o"senordnung besitzt, wie die Gitterkonstante des Films.
 
\section{Kommensurables Substrat}
Wir beschr"anken uns hier auf die Untersuchung einer kommensurablen Situation, bei der das Substrat ein Potential erzeugt, dessen Minima dieselbe Geometrie wie der kristalline Film aufweisen, das hei"st im Fall eines kristallinen Films mit Quadratgitter-Struktur sind die Minima ebenso als Quadratgitter angeordnet, und im Fall eines Dreiecksgitters bilden auch die Potentialminima ein Dreiecksgitter. Ist $a$ die Gitterkonstante des kristallinen Films, so gelte f"ur die Gitterkonstante $b$ des aus den Potentialminima gebildeten Gitters
\begin{equation}
  \label{501a}
  b = \frac{a}{n},
\end{equation}
wobei $n$ eine nat"urliche Zahl sei. Liegt ein solches Potential vor, ist es m"oglich alle Gitterbausteine des kristallinen Films in einem Potentialminimum zu plazieren, ferner ist kein Potentialminimum mehrfach besetzt. Es ist wesentlich zwischen der Gitterkonstanten $b$ des aus den Potentialminima gebildeten Gitters und der Gitterkonstanten $a_S$ des Substrats zu unterscheiden, da $b \neq a_S$ gelten kann. Ein Beispiel hierf"ur zeigt Abbildung 6.2, wo die Potentialminima ein Dreiecksgitter bilden, die das Potential erzeugenden Substratatome k"onnen allerdings bei den Potentialmaxima sitzen. In diesem Fall w"aren sie in einem hexagonalen Gitter mit Gitterkonstanten $a_G = \frac{b}{\sqrt{3}}$ angeordnet.
 
F"ur $T = 0$ und $\sigma = 0$ werden daher alle Kristallbausteine in den Potentialminima eingelagert sein und ein maximaler Energiegewinn vorliegen. Erst bei h"oheren Temperaturen oder st"arkerer Unordnung wird es m"oglich sein, die  Wechselwirkung zwischen Substrat und Film auf gro"sen L"angenskalen zu vernachl"assigen.
 
Im Fall des Quadratgitters sind die primitiven Gittervektoren des kristallinen Films
\begin{equation}
  \label{502a}
  {\bf b_1} = \left(\begin{array}{c} a \\ 0 \end{array} \right) \quad \quad \quad {\bf b_2} = \left(\begin{array}{c} 0 \\ a \end{array} \right),
\end{equation}
f"ur die primitiven Vektoren des reziproken Gitters folgt hieraus
\begin{equation}
  \label{503a}
  {\bf G_1} = \left(\begin{array}{c} \frac{2 \pi}{a} \\ 0 \end{array} \right) \quad \quad \quad {\bf G_2} = \left(\begin{array}{c} 0 \\  \frac{2 \pi}{a} \end{array} \right).
\end{equation}
Diese Vektoren erf"ullen die Relation ${\bf b_i} \, {\bf G_j} = 2 \pi \delta_{ij}$. Analog hat man f"ur das Dreiecksgitter
\begin{equation}
  \label{504a}
  {\bf b_1} = \left(\begin{array}{c} a \\ 0 \end{array} \right) \quad \quad \quad {\bf b_2} = \left(\begin{array}{c} - \frac{a}{2} \\ a \frac{\sqrt{3}}{2} \end{array} \right),
\end{equation}
und
\begin{equation}
  \label{505a}
  {\bf G_1} = \left(\begin{array}{c} \frac{2 \pi}{a} \\ \frac{2 \pi}{a \sqrt{3}} \end{array} \right) \quad \quad \quad {\bf G_2} = \left(\begin{array}{c} 0 \\  \frac{4 \pi}{a \sqrt{3}} \end{array} \right).
\end{equation}
 
Nun soll die periodische Wechselwirkungsenergie $V({\bf u})$ eines Kristallbausteins mit dem Substrat in Abh"angigkeit der Verschiebung {\bf u} aus seiner Gleichgewichtsposition in einem Potentialminimum dargestellt werden. Eine geeignete allgemeine Formulierung, die die oben aufgef"uhrten Bedingungen an das Potential erf"ullt, liefert die Darstellung in Form einer Fourierreihe 
\begin{equation}
   \label{505b}
   V({\bf u}) = - \sum_{k,l} C_{kl} \, \exp(i \, n (k {\bf G_1} + l {\bf G_2}) {\bf u}).
\end{equation}
Hierbei sind $k$ und $l$ ganze Zahlen. F"ur den Vorfaktor mu"s $C_{kl} = C_{-k-l}$ gelten, damit das Potential $V$ reell bleibt, ferner fordern wir $C_{kl} \ge 0$, um Potentialminima garantieren zu k"onnen. Setzt man ${\bf G_{kl}} = n (k {\bf G_1} + l {\bf G_2})$, so ist ${\bf G_{kl}}$ immer ein Vektor des reziproken Gitters. Es ist also
\begin{equation}
   \label{505c}
   V({\bf u}) = - \sum_{k,l} C_{kl} \, \exp(i \, {\bf G_{kl}} {\bf u}),
\end{equation}
 
die Summation entspricht hierbei einer Summation "uber alle reziproken Gittervektoren. Zur beispielhaften Darstellung solcher Potentiale beschr"anken wir uns auf jene Summanden aus \refformel{505c}, bei denen ${\bf G_{kl}}$ einen minimalen, nicht verschwindenden Betrag besitzt. Wie \refformel{515a} zeigen wird, tragen gerade diese Terme wesentlich zur Hamiltonfunktion bei, Terme mit gr"o"seren Betr"agen von ${\bf G_{kl}}$ k"onnen f"ur die Fragestellung dieses Kapitels vernachl"assigt werden.
 
F"ur das Quadratgitter haben ($\pm {\bf G_{10}} = \pm n {\bf G_1}$) und ($\pm {\bf G_{01}} = \pm n {\bf G_2}$) minimalen Betrag. Setzen wir mit $C_{10} = C_{01} = v$ ein bez"uglich der Rotation um $\frac{\pi}{2}$ symmetrisches Potential an, so resultiert
\begin{align}
   \label{506a}
   V({\bf u}) = & - v \left[ \cos \left(n \, {\bf G_1 u} \right) +  \cos \left( n \, {\bf G_2 u} \right) \right] \\
\notag \\
   \label{507a}
   = & - v \left[ \cos \left( n \, \frac{2 \pi}{a} \, u_x \right) +  \cos \left( n \, \frac{2 \pi}{a} \, u_y \right) \right].
\end{align}
Dieses Potential wird in Abbildung 5.1 gezeigt.
  
{\bf A separate downloading of this figure is possible.}

Abbildung 5.1: {\it Einfaches Potential einer Wechselwirkung zwischen einem Kristallbaustein und dem Substrat f"ur das Quadratgitter. Dunkle Regionen entsprechen Potentialminima, helle den Potentialmaxima}
 
F"ur das Dreiecksgitter lauten die ${\bf G_{kl}}$ mit minimalem Betrag ($\pm {\bf G_{10}} = \pm n {\bf G_1}$), ($\pm {\bf G_{01}} = \pm n {\bf G_2}$) und ($\pm {\bf G_{-11}} = \pm n ({\bf G_2} - {\bf G_1})$). Die analoge Darstellung des Potentials (vgl. Abbildung 5.2) ist:
\begin{align}   
    \label{508a}
    V({\bf u}) = & - v \left[ \cos \left(n \, {\bf G_1 u} \right) +  \cos \left( n \, {\bf G_2 u} \right) +  \cos \left( n \, ({\bf G_2} - {\bf G_1}) {\bf  u} \right) \right] \\
\notag \\
   \label{509a}
   = & - v \bigg[ \cos \left( n \, \frac{4 \pi}{a \sqrt{3}} \, u_y \right) +  \cos \left( n \, \frac{2 \pi}{a} \, u_x + n \, \frac{2 \pi}{a \sqrt{3}} \, u_y \right)  \notag  \\
   & + \cos \left( - n \, \frac{2 \pi}{a} \, u_x + n \, \frac{2 \pi}{a \sqrt{3}} \, u_y \right) \bigg].
\end{align} 
 
{\bf A separate downloading of this figure is possible.}

Abbildung 5.2: {\it Einfaches Potential einer Wechselwirkung zwischen einem Kristallbaustein und dem Substrat f"ur das Dreiecksgitter. Dunkle Regionen entsprechen Potentialminima, helle den Potentialmaxima}
 
Im folgenden kehren wir wieder zur allgemeinen Formel \refformel{505c} zur"uck. Betrachtet man nun nicht nur die Wechselwirkung eines Kristallbausteins mit dem Substrat, sondern die Wechselwirkung aller Bausteine, so ergibt sich in der Kontinuumsform folgender Beitrag zur Hamiltonfunktion
\begin{equation}
   \label{511a}
   H_{Sub} = - v \sum_{kl} \tilde{C}_{kl} \int d^2r \, \exp (i {\bf G_{kl}} {\bf u(r)}).
\end{equation}
Hierbei ist $v:= \mbox{max} \{ C_{kl} \}$. Die $C_{kl}$ aus \refformel{505c} m"ussen ein solches Maximum haben, da sonst ein Kristallbaustein durch die Wechselwirkung mit dem Substrat beliebig viel Energie gewinnen k"onnte, was unphysikalisch ist. Ferner gilt $\tilde{C}_{kl}:= \frac{C_{kl}}{v}$, woraus $0 \le \tilde{C}_{kl} \le 1$ folgt.

Die Kontinuumsform in \refformel{511a} ist gerechtfertigt, wenn ein kontinuierliches Verschiebungsfeld {\bf u(r)} vorliegen kann. Dies ist bei sehr gro"sen Werten von $v$ nicht mehr der Fall. Die starke Bindung der Kristallelemente in den Potentialminima w"urde dazu f"uhren, da"s {\bf u(r)} nur noch diskrete Werte annehmen kann, da jede Kristallelement-Verschiebung {\bf u} ein Gittervektor des Potentialminima-Gitters sein m"u"ste.
 
Eine Bedingung f"ur $v$ l"a"st sich wiefolgt absch"atzen. F"ugt man einem Kristall, der kommensurabel (mit $n=1$) auf einem Substrat aufliegt, eine zus"atzliche Reihe von Kristallbausteinen hinzu, so lagert sich diese Reihe im Falle sehr gro"ser Werte von $v$ bei einer anderen, bereits in Potentialminima liegenden Reihe an. Diese Minima w"aren somit doppelt besetzt, w"ahrend links und rechts davon die normale kommensurable Situation erhalten bliebe. Der St"orbereich, den man auch Dom"anenwand nennt, h"atte somit eine Breite von $l_0 = b$. Im Falle kleinerer $v$ hingegen w"urde sich die Breite der Dom"anenwand vergr"o"sern. Ein Vergleich der elastischen Energiekosten aufgrund von {\bf u(r)} und des Energiegewinns durch die Wechselwirkung mit dem Substrat zeigt, da"s ein Energieminimum bei einer Dom"anenwandbreite von 
\begin{equation}
   \label{511b}
   l_0 \propto b \, \sqrt{\frac{4 \mu (\mu + \lambda)}{(2 \mu + \lambda) v}}.
\end{equation}
vorliegt \verweistext{pok1}. Da die Bedingung f"ur die Kontinuumsform $l_0 \gg b$ lauten mu"s, folgt f"ur die St"arke des Potentials $v \ll \frac{4 \mu (\mu + \lambda)}{2 \mu + \lambda}$. 
 
Der Beitrag von $H_{Sub}$ l"a"st sich dann als St"orung der Hamiltonfunktion auffassen. St"orungsrechnung in erster Ordnung liefert
\begin{equation}
   \label{512a}
   H_{Sub} = -v \sum_{kl} \tilde{C}_{kl} \int d^2r \left[ \langle \exp (i {\bf G_{kl}} {\bf u(r)}) \rangle \right]_D,
\end{equation}
wobei $<...>$ eine thermische Mittelung mit dem ungest"orten Hamiltonian und $[...]_D$ eine Mittelung "uber die Unordnung symbolisieren. Da in der kristallinen Phase auf sehr gro"sen L"angenskalen die Versetzungen keine Rolle mehr spielen, ist eine Mittelung mit einem rein phononischen Verschiebungsfeld m"oglich, welches eine Gau"ssche Wahrscheinlichkeitsverteilung zeigt. Berechnungen von Nelson \verweistext{nel83} verwendend (vgl. Anhang D), finden wir
\begin{equation}
   \label{513a}
    \left[ \langle \exp (i {\bf G} {\bf u(r)}) \rangle \right]_D = R^{-\frac{T G^2 (3 \mu + \lambda)}{8 \pi \mu (\lambda + 2 \mu)}} \, R^{-\frac{\sigma G^2 \Omega_0^2 (\mu + \lambda)^2}{8 \pi (2 \mu + \lambda)^2}}.
\end{equation}
Hierbei sind $R$ die Systemgr"o"se, $\Omega_0$ die in Kapitel 2 erl"auterte "Anderung der Kristallfilmfl"ache aufgrund von Verunreinigungen und $\lambda$, $\mu$ die vollst"andig renormierten Lam\'e-Koeffizienten, die zu verwenden sind, da obige Rechnung ja sehr gro"se L"angen\-skalen bedingt. Betrachtet man \refformel{512a}, so stellt man fest, da"s der Beitrag des Summanden $kl$ mit
\begin{equation}
   \label{514a}
   d - \frac{T G_{kl}^2 (3 \mu + \lambda)}{8 \pi \mu (\lambda + 2 \mu)} - \frac{\sigma G_{kl}^2 \Omega_0^2 (\mu + \lambda)^2}{8 \pi (2 \mu + \lambda)^2}
\end{equation}
skaliert (d ist hier die Dimension). Der elastische Anteil der Hamiltonfunktion skaliert wegen der dort enthaltenen Ableitungen mit $d - 2$. Im vorliegenden zweidimesionalen Fall ($d=2$) folgt daraus, da"s bei gro"sen Systemen der Summand $kl$ somit nur dann wesentlich zur Hamiltonfunktion beitr"agt, wenn 
\begin{equation}
   \label{515a}
    2 \ge  \frac{T G_{kl}^2 (3 \mu + \lambda)}{8 \pi \mu (\lambda + 2 \mu)} + \frac{\sigma G_{kl}^2 \Omega_0^2 (\mu + \lambda)^2}{8 \pi (2 \mu + \lambda)^2}.
\end{equation}
erf"ullt ist.
 
Ist diese Ungleichung f"ur die Summanden mit den kleinsten Betr"age von ${\bf G_{kl}}$ verletzt, so gilt sie f"ur keinen Summanden. Daher sind nur diese Summanden f"ur unsere Betrachtung wesentlich (vgl. Bemerkung vorne). Beim Quadratgitter liefern die ${\bf G_{kl}}$ mit den kleinsten Betr"agen (${\bf G_{10}}$ und ${\bf G_{01}}$) $G_{10}^2 = G_{01}^2 = n^2 \frac{4 \pi^2}{a^2}$, f"ur das Dreieckgitter gilt analog $G_{10}^2 = G_{01}^2 = G_{-11}^2 = n^2 \frac{16 \pi^2}{3a^2}$. 
 
Eine Verletzung der Ungleichung \refformel{515a} f"ur alle Summanden bedeutet, da"s die Wechselwirkung mit dem Substrat vernachl"assigt werden kann. Die in den vorangehenden Kapiteln untersuchte Theorie des Schmelzens ist dann weiterhin anwendbar. Die Bedingung f"ur die Existenz eines Kosterlitz-Thouless-Phasen"ubergangs vom kristallinen Zustand mit quasi-langreichweitiger Ordnung in den nicht-kristal\-linen Zustand lautet im Falle eines unterliegenden kommensurablen Substrats daher:
\begin{equation}
   \label{516a}
    \frac{T (3 \mu + \lambda)}{8 \pi \mu (\lambda + 2 \mu)} + \frac{\sigma \Omega_0^2 (\mu + \lambda)^2}{8 \pi (2 \mu + \lambda)^2} > 
    \begin{cases}
    \: \frac{2 a^2}{4 \pi^2 n^2} & \quad \text{f"ur Quadratgitter}  \\ 
    \: \frac{6 a^2}{16 \pi^2 n^2} & \quad \text{f"ur Dreiecksgitter}  
    \end{cases}
\end{equation} 
 
Dieses Ergebnis wurde f"ur Systeme ohne Unordnung bereits von Pokrovsky und Talapov \verweistext{pok2} gefunden. 
 
{\it Kritischer Wert von n}
 
Im folgenden betrachten wir lediglich das Dreiecksgitter, da die benutzte Elastizit"atstheorie exakt nur f"ur diese Gitterform gelten und man die analogen Ergebnisse f"ur das Quadratgitter durch Austausch von konstanten Faktoren erh"alt. 
 
Es ist nun notwendig die Bedingung \refformel{516a} mit der in Abbildung 4.5 dargestellten und durch Formel \refformel{414a} beschriebenen Phasengrenze zu vergleichen. Mit der Kopplungskonstanten
\begin{equation}
   \label{517a}
   J = \frac{4 \mu (\mu + \lambda) a^2}{2 \mu + \lambda}
\end{equation}
und unter Verwendung von $\bar{\sigma} = \sigma \frac{\Omega_0^2}{a^2}$ wird \refformel{516a} zu
\begin{equation}
   \label{518a}
   \frac{T}{J} \, \frac{(3 \mu + \lambda)(\mu + \lambda)}{(2 \mu + \lambda)^2}  + \bar{\sigma} \, \frac{(\mu + \lambda)^2}{4 (2 \mu + \lambda)^2} > \frac{3}{4 \pi n^2}.
\end{equation}
 
Beachtet man die physikalischen Bedingung, da"s der {\it Schubmodul} $\mu$ und der {\it Kompressionsmodul} $ \mu + \lambda$ beide positiv sein m"ussen \verweistext{lanlif}, so findet man, da"s die Gr"o"sen $\frac{(3 \mu + \lambda)(\mu + \lambda)}{(2 \mu + \lambda)^2}$ und $\frac{(\mu + \lambda)^2}{(2 \mu + \lambda)^2}$ aus \refformel{518a} nur Werte innerhalb des Intervalls [0, 1] annehmen k"onnen. Das Maximum 1 wird im Fall $\lambda \gg \mu$ erreicht. Da ein unterer kritischer Wert von $n$ berechnet werden soll, sind diese beiden Werte maximal zu setzen. Somit gilt
\begin{equation}
   \label{519a}
   \frac{T}{J} + \frac{\bar{\sigma}}{4} > \frac{3}{4 \pi n^2}.
\end{equation}
Der kritische Wert von $\frac{T}{J}$, bis zu dem die kristalline Phase noch existieren kann ist $\frac{1}{16 \pi}$, der entsprechende kritische Wert der Unordnungsst"arke ist $ \bar{\sigma}_c = \frac{1}{16 \pi}$. Um die kristalline Phase mit QLRO auch bei unterliegendem Substrat beobacheten zu k"onnen, mu"s
\begin{equation}
   \label{520a}
   \frac{1}{16 \pi} + \frac{1}{64 \pi} > \frac{3}{4 \pi n^2}.
\end{equation}
erf"ullt sein, was der notwendigen Bedingung $n \ge 4$ entspricht. 

Es bleibt noch zu zeigen, da"s $n \ge 4$ auch eine hinreichende Bedingung ist, das hei"st, da"s es im Fall $\lambda \gg \mu$ tats"achlich einen Bereich der geordneten Phase gibt f"ur den \refformel{516a} f"ur $n = 4$ erf"ullt ist ($\frac{T}{J} = \frac{1}{16 \pi}$ und $ \bar{\sigma} = \frac{1}{16 \pi}$ sind ja nie gleichzeitig m"oglich). 

Nur im System ohne Unordnung liegt der Phasen"ubergang beim Maximalwert $\frac{T_m}{J} = \frac{1}{16 \pi}$. \refformel{519a} reduziert sich hier zu
\begin{equation}
   \label{521a}
   \frac{T}{J} > \frac{3}{4 \pi n^2}.
\end{equation}
F"ur $n = 4$ ist diese Ungleichung erf"ullt, somit ist $n \ge 4$ auch eine hinreichende Bedingung f"ur die Existenz der kristallinen Phase mit QLRO. 
 
Abbildung 5.3 verdeutlicht dieses Ergebnis, indem die Bedingung \refformel{518a} graphisch gezeigt wird. Hierbei wird ein System ohne Unordnung mit $\lambda = 2 \mu$ angesetzt. Man sieht deutlich, da"s nur f"ur $n \ge 4$ die kristalline Phase mit QRLO existiert.
  
\graphik{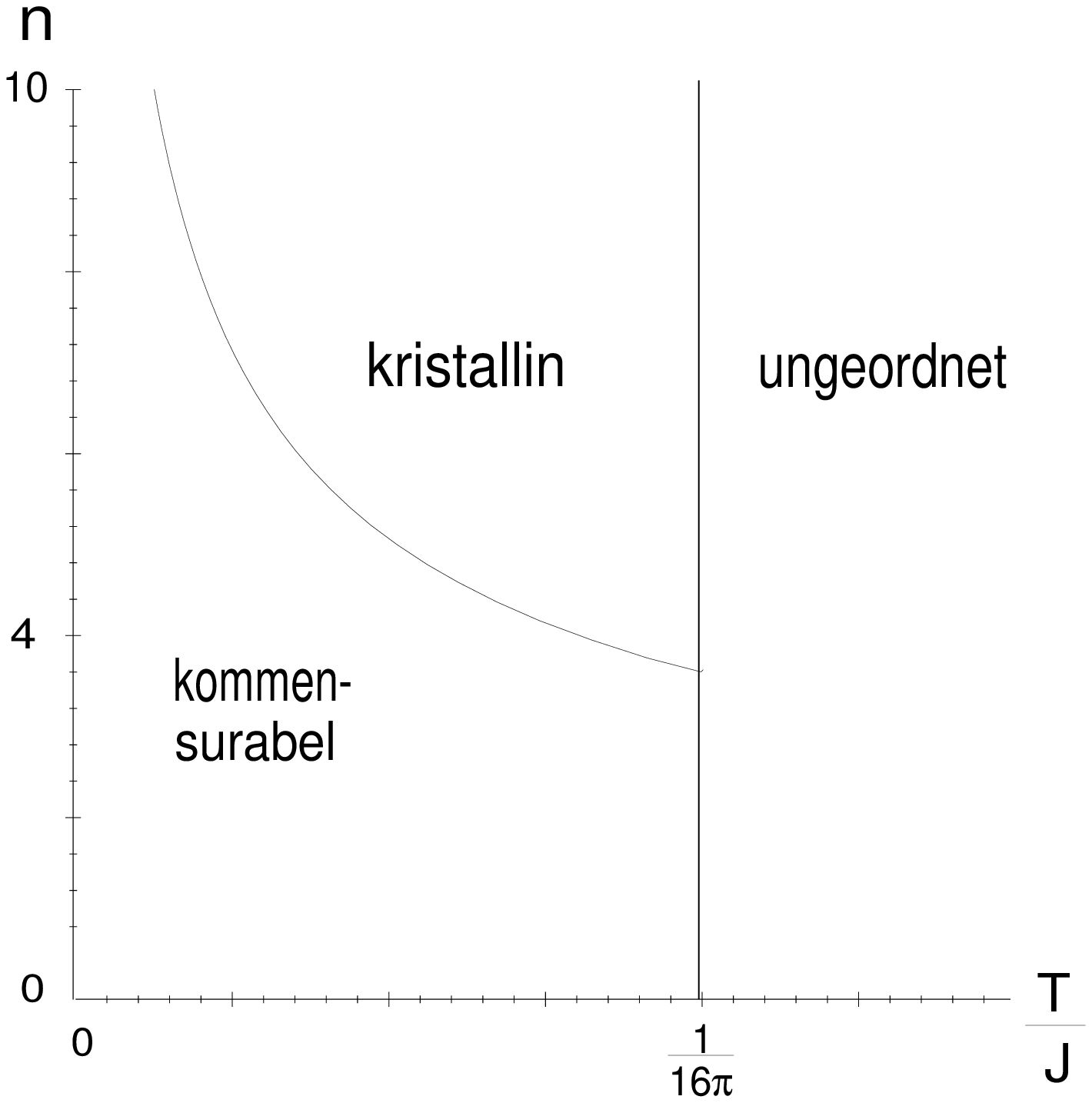}{5cm}{Beispielhafte graphische Darstellung von \refformel{518a} f"ur $\lambda = 2 \mu$ und $ \bar{\sigma} = 0$.}{503}
  
Insgesamt ist also festzustellen, da"s bei einem unterliegenden kommensurablen Substrat, der beschriebene Kosterlitz-Thouless-Schmelz"ubergang nur dann gefunden werden kann, wenn die Gitterkonstante des Substrats die Relation $b < \frac{a}{4}$ erf"ullt. Ist dies nicht der Fall, f"allt die gesamte kristalline Phase mit QLRO weg, da sie in dem Bereich l"age, in dem die Wechselwirkung mit dem Substrat auch auf gro"sen L"angenskalen wichtig ist. Daher tritt auch der Kosterlitz-Thouless-"Ubergang von der kristallinen in die hexatische Phase nicht mehr auf. Als kritischen Wert f"ur $n$ haben wir somit
\begin{equation}
   \label{522a}
   n_{krit} = 4
\end{equation}
F"uhrt man obige Betrachtungen v"ollig analog f"ur das Quadratgitter durch, so findet man denselben kritschen Wert f"ur $n$. Das Ergebnis deckt sich mit der qualitativen Aussage von Lyuksyutov, Naumovets und Pokrovsky, da"s die kristalline Phase mit quasi-langreichweitiger Ordnung nur f"ur gr"o"sere Werte von $n$ gefunden werden kann \verweistext{pok1}. Ziel des folgenden Abschnitts ist es, bei festem $n$ ein Phasendiagramm f"ur einen kristallinen Film zu finden, der auf einem kommensurablen Substrat aufgebracht ist.
 
\vspace{3cm}
 
\section{Phasendiagramm bei unterliegendem Substrat}
 
Sowohl der Schub- als auch der Kompressionsmodul sind positiv. Dies entspricht f"ur die Lam\'e-Koeffizienten den Bedingungen $\mu > 0$ und $- \mu < \lambda < \infty$. Setzen wir
\begin{equation}
   \label{523a}
   \kappa = \frac{\lambda}{\mu} 
\end{equation}
so liegt die Konstante $\kappa$ im Intervall $[-1, \infty]$. Messungen an realen physikalischen Systemen zeigen, da"s die Werte f"ur $\kappa$ tats"achlich in einem Intervall $[-1, 10]$ zu finden sind. 
 
M"ochte man bei festem $n$ den Bereich der geordneten Phase des in Kapitel 4 gewonnen Phasendiagramms absch"atzen, der bei unterliegendem Substrat noch beobachtbar ist, so ist es sinnvoll diese Absch"atzung in Abh"angigkeit der Konstanten $\kappa$ durchzuf"uhren. F"uhrt man $\kappa$ in \refformel{518a} f"ur das Dreiecksgitter ein, so erh"alt man:
\begin{equation}
   \label{524a}
   \frac{T}{J} \, \frac{(3 + \kappa)(1 + \kappa)}{(2 + \kappa)^2}  + \bar{\sigma} \, \frac{(1 + \kappa)^2}{4 (2 + \kappa)^2} > \frac{3}{4 \pi n^2}.
\end{equation}
 
F"ugt man dem Phasendiagramm aus Abbildung 4.5 noch eine dritte $\kappa$-Achse hinzu, so lassen sich sowohl die Phasengrenze des Kosterlitz-Thouless-Schmelz"ubergangs, die keine $\kappa$-Abh"angigkeit zeigt, als auch die Bedingung \refformel{524a} als Fl"achen in diesem Phasendiagramm darstellen. So kann der Bereich des Phasendiagramms verdeutlicht werden, in dem die Wechselwirkung mit dem Substrat wichtig wird und die in Kapitel 4 erhaltenen Resultate daher nicht mehr zutreffen.
 
Im folgenden wird das Phasendiagramm f"ur das Dreiecksgitter bei $n = 6$ dargestellt und erl"autert. 
 
\graphik{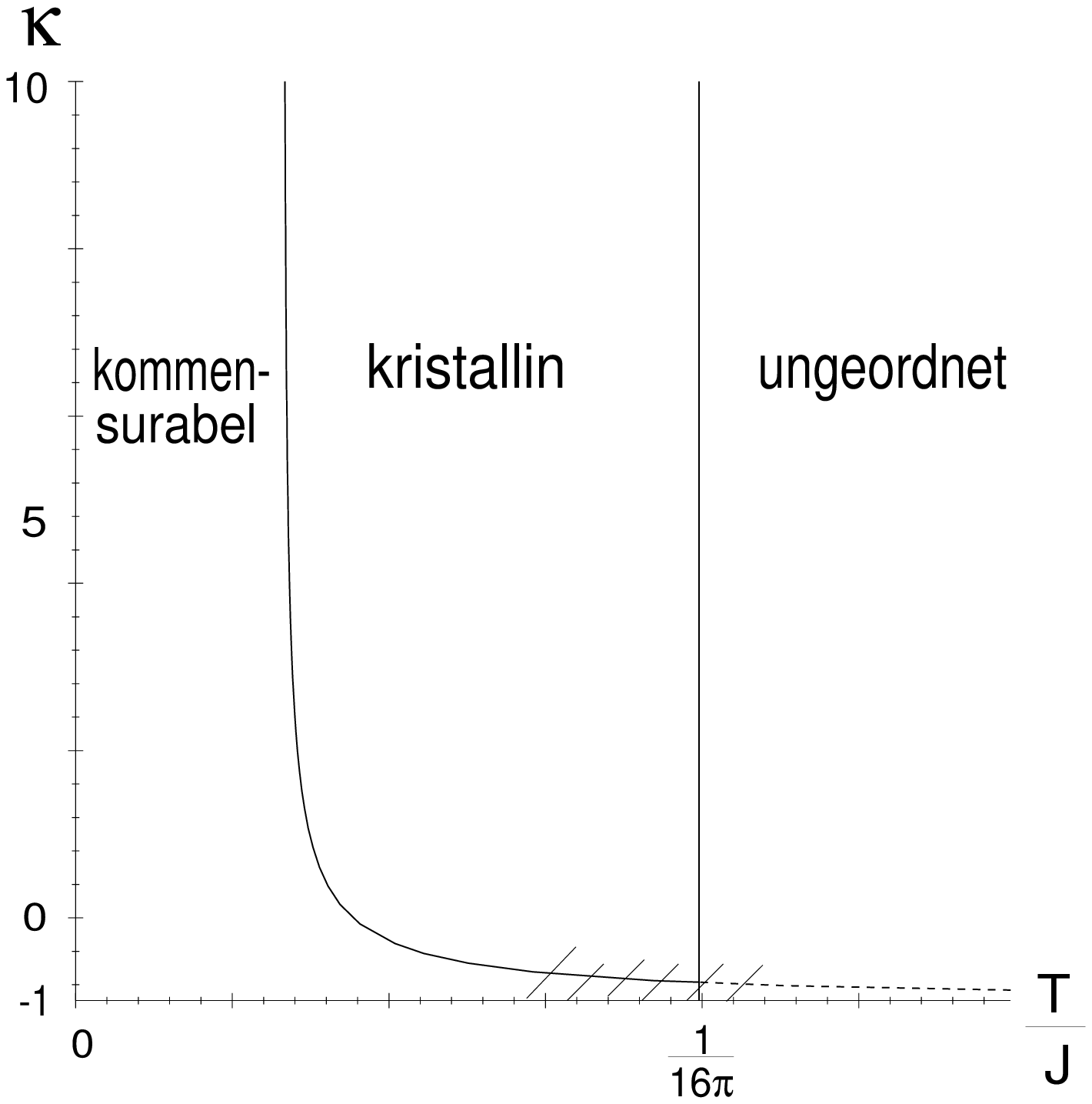}{5cm}{Phasendiagramm des Dreiecksgitters bei $n=6$: $\frac{T}{J}$ in Abh"angigkeit von $\kappa$ ohne Unordnung ($\bar{\sigma} = 0$).}{504} 
  
Abbildung 5.4 zeigt die Abh"angigkeit der kritischen Temperatur von $\kappa$ bei $\bar{\sigma} = 0$. Da, wie bereits erl"autert, der Wert f"ur $\frac{T_m}{J}$ des Kosterlitz-Thouless-Phasen"ubergangs keine $\kappa$-Abh"angigkeit besitzt, liegt die Phasengrenze zwischen fl"ussiger Phase ohne Translationsordnung und fester kristalliner Phase mit quasi-lang\-reich\-weitiger Translationsordnung konstant bei $\frac{1}{16 \pi}$. Die andere eingezeichnete Kurve separiert den Bereich, in dem \refformel{516a} erf"ullt ist, von jenem in dem diese Bedingung nicht gilt. Rechts dieser Kurve (hier gilt \refformel{516a}) spielt die Wechselwirkung mit dem Substrat auf gro"sen L"angenskalen somit keine Rolle. Hier existieren die kristalline Phase mit QLRO und der Kosterlitz-Thouless-"Ubergang zur ungeordneten Phase. Im linken Bereich ist die Wechselwirkung mit dem Potential wesentlich. Wir erwarten hier, da"s die Gitterbausteine in die Potentialminima eingebettet sind und {\it langreichweitige} Ordnung zeigen. Es liegt also eine kommensurable Phase vor. Somit bildet die Bedingung \refformel{516a} die Phasengrenze zwischen der kommensurablen Phase mit langreichweitiger Ordnung und der "ublichen kristallinen Phase mit quasi-langreichweitiger Ordnung. Dieser Phasen"ubergang ist kontinuierlich. Da \refformel{513a} in der ungeordneten Phase nicht gilt, kann nicht erwartet werden, da"s die Phasengrenze der kommensurablen Phase in diesen Bereich fortsetzbar ist. Sie ist daher dort nur gestrichelt eingezeichnet. Im schraffierten Bereich ist ein Phasen"ubergang von der kommensurablen zur ungeordneten Phase zu erwarten, dessen genaue Lage und Ordnung hier nicht diskutiert werden soll, ebenso wie die Frage, ob in diesem Bereich eine m"ogliche Zwischenphase existiert. 
 
\graphik{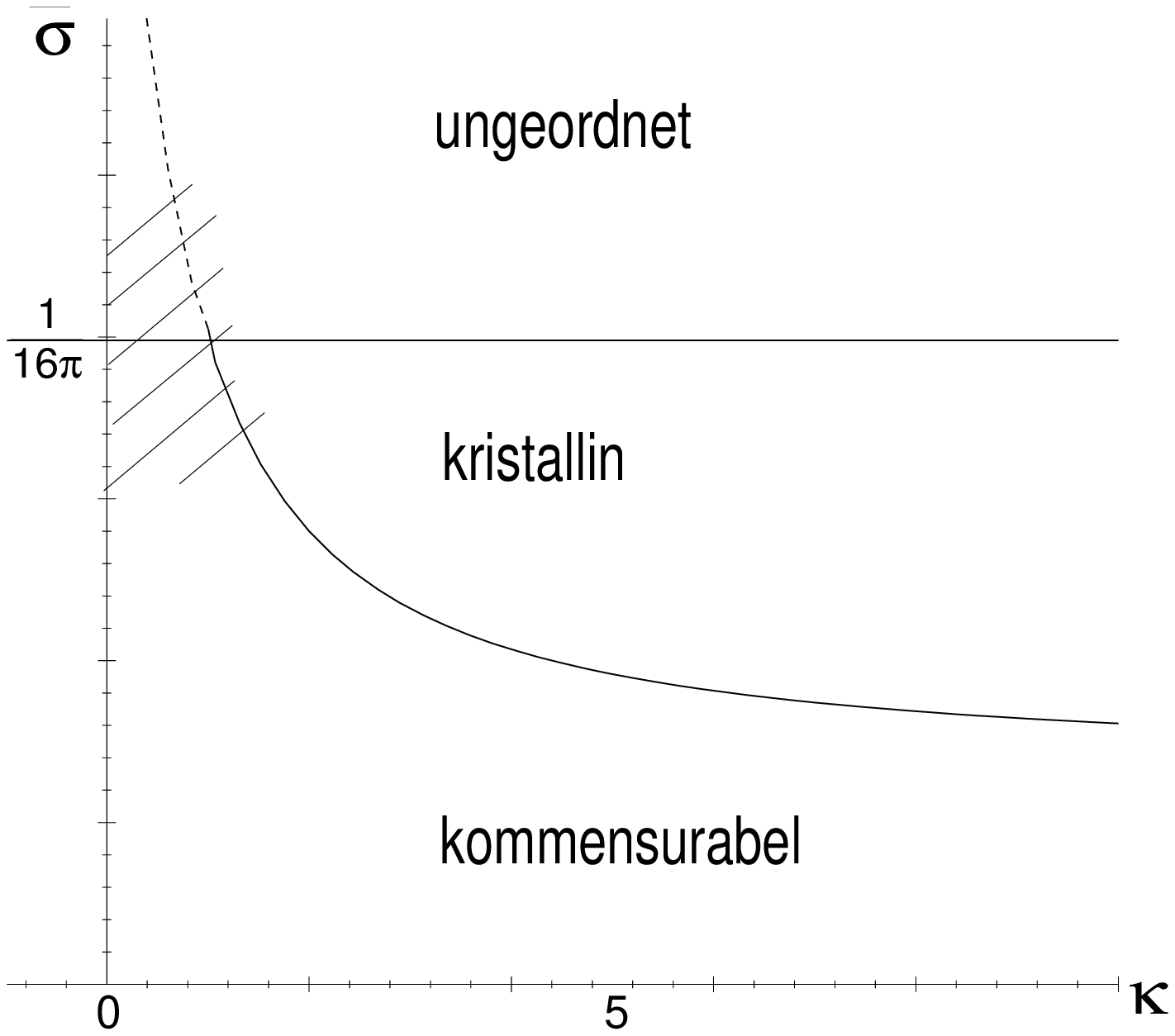}{5cm}{Phasendiagramm des Dreiecksgitters bei $n=6$: $\bar{\sigma}$ in Abh"angigkeit von $\kappa$ bei gegebener Temperatur $T = \frac{J}{64 \pi}$.}{505}
  
Analog zur vorhergehenden Abbildung zeigt Abbildung 5.5 nun den Phasenverlauf in einem $\bar{\sigma}$-$\kappa$-Diagramm. Hier ist $T = \frac{J}{64 \pi}$ fest gew"ahlt. Man sieht, da"s auch hier bei gen"ugend niedriger Unordnungsst"arke eine kommensurable Phase existiert, die bei niedrigen Werten von $\kappa$ in die ungeordnete Phase "ubergehrt, w"ahrend bei steigendem $\kappa$ die kristalline Zwischenphase zu finden ist.
   
\graphik{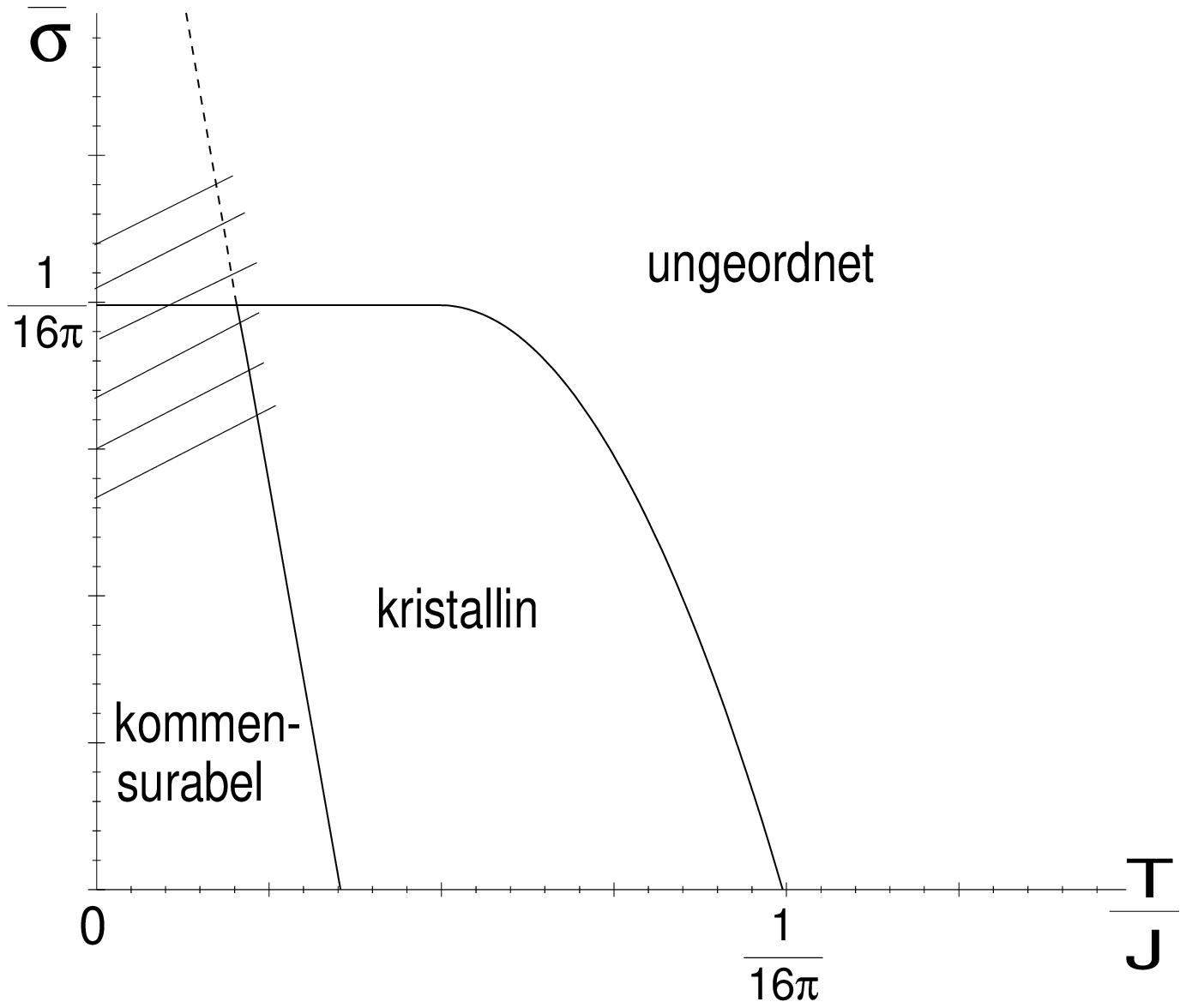}{5cm}{Phasendiagramm des Dreiecksgitters bei $n=6$: Unordnungsst"arke $\bar{\sigma}$ in Abh"angigkeit der Temperatur, $\kappa = 2$ ist hier fest gew"ahlt.}{506}
  
Darstellung 5.6 zeigt das $\bar{\sigma}$-$T$-Diagramm f"ur ein festes Verh"altnis zwischen den Lam\'e-Koeffizienten $\kappa = 2$, welches von realen Systemen durchaus angenommen werden kann. Dieses Diagramm ist mit Abbildung 4.5 zu vergleichen. Die kommensurable Phase, die bei tiefen Temperaturen und geringer Unordnungsst"arke auftritt, ist durch eine gerade Phasengrenze von der kristallinen Phase separiert. Bei sehr niedrigen Temperaturen geht sie direkt in die ungeordnete Phase "uber. Lehnen \verweistext{ml97} hat ein solches Phasendiagramm f"ur ein ungeordnetes XY-Spinsystem gefunden, welches nicht durch ein Substrat, sondern durch ein "au"seres Magnetfeld gest"ort wird. Neben den auch hier gefundenen Phasen mit vergleichbaren Phasengrenzen, diskutiert er im schraffierten Bereich die Existenz einer zus"atzlichen superrauhen Zwischenphase ($\ln^2 r$-Abh"angigkeit der Korrelationsfunktion), deren Grenzen aber nicht genau bestimmt werden k"onnen.   
  
Festzuhalten bleibt, da"s sich die Grenze der kommensurablen Phase mit steigendem $n$ nach links verschiebt, der schraffierte Bereich entf"allt dann. Bei niedrigeren Werten von $n$ allerdings verschiebt sich diese Grenze nach rechts. F"ur $n < 4$ befindet sie sich soweit rechts, da"s die kristalline Phase nicht mehr auftritt. Hier existiert dann der vorne beschriebene Kosterlitz-Thouless-Schmelz"ubergang nicht mehr, sondern lediglich die kommensurable und die ungeordnete Phase. Somit ist insbesondere f"ur $a = b$ kein Kosterlitz-Thouless-Schmelz"ubergang zu erwarten.
  
Abbildung 5.7 zeigt abschlie"send nocheinmal eine dreidimensionale Darstellung der oben diskutierten Phasengrenzen. Im linken Bereich bei kleinen Werten von $\bar{\sigma}$ und $\frac{T}{J}$ befindet sich die kommensurable Phase, w"ahrend im rechten Bereich, bei hohen Werten von $\frac{T}{J}$ und $\kappa$, die ungeordnete Phase vorliegt. Die kristalline Phase mit quasi-langreichweitiger Ordnung befindet sich zwischen den beiden Phasengrenzfl"achen. 
 
{\bf A separate downloading of this figure is possible.}

Abbildung 5.7: {\it Dreidimensionales Phasendiagramm des Dreiecksgitters bei $n=6$: Hier sind die Unordnungsst"arke $\bar{\sigma}$, $\frac{T}{J}$ und $\kappa$ gegeneinander aufgetragen. Die Fortsetzung der Phasengrenze zwischen kommensurabler und kristalliner Phase in den Bereich der ungeordneten Phase hinein ist, wie bereits erw"ahnt, nicht korrekt, aus Gr"unden der Anschaulichkeit hier aber dennoch eingezeichnet.}
   
Neben den beschriebenen kommensurablen Situationen, in denen die primitiven Gittervektoren des aufliegenden Films und des Potentialminimagitters bis auf einen Faktor $n$ gleich sind, existieren auch kommensurable Situationen, bei denen dies nicht der Fall ist. Abbildung 5.8 zeigt ein solches Beispiel. Hier ist das Dreiecksgitter der Potentialminima gegen"uber dem Dreiecksgitter des Films (wei"se Punkte) um 30 Grad gedreht. 
 
{\bf A separate downloading of this figure is possible.}

Abbildung 5.8: {\it Potential der Wechselwirkung zwischen einem Film (wei"se Punkte) und einem kommensurablen Substrat. Die beiden  Dreiecksgitter sind um 30 Grad zueinenader gedreht.} 

Hier ist $a = \sqrt{3}b$. M"ochte man die Situation analog zu \refformel{501a} verallgemeinern, so gilt
\begin{equation}
  \label{524b}
  a = n \, \sqrt{3} b.
\end{equation}
Geht man wie in Abschnitt 5.1 vor, so findet man anstelle der Bedingung \refformel{516a} in diesem Fall:
\begin{equation}
   \label{524c}
   \frac{T (3 \mu + \lambda)}{8 \pi \mu (\lambda + 2 \mu)} + \frac{\sigma \Omega_0^2 (\mu + \lambda)^2}{8 \pi (2 \mu + \lambda)^2} > \frac{2 a^2}{16 \pi^2 n^2}.
\end{equation}
Als kritischer Wert ergibt sich daraus $n_{krit} = 2$.
 
In analoger Weise lassen sich alle kommensurablen Situationen behandeln, bei denen sich nicht mehr als ein Kristallbaustein in einem Potentialminimum befindet.
 
Nicht diskutiert wurden in diesem Kapitel
\begin{itemize}
\item Kommensurable Substrate h"oherer Ordnung mit einer Gitterkonstanten
\begin{equation}
   \label{525a}
   b = p \, a,
\end{equation}
wobei p eine nat"urliche Zahl ist. Hier befinden sich dann $p$ Kristallbausteine in einer Potentialmulde. Nach \verweistext{pok2} wird in diesem Fall die rechte Seite der Ungleichung \refformel{516a} proportional zu $\frac{b^2}{p^2}$. Insgesamt ist also wiederum eine Proportionalt"at zu $a^2$ festzustellen, wie schon im Fall $b = a$. Es kann also bei diesen Substraten ebenfalls kein Kosterlitz-Thouless-Schmelzen auftreten.
 
\item Inkommensurable Substrate mit einer Gitterkonstanten der Form
\begin{equation}
   \label{526a}
   b = \frac{p}{n} (a - \delta),
\end{equation}
hierbei ist $\delta = \frac{b - a}{a}$ der {\it Misfit} zwischen den beiden Gitterkonstanten. Entsprechend \verweistext{pok3} ist mit steigendem Misfit ein Absenken der Phasengrenze der kommensurablen Phase zu niedrigeren Temperaturen und Unordnungsst"arken zu erwarten, bis zu einem kritischen Wert von $\delta$, bei dem die kommensurable Phase vollst"andig verschwindet.
 
\item Substrate mit beliebiger Gitterkonstante, die aber gegen"uber dem Gitter des kristallinen Films gedreht sind; die Symmetrieachsen der beiden Gitter sind also nicht parallel. Sofern eine kristalline Phase mit quasi-langreichweitiger Ordnung und damit ein Kosterlitz-Thouless-Schmelz"ubergang in die hexatische Phase existieren, findet man in \refformel{245a} zwei verschiedene Kopplungskonstanten f"ur den Logarithmus-Term und den Winkelterm, die unterschiedlich renormieren \verweistext{nh79}.
  
\end{itemize}

\chapter{Zusammenfassung}
 
In dieser Arbeit wurde das Schmelzen zweidimensionaler kristalliner Filme bei vorliegenden eingefrorenen Verunreinigungen untersucht. Im Rahmen der Koster\-litz-Thouless-Theorie wird ein zweistufiges Schmelzen mittels zwei kontinuierlicher Phasen"uberg"ange erwartet. Zun"achst geht die kristalline Phase mit quasi-langreichweitiger Translationsordnung (algebraische abfallende Translations-Korrelationsfunktion) in eine hexatische bzw. tetratische Fl"ussigkristall-Phase "uber, die keine Translationsordnung mehr aufweist, aber noch  quasi-langreichweitige Orientierungsordnung zeigt. Beim zweiten Phasen"ubergang in die isotrop fl"ussige Phase geht auch diese verloren. Gegenstand dieser Arbeit ist lediglich der erste Phasen"ubergang in die Fl"ussigkristall-Phase.
 
Dieser ist bereits von Nelson \verweistext{nel83} im Falle glatter Substrate betrachtet worden. Er fand bei tiefen Temperaturen und Unordnungsst"arken $\bar{\sigma} > 0$ einen Wiedereintritt in die Fl"ussigkristall-Phase ohne Translationsordnung. Im Bereich $T < \frac{J \bar{\sigma}}{2}$, der den gesamten Wiedereintritt umfa"st, f"uhren die von ihm gefundenen Renormierungsflu"sgleichungen der Kopplungskonstanten $J$ und der Fugazit"at $y$ allerdings zu einem Absinken der Entropie bei Renormierung und zu negativen Entropiewerten auf gro"sen L"angenskalen. Grund hierf"ur war, da"s Nelson f"ur die Wahrscheinlichkeit einen Versetzungsdipol vorzufinden eine Boltzmann-Verteilung angesetzt hat, die im Bereich des Wiedereintritts sehr gro"se Werte annimmt. Damit wird dort aber auch die Fugazit"at sehr gro"s und die "ubliche N"aherung f"ur kleine Fugazit"aten kann nicht mehr angewendet werden.
 
In dieser Arbeit wird eine {\it normierte} Dipolwahrscheinlichkeit angesetzt, die Versetzungen als nicht wechselwirkende Fermionen behandelt. Dies geschieht in Anlehnung an eine Arbeit von Nattermann, Scheidl, Li und Korshunov \verweistext{nat95}, die dieses Konzept f"ur das ungeordnete XY-Spinsystem entwickelt haben, wo Vortices als topologische Defekte auftreten. Renormierung mit Hilfe eines dielektrischen Formalismus liefert Flu"sgleichungen f"ur die Kopplungskonstante und die Fugazit"at, die f"ur $T > \frac{J \bar{\sigma}}{2}$ mit den von Nelson gefundenen Gleichungen "ubereinstimmen. Eine Renormierung der Unordnungsst"arke wird nicht untersucht. Das aus den Flu"sgleichungen resultierende Phasendiagramm zeigt keinen Wiedereintritt, im Bereich $T < \frac{J \bar{\sigma}}{2}$ verl"auft die Phasengrenze zwischen fester Phase und Fl"ussigkristall-Phase parallel zur Temperaturachse bei einem Wert von $\bar{\sigma}_c = \frac{1}{16 \pi}$. Abbildung 4.5 zeigt das Phasendiagramm noch einmal. Dieses Resultat stimmt mit den Absch"atzungen der Arbeit von Cha und Fertig \verweistext{cf95} "uberein. Ein Absinken der Entropie bei Renormierung wird nicht mehr gefunden. 
    
Zus"atzlich wurde der Einflu"s eines periodischen Substratpotentials nicht-glatter Substrate untersucht. Diese Analyse beschr"ankte sich auf kommensurable Situationen, bei denen die Potentialminima den selben Gittertyp bilden wie der aufliegende kristalline Film. Hat der Film die Gitterkonstante $a$ und das Gitter der Potentialminima die Gitterkonstante $b$, so wurde $a = n \, b$ gefordert, wobei $n$ eine nat"urliche Zahl ist. Es konnte gezeigt werden, da"s die Wechselwirkung mit dem Substrat nur dann die Ausbildung einer kristallinen Phase mit quasi-langreichweitiger Translationsordnung zul"a"st, wenn $n \ge 4$ erf"ullt ist, das Substrat somit schnell moduliert (vgl. Abbildung 5.3). Anderenfalls, also auch f"ur $a = b$, verschwindet diese Phase v"ollig und damit auch der oben beschriebene Phasen"ubergang. An ihre Stelle tritt eine kommensurable Phase mit langreichweitiger Ordnung, die direkt schmilzt. F"ur $n \ge 4$ existiert diese kommensurable Phase zwar auch, geht aber zun"achst in die kristalline Phase mit quasi-langreichweitiger Ordnung "uber, bevor dann der in dieser Arbeit behandelte Phasen"ubergang eintritt (vgl. Abbildung 5.5). 
 
Ein Vergleich mit Experimenten und Simulationen bringt keine direkte Best"atigung der Ergebnisse dieser Arbeit, da sich sowohl Experimente als auch Simulationen bisher auf die Frage nach der Richtigkeit der Kosterlitz-Thouless-Beschreibung beschr"anken und daher noch keine Arbeiten vorliegen, die versuchen den Einflu"s von Unordnung oder verschiedenen Substratpotentiale auf den Schmelzvorgang explizit zu untersuchen. Dennoch scheint das Verhalten von adsorbiertem Krypton auf einem kommensurablen Graphitgitter die hier erzielten Resultate zu best"atigen.
 
Mit dieser Arbeit konnte somit das in j"ungster Zeit gefundene Verhalten des ungeordneten XY-Modells auf das verwandte Modell zweidimensionaler Gitter mit eingefrorenen Verunreinigungen "ubertragen werden.

\begin{appendix}
  
  \chapter{Coulombgas-Beschreibung}
\section{Abbildung auf ein Coulombgas}
Ausgehend von \refformel{244a} und unter Verwendung der Formeln \refformel{240a} und \refformel{243a} erh"alt man:
\begin{equation}
   \label{a001a}
   H_{el} = \frac{J}{16 \pi} \int \int_{|{\bf r} - {\bf r'}|> 1} d^2r d^2r' b_i({\bf r}) \epsilon_{ij} \nabla_j b_k({\bf r'}) \epsilon_{kl} {\nabla'}_l ({\bf r} - {\bf r'})^2 \left( \ln |{\bf r} - {\bf r'}| + C \right)
\end{equation}
Benutzt man ${\bf b(r)} = \sum_{\alpha} {\bf b_{\alpha}} \delta({\bf r} - {\bf r_{\alpha}})$, so resultiert mit einer modifizierten von Nelson \verweistext{nel78}  benutzten Konstanten $\tilde{C} = C + \frac{1}{2}$ 
\begin{eqnarray}
  \label{a002a}
  \begin{array}{rcl}
   H_{el} & = & - \frac{J}{8 \pi} \sum_{\alpha \neq \beta} b_{\alpha i} \epsilon_{ij} \nabla_{\alpha j} b_{\beta k} \epsilon_{kl} (x_{\alpha l} - x_{\beta l}) \left( \ln |{\bf r_{\alpha}} - {\bf r_{\beta}}| + \tilde{C} \right) \vspace{0.2cm} \\
   \mbox{} & = & - \frac{J}{8 \pi} \sum_{\alpha \neq \beta} b_{\alpha i} b_{\beta i} \left( \ln |{\bf r_{\alpha \beta}}| + \tilde{C} \right) + \frac{b_{\alpha i} \epsilon_{ij} (x_{\alpha j} - x_{\beta j}) b_{\beta k} \epsilon_{kl} (x_{\alpha l} - x_{\beta l})}{({\bf r_{\alpha}} - {\bf r_{\beta}})^2} \vspace{0.2cm} \\
   \mbox{} & = & - \frac{J}{8 \pi} \sum_{\alpha \neq \beta} b_{\alpha i} b_{\beta i} \left( \ln |{\bf r_{\alpha \beta}}| + \tilde{C} \right) + b_{\alpha i} b_{\beta i} \\ 
 & & - \frac{\left(b_{\alpha i} (x_{\alpha i} - x_{\beta i})\right) \left(b_{\beta j} (x_{\alpha j} - x_{\beta j})\right)}{({\bf r_{\alpha}} - {\bf r_{\beta}})^2}
 \end{array}
\end{eqnarray}
Unter Benutzung der Ladungsneutralit"at $\sum {\bf b_{\alpha}} = 0$ gilt die Beziehung
\begin{equation}
   \label{a003a}
   -(\tilde{C} + 1) \sum_{\alpha \neq \beta} {\bf b_{\alpha}} {\bf b_{\beta}} = (\tilde{C} + 1) \sum_{\alpha} {\bf b_{\alpha}}^2.
\end{equation}
Definiert man nun $E_C = (\tilde{C} + 1) J/ 8 \pi$ als Energie eines Versetzungskerns, so erh"alt man direkt Formel \refformel{245a}, welche zu zeigen war:
\begin{equation}
   \label{a004a}
   H_{el} = - \frac{J}{8 \pi} \sum_{\alpha \neq \beta} \left({\bf b_{\alpha} b_{\beta}} \ln |{\bf r_{\alpha \beta}}| - \frac{({\bf b_{\alpha} r_{\alpha \beta}}) ({\bf b_{\beta} r_{\alpha \beta}})}{r_{\alpha \beta}^2} \right) + E_C \sum _{\alpha} |{\bf b_{\alpha}}|^2
\end{equation}
 
Dies ist exakt das Ergebnis von Nelson \verweistext{nel83, nel78}. Young \verweistext{you79} hat eine leicht andere Definition von $E_C$ benutzt und daher mit einem zus"atzlichen Term in der Hamiltonfunktion gearbeitet. Wir folgen im weiteren aber dem Formalismus von Nelson. $\tilde{C}$ ist eine positive Konstante, deren Wert von der Gitterstruktur abh"angt, sie mi"st das Verh"altnis von Kerndurchmesser zu Gitterkonstante. Ihr Wert ist von der Ordnung $O(1)$, f"ur das Quadratgitter wurde $\tilde{C} \approx \frac{\pi}{2}$  berechnet. Da f"ur die Erzeugung eines Versetzungsdipols mit Separation 1 (a = 1: Gitterkonstante) maximal die Energie $2 E_C$ aufgewendet werden mu"s, kann $E_C$ tats"achlich als Energie eines Versetzungskerns bezeichnet werden.

\twocolumn[ 
\section{Vergleich mit der Elektrodynamik}
 
In diesem Abschnitt sollen die in dieser Arbeit auftretenden elastizit"atstheoretischen Gr"o"sen mit den analogen elektrodynamischen Gr"o"sen verglichen werden. Die folgende Tabelle zeigt die wesentlichen Relationen \\
\\]
 
{\bf Elastizit"atstheorie} 
 
{\footnotesize 
{\it Quellenfunktion}
\begin{align*}
& \eta({\bf r}) = \epsilon_{ij} \nabla_i \sum_{\alpha} b_{\alpha, i} \, \delta({\bf r} - {\bf r_{\alpha}}) \notag \\
\\
\intertext{\it Greenfunktion}
& g({\bf r}) = \frac{1}{8 \pi} r^2 \ln |{\bf r}| \notag \\
\\
\intertext{\it Verzerrungsfunktion}
& \chi({\bf r}) = J \int d^2r' \, \eta({\bf r'}) \, g({\bf r} - {\bf r'}) \notag \\
\\
\intertext{\it elastische Energie}
& H_{el} = \frac{1}{2} \int d^2r \, \eta({\bf r}) \, \chi({\bf r}) \notag \\
\\
\intertext{und}
& \nabla^4 \chi({\bf r}) = J \eta({\bf r}) = \frac{4 \pi \, \eta({\bf r})}{\varepsilon} \notag \\
\end{align*}
mit $\varepsilon = \frac{4 \pi}{J}$.

}

\newpage
 
{\bf Elektrodynamik}
 
{\footnotesize 
{\it Ladungsdichte}
\begin{align*}
& \varrho(\bf r) = \sum_{\alpha} \varrho_{\alpha} \delta({\bf r} - {\bf r_{\alpha}}) \notag \\
\\
\intertext{\it Greenfunktion}
& g({\bf r}) = \ln |{\bf r}| \notag \\
\\[2ex]
\intertext{\it elektrostat. Potential}
& \Phi({\bf r}) = \int d^2r' \, \varrho({\bf r'}) \, g({\bf r} - {\bf r'}) \notag \\
\\
\intertext{\it Energie}
& H = \frac{1}{2} \int d^2r \, \varrho({\bf r}) \, \Phi({\bf r}) \notag \\
\\ 
\intertext{\it Poisson-Gleichung}
& \nabla^2 \Phi({\bf r}) = - \frac{4 \pi \,  \varrho({\bf r})}{\varepsilon}  \notag \\
\end{align*}}

\onecolumn
\twocolumn[
Im folgenden wird die elektrodynamische Mittelung "uber atomare bzw. molekulare Strukturen verglichen mit der Mittelung "uber Dipole aus Burgers-Vektoren. Die elektrodynamische Vorgehensweise entspricht der von Jackson (1963). \\
\\]
 
{\bf Elastizit"atstheorie}
 
{\footnotesize Die Quellenfunktion eines Dipols $\Gamma$ ist in inneren Koordinaten ${\pmb \rho^{(\Gamma)}}$ \refformel{304a}:
\begin{equation*}
  \tilde{\eta}_{\Gamma}({\pmb \rho^{(\Gamma)}})
\end{equation*}
 
Die Verzerungsfunktion bei {\bf r} erzeugt durch Dipol $\Gamma$, dessen Zentrum bei ${\bf r_{\Gamma}}$ liegt, ist 
\begin{equation*}
  \chi_{\Gamma}({\bf r}) = J \int d^2 \rho^{(\Gamma)} \tilde{\eta}_{\Gamma}({\pmb \rho^{(\Gamma)}}) g({\bf r} - {\bf r_{\Gamma}} - {\pmb \rho^{(\Gamma)}})
\end{equation*}
Entwicklung dieses Ausdrucks \\
 f"ur $|{\pmb \rho^{(\Gamma)}}| \ll |{\bf r} - {\bf r_{\Gamma}}|$:
\begin{equation*}
\begin{split}
   &\chi_{\Gamma}({\bf r}) =  \, J \int d^2 \rho^{(\Gamma)} \, \tilde{\eta}_{\Gamma}({\pmb \rho^{(\Gamma)}}) \, g({\bf r} - {\bf r_{\Gamma}}) \\
   & + \frac{J}{2} \int d^2 \rho^{(\Gamma)} \, \tilde{\eta}_{\Gamma}({\pmb \rho^{(\Gamma)}}) \, \rho^{(\Gamma)}_i \, \rho^{(\Gamma)}_j \frac{\partial^2 g({\bf r} - {\bf r_{\Gamma}})}{\partial x_{\Gamma i} \partial x_{\Gamma j}}
\end{split}
\end{equation*}
(kein Term erster Ordnung)
 
Gesamtverzerrungsfunktion durch Summation "uber alle Dipole
\begin{equation*}
  \chi({\bf r}) = \sum_{\Gamma} \chi_{\Gamma}({\bf r})
\end{equation*}
 
nur 1. Term
\begin{equation*}
\begin{split}
 \chi_1 ({\bf r}) & = J \sum_{\Gamma} \int d^2 \rho^{(\Gamma)} \, \tilde{\eta}_{\Gamma}({\pmb \rho^{(\Gamma)}}) \, g({\bf r} - {\bf r_{\Gamma}}) \\
 & = J \int d^2r'' \eta^{Dip} ({\bf r''}) g({\bf r} - {\bf r''})
\end{split}
\end{equation*}
 
mit 
\begin{equation*}
  \eta^{Dip} ({\bf r''}) = \sum_{\Gamma} \int d^2 \rho^{(\Gamma)} \, \tilde{\eta}_{\Gamma}({\pmb \rho^{(\Gamma)}}) \, \delta({\bf r''} - {\bf r_{\Gamma}})
\end{equation*}}

\newpage
 
{\bf Elektrodynamik}
 
{\footnotesize Die Ladungsdichte des j-ten Molek"uls ist in internen Koordinaten ${\bf r'}$:
\begin{equation*}
  \varrho'_j ({\bf r'})  
\end{equation*}
 
Das Potential bei {\bf r} erzeugt durch das j-te Molek"ul mit Zentrum bei ${\bf r_{j}}$ ist 
\begin{equation*}
  \Phi_{j}({\bf r}) = \int d^3r' \, \varrho'_j ({\bf r'})  g({\bf r} - {\bf r_{j}} - {\bf r'}).
\end{equation*}
 
Entwicklung dieses Ausdrucks \\
 f"ur $|{\bf r'}| \ll |{\bf r} - {\bf r_{j}}|$:
\begin{equation*}
\begin{split}
   \Phi_{j}({\bf r}) & =  \int d^3r' \, \varrho'_j ({\bf r'})  \, g({\bf r} - {\bf r_{j}}) \\
   & + \int d^3r' \, \varrho'_j ({\bf r'}) \, {x'}_k \frac{\partial g({\bf r} - {\bf r_{j}})}{\partial x_{j, k}}
\end{split}
\end{equation*}
 
\vspace{0.6cm}
Gesamtpotential durch Summation "uber alle Molek"ule
\begin{equation*}
  \Phi({\bf r}) = \sum_{j} \Phi_{j}({\bf r})
\end{equation*}
 
nur 1. Term
\begin{equation*}
\begin{split}
 \Phi_1 ({\bf r}) & =\sum_{j} \int d^3r' \varrho'_j({\bf r'}) \, g({\bf r} - {\bf r_{j}}) \\
 & = \int d^2r' \varrho_{mol}({\bf r''}) g({\bf r} - {\bf r''})
\end{split}
\end{equation*}
 
mit 
\begin{equation*}
  \varrho_{mol} ({\bf r''}) = \sum_{j} \int d^3 r' \, \varrho'_j({\bf r'}) \, \delta({\bf r''} - {\bf r_{j}})
\end{equation*}}

\newpage
 
{\footnotesize Mittelung "uber Fl"achen, die gr"o"ser sind als quadrierte  maximale Dipolseparation \\
$\Delta A \gg \zeta^2$:
\begin{equation*}
\begin{split}
  \langle \chi_1 ({\bf r}) \rangle & = \frac{1}{\Delta A} \int_{\Delta A} d^2 \xi \, \chi_1 ({\bf r} + {\pmb \xi}) \\
   & = \frac{J}{\Delta A} \int_{\Delta A} d^2 \xi \int d^2r''  \eta^{Dip} ({\bf r''} + {\pmb \xi}) \\
   & \hspace{3cm}  \times g({\bf r} - {\bf r''}) \\ 
\\
   & = J \int d^2 r'' \langle \eta^{Dip} ({\bf r''}) \rangle g({\bf r} - {\bf r''})
\end{split}
\end{equation*}
 
Damit makroskopische Quellenfunktion
\begin{equation*}
   \langle \eta^{Dip} ({\bf r}) \rangle = \langle \eta^{Dip} ({\bf r}) \rangle_{\Delta N}  \, n ({\bf r})
\end{equation*}
wobei
\begin{equation*}
\begin{split}
   \langle \eta^{Dip} ({\bf r}) \rangle_{\Delta N} :& \text{ Durchschnittsquellen-}\\
   & \text{ funktion pro Dipol} \\
    n ({\bf r}) : & \text{ makroskop. Dipoldichte}
\end{split}
\end{equation*}
 
d.h, wenn in $\Delta A$ durchschnittlich $\Delta N$ Dipole:
\begin{equation*}
   \langle \eta^{Dip} ({\bf r}) \rangle_{\Delta N} = \frac{1}{\Delta N} \int_{\Delta A} d^2 \xi \, \eta^{Dip} ({\bf r} + {\pmb \xi})
\end{equation*}
 
Erlaube nun Dipole mit gro"ser Separation $r > \zeta$. Die makroskopische Quellenfunktion wird zu:
\begin{equation*}
   \eta^{\zeta} ({\bf r}) =  \langle \eta^{Dip} ({\bf r}) \rangle + \eta_{ex}({\bf r})
\end{equation*}
 
Analoge Mittelung f"ur zweiten Term der entwickelten Verzerrungsfunktion liefert makroskop. Dipoltensor $C^{\zeta}_{ij} ({\bf r})$. Damit wird die makroskop. Gesamtverzerrungsfunktion
\begin{equation*}
\begin{split}
   \chi ({\bf r}) = & J \int d^2r' \eta^{\zeta} ({\bf r'}) g({\bf r} - {\bf r'}) \\
   & + J \int d^2r' C_{ij}^{\zeta}({\bf r'}) \frac{\partial^2 g({\bf r} - {\bf r'})}{\partial {x'}_i \partial {x'}_j}
\end{split}
\end{equation*}}
 
\newpage
 
{\footnotesize Mittelung "uber Volumina, die gr"o"ser sind als die maximalen Molek"ulvolumina \\
$\Delta V \gg V_{mol}$:
\begin{equation*}
\begin{split}
  \langle \Phi_1 ({\bf r}) \rangle & = \frac{1}{\Delta V} \int_{\Delta V} d^3 \xi \, \Phi_1 ({\bf r} + {\pmb \xi}) \\
   & = \frac{1}{\Delta V} \int_{\Delta V} d^3 \xi \int d^3r''  \varrho_{mol} ({\bf r''} + {\pmb \xi}) \\
   & \hspace{3cm} \times g({\bf r} - {\bf r''}) \\ 
\\
   & = \int d^3 r'' \langle \varrho_{mol} ({\bf r''}) \rangle g({\bf r} - {\bf r''})
\end{split}
\end{equation*}
 
Damit makroskopische Ladungsdichte
\begin{equation*}
   \langle \varrho_{mol} ({\bf r}) \rangle = \langle e_{mol} ({\bf r}) \rangle  \, n ({\bf r})
\end{equation*}
wobei
\begin{equation*}
\begin{split}
   \langle e_{mol} ({\bf r}) \rangle : & \text{ Durchschnittsladung}\\
   & \text{ pro Molek"ul} \\
    n ({\bf r}) : & \text{ makroskop. Molek"uldichte}
\end{split}
\end{equation*}
 
d.h, wenn in $\Delta V$ durchschnittlich $\Delta N$ Molek"ule:
\begin{equation*}
   \langle e_{mol} ({\bf r}) \rangle = \frac{1}{\Delta N} \int_{\Delta V} d^3 \xi \, \varrho_{mol} ({\bf r} + {\pmb \xi})
\end{equation*}
 
Erlaube nun freie, von Molek"ulen unabh"angige, Ladungen. Die makroskopische Ladungsdichte wird zu:
\begin{equation*}
   \varrho ({\bf r}) =  \langle \varrho_{mol} ({\bf r}) \rangle + \varrho_{ex}({\bf r})
\end{equation*}
 
Analoge Mittelung  f"ur zweiten Term des entwickelten PotentialsV liefert makroskopisches Dipolmoment  ${\bf p} ({\bf r})$. Damit wird das makroskopische Gesamtpotential
\begin{equation*}
\begin{split}
   \Phi ({\bf r}) = & \int d^3r' \varrho ({\bf r'}) g({\bf r} - {\bf r'}) \\
   & + \int d^3r' p_{i} ({\bf r'}) \frac{\partial g({\bf r} - {\bf r'})}{\partial {x'}_i}
\end{split}
\end{equation*}}
 
\newpage
 
{\footnotesize Partielle Integration und Ableiten liefern
\begin{equation*}
\begin{split}
   \frac{\nabla^4 \chi ({\bf r})}{J} & = \eta^{\zeta}({\bf r}) + \int d^2r' \frac{\partial^2 C_{ij}^{\zeta} ({\bf r'})}{\partial {x'}_i \partial {x'}_j} \delta ({\bf r} - {\bf r'}) \\
    & = \eta^{\zeta}({\bf r}) + \frac{\partial^2}{\partial x_i \partial x_j} C_{ij}^{\zeta} ({\bf r})
\end{split}
\end{equation*}
 
Mit Linear Response geben die Symmetrien des Dreiecksgitters
\begin{equation*}
\begin{split}
   C_{ij}^{\zeta} ({\bf r}) = &\sum_{kl} a_{ijkl}(\zeta) \frac{\partial^2 \chi ({\bf r})}{\partial x_k \partial x_l} \\
  = & \frac{1}{2} \, C_1({\zeta}) \left(\frac{\partial^2 \chi ({\bf r})}{\partial x_i \partial x_j} + \frac{\partial^2 \chi ({\bf r})}{\partial x_j \partial x_i} \right) \\
  & + C_2({\zeta}) \delta_{ij} \frac{\partial^2 \chi ({\bf r})}{\partial^2 x_k \partial x_l}
\end{split}
\end{equation*}
 
Nun kann die dielektrische Konstante berechnet werden
\begin{equation*}
\begin{split}
  \varepsilon({\zeta}) \, \nabla^4 \chi ({\bf r}) =  & 4 \pi \eta^{\zeta} ({\bf r}) \\
   = & \frac{4 \pi}{J} \nabla^4 \chi ({\bf r}) \\
   & - 4 \pi (C_1({\zeta}) + C_2({\zeta})) \, \nabla^4 \chi ({\bf r})
\end{split}
\end{equation*}
 
Somit gilt f"ur $\varepsilon$
\begin{equation*}
   \varepsilon = \varepsilon_0 - 4 \pi (C_1 + C_2)
\end{equation*}
 
Die elektrische Suszeptibilt"at $\kappa$ ist also
\begin{equation*}
   \kappa = - C_1 - C_2.
\end{equation*}
}
 
\newpage

{\footnotesize Partielle Integration und Ableiten liefern
\begin{equation*}
\begin{split}
   \frac{\nabla^2 \Phi ({\bf r})}{4 \pi} & = - \varrho ({\bf r}) + \int d^3r' \frac{\partial p_{i} ({\bf r'})}{\partial {x'}_i} \delta ({\bf r} - {\bf r'}) \\
    & = - \varrho ({\bf r}) + \frac{\partial}{\partial x_i} p_{i} ({\bf r})
\end{split}
\end{equation*}
 
Mit Linear Response geben die Symmetrien des isotropen Systems
\begin{equation*}
\begin{split}
   p_{i} ({\bf r}) = &\sum_{k} a_{ik} \frac{\partial \Phi ({\bf r})}{\partial x_k} \\
  = - \chi_{el} \frac{\partial \Phi ({\bf r})}{\partial x_i}\\
\\
\end{split}
\end{equation*}
 
\vspace{1cm}
Nun kann die dielektrische Konstante berechnet werden
\begin{equation*}
\begin{split}
  \varepsilon \, \nabla^2 \Phi ({\bf r}) =  & - 4 \pi \varrho ({\bf r}) \\
   = &  \nabla^2 \Phi ({\bf r}) + 4 \pi \underbrace{\frac{\partial^2}{\partial {x_i}^2} \Phi ({\bf r})}_{\nabla^2 \Phi} \, \chi_{el}
\end{split}
\end{equation*}
 
Somit gilt f"ur $\varepsilon$
\begin{equation*}
   \varepsilon = 1 +\ 4 \pi \,  \chi_{el}
\end{equation*}
 
Die elektrische Suszeptibilt"at ist also $\chi_{el}$.

}
\onecolumn

  \chapter{Unordnungswechselwirkung}
\section{Varianz bei einer einzelnen freien Versetzung}
Entsprechend Formel \refformel{262a} betr"agt die Varianz der Wechselwirkung mit der Unordnung unter Ber"ucksichtigung der Translationsinvarianz des Systems
\begin{equation}
   \label{b01a}
   \varepsilon(R) = \left[V({\bf r_{\alpha}})^2 \right]_D = \left[V({\bf 0})^2 \right]_D.
\end{equation}
Nach \refformel{257a} gilt
\begin{equation}
   \label{b02a}
   V({\bf 0}) =  \frac{J}{4 \pi} \Omega_0 b_i \epsilon_{ij} \int d^2r \delta c({\bf r}) \nabla_j \ln |{\bf r}|.
\end{equation}
Mit Hilfe der Relationen $[\delta c({\bf r}) \delta c ({\bf r'})]_D = \sigma \delta ({\bf r} - {\bf r'})$ und $\bar{\sigma} = \sigma \Omega_0^2$ ergibt sich
\begin{equation}
   \begin{split}
   \label{b03a}
    \left[V({\bf 0})^2 \right]_D & =  \left[\frac{J^2 \Omega_0^2}{16 \pi^2} b_i \epsilon_{ij} \int d^2r \delta c({\bf r}) \nabla_j \ln |{\bf r}| b_k \epsilon_{kl} \int d^2r' \delta c({\bf r'}) \nabla'_l \ln |{\bf r'}| \right]_D  \vspace{0.8cm} \\
    & = \frac{J^2 \bar{\sigma}}{16 \pi^2} b_i \epsilon_{ij} b_k \epsilon_{kl} \int d^2r \nabla_j \ln |{\bf r}| \nabla_l \ln |{\bf r}| \vspace{0.8cm} \\
    & = \frac{J^2 \bar{\sigma}}{16 \pi^2} \left[ b_x^2 \int d^2r \frac{y^2}{r^4} + b_y^2 \int d^2r \frac{x^2}{r^4} - 2 b_x b_y \int d^2r \frac{xy}{r^4} \right].
   \end{split}
\end{equation}
 
Unter Verwendung des Cut-Offs 1 f"ur die Integration folgt schlie"slich die Behauptung aus \refformel{262a}
\begin{equation}
   \begin{split}
   \label{b04a}  
   \left[V({\bf 0})^2 \right]_D & = \frac{J^2 \bar{\sigma}}{16 \pi^2} \left[b_x^2 \ln R \int_0^{2 \pi} d \phi \sin^2 \phi + b_y^2 \pi \ln R - 0 \right] \vspace{0.8cm} \\
   & = \frac{J^2}{16 \pi} \bar{\sigma} \ln R.
   \end{split}
\end{equation}

\section{Varianz bei einem Versetzungsdipol}

Nach \refformel{268a} gilt f"ur die Varianz der Unordnungswechselwirkung eines Versetzungdipols
\begin{equation}
   \label{b05a}
   \Delta^2(r, \theta) = \left[(V({\bf r}) + V({\bf 0}))^2 \right]_D = 2 \left[V({\bf 0})V({\bf 0}) + V({\bf r})V({\bf 0}) \right]_D.
\end{equation}
Aufgrund der Berechnungen aus B.1 ist nur noch notwendig $\left[ V({\bf r})V({\bf 0}) \right]_D$ zu untersuchen. Mit \refformel{257a} hat man hier bei analogem Vorgehen zu B.1
\begin{equation}
   \begin{split}
   \label{b06a}
    \left[V({\bf r}) V({\bf 0}) \right]_D & =  \biggl[\frac{J^2 \Omega_0^2}{16 \pi^2} b_{\alpha i} \epsilon_{ij} \int d^2r' \delta c({\bf r'}) \nabla'_j \ln |{\bf r'} - {\bf r}| \\
 & \quad \times b_{\beta k} \epsilon_{kl} \int d^2r'' \delta c({\bf r''}) \nabla''_l \ln |{\bf r''}| \biggr]_D  \vspace{0.8cm} \\
    & = \frac{J^2 \bar{\sigma}}{16 \pi^2} b_{\alpha i} \epsilon_{ij} b_{\beta k} \epsilon_{kl} \int d^2r' \nabla'_j \ln |{\bf r} - {\bf r'}| \, \nabla'_l \ln |{\bf r'}|.
   \end{split}
\end{equation}
Da f"ur einen Versetzungsdipol ${\bf b_{\alpha}} = - {\bf b_{\beta}}$ gilt, kann auf die Unterscheidung verzichtet werden, wenn man obige Gleichung mit $(-1)$ multipliziert. Verwendet man die Relation $\Delta ({\bf r}^2 \ln |{\bf r}|) = 4 + 4 \ln |{\bf r}|$, so ergibt sich, da die Konstante bei nochmaliger Differentiation wegf"allt
\begin{equation}
   \label{b07a}
    \left[V({\bf r}) V({\bf 0}) \right]_D = - \frac{1}{4} \frac{J^2 \bar{\sigma}}{16 \pi^2} b_i \epsilon_{ij} b_k \epsilon_{kl} \int d^2r' \nabla'_j \ln |{\bf r} - {\bf r'}| \, \nabla'_l \Delta' ({\bf r'}^2 \ln |{\bf r'}|).
\end{equation}
 
Zweifache partielle Integration liefert als Haupterm $H$
\begin{equation}
   \begin{split}
   \label{b08a}
   H({\bf r},{\bf 0}) & = - \frac{J^2 \bar{\sigma}}{64 \pi^2}  b_i \epsilon_{ij} b_k \epsilon_{kl} \int d^2r' \nabla'_j \Delta' \ln |{\bf r} - {\bf r'}| \, \nabla'_l ({\bf r'}^2 \ln |{\bf r'}|) \vspace{0.8cm} \\
   & = - \frac{J^2 \bar{\sigma}}{32 \pi}  b_i \epsilon_{ij} b_k \epsilon_{kl} \nabla_j \nabla_l ({\bf r}^2 \ln |{\bf r}|),
   \end{split}
\end{equation}
wobei letzteres nur f"ur Separationen $r \gg 1$ gilt. Wendet man hierauf das in \refformel{a001a} bis \refformel{a004a} beschriebene Vorgehen an, so erh"alt man unter Vernachl"assigung von konstanten Beitr"agen
\begin{equation}
   \label{b09a}
   H({\bf r},{\bf 0}) = \frac{J^2}{16 \pi} \bar{\sigma} \left( \ln |{\bf r}| - \cos^2 \theta \right).
\end{equation}
 
Es gilt nun die Randterme der partiellen Integration zu betrachten. In zwei Dimensionen hat man f"ur den Laplace-Operator dargestellt in Polarkoordinaten $\Delta = \frac{1}{r} \frac{\partial}{\partial r} r \frac{\partial}{\partial r} + \frac{1}{r^2} \frac{\partial^2}{\partial \phi^2}$. Beschr"ankt man sich zun"achst auf die erste partielle Integration f"ur den ersten Term des Laplace-Operators, so ist der Randtermbeitrag $R_{r1}$
\begin{equation}
   \label{b10a}
   R_{r1}({\bf r},{\bf 0}) = - \frac{J^2 \bar{\sigma}}{64 \pi^2}  b_i \epsilon_{ij} b_k \epsilon_{kl} \int d \phi r' \nabla'_j \ln |{\bf r'} - {\bf r}| \, \nabla'_l \frac{\partial}{\partial r'} ({r'}^2 \ln r') \bigg\arrowvert_{1}^{R}.
\end{equation}
Mit Hilfe der Relationen
\begin{equation}
   \begin{split}
   \label{b11a}
   & \nabla'_j \ln |{\bf r'} - {\bf r}| = \frac{x'_j - x_j}{({\bf r'} - {\bf r})^2} \vspace{0.8cm} \\
   & \nabla'_l \frac{\partial}{\partial r'} ({r'}^2 \ln r') = \frac{1}{r'} (3 x'_l + 2 x'_l \ln r'),
   \end{split}
\end{equation}
erh"alt man somit
\begin{equation}
   \begin{split}
   \label{b12a}
   R_{r1}({\bf r},{\bf 0}) & = - \frac{J^2 \bar{\sigma}}{64 \pi^2}  b_i \epsilon_{ij} b_k \epsilon_{kl} \int_0^{2 \pi} d \phi \frac{x'_j - x_j}{({\bf r'} - {\bf r})^2} (3 x'_l + 2 x'_l \ln r') \bigg\arrowvert_1^R \vspace{0.8cm} \\
   & = - \frac{J^2}{32 \pi} \bar{\sigma} \ln R,
   \end{split}
\end{equation}
wobei von $x' = r'\cos \phi$ und $y' = r'\sin \phi$ Gebrauch gemacht wurde. 
 
Die zweite partielle Integration des ersten Laplace-Terms hat den Randterm
\begin{equation}
   \label{b13a}
   R_{r2}({\bf r},{\bf 0}) = \frac{J^2 \bar{\sigma}}{64 \pi^2}  b_i \epsilon_{ij} b_k \epsilon_{kl} \int d \phi r' \nabla'_j \frac{\partial}{\partial r'} \ln |{\bf r'} - {\bf r}| \, \nabla'_l ({r'}^2 \ln r') \bigg\arrowvert_{1}^{R}.
\end{equation}
Hier gelten die Beziehungen
\begin{equation}
   \begin{split}
   \label{b14a}
   & \nabla'_j \frac{\partial}{\partial r'} \ln |{\bf r'} - {\bf r}| \approx -\frac{x'_j - x_j}{({\bf r'} - {\bf r})^2 r'} \vspace{0.8cm} \\
   & \nabla'_l ({r'}^2 \ln r') = x'_l + 2 x'_l \ln r',
   \end{split}
\end{equation}
die analog zu oben eingesetzt folgendes ergeben
\begin{equation}
   \begin{split}
   \label{b15a}
   R_{r2}({\bf r},{\bf 0}) & = \frac{J^2 \bar{\sigma}}{64 \pi^2}  b_i \epsilon_{ij} b_k \epsilon_{kl} \int_0^{2 \pi} d \phi \frac{x'_j - x_j}{({\bf r'} - {\bf r})^2} (x'_l + 2 x'_l \ln r') \bigg\arrowvert_1^R \vspace{0.8cm} \\
   & = - \frac{J^2}{32 \pi} \bar{\sigma} \ln R.
   \end{split}
\end{equation}  
Die beiden Randterme liefern also den gleichen Beitrag. Die Beitr"age durch die Randterme der partiellen Intergration f"ur den zweiten Teil des Laplace-Operators ($\frac{1}{r^2} \frac{\partial^2}{\partial \phi^2}$) sind 0, da hier nur Integrale der Form
\begin{equation}
   \label{b16a}
   \int_1^R dr f(r, \cos \phi, \sin \phi) \bigg\arrowvert_0^{2 \pi}
\end{equation}
auftreten. Es gilt also insgesamt
\begin{equation}
   \begin{split}
   \label{b17a}
   \left[V({\bf r}) V({\bf 0}) \right]_D & = H({\bf r},{\bf 0}) + R_{r1}({\bf r},{\bf 0}) + R_{r2}({\bf r},{\bf 0}) \vspace{0.8cm} \\
   & = \frac{J^2}{16 \pi} \bar{\sigma} (\ln r - \cos^2 \theta - \ln R).
   \end{split}
\end{equation}
 
Abschlie"send kann festgestellt werden, da"s die Varianz f"ur Separationen $ r \gg 1$ den in \refformel {268a} angegebenen Wert annimmt 
\begin{equation}
   \begin{split}
   \label{b18a}
   \Delta^2(r, \theta) & =  2 \left[V({\bf 0})V({\bf 0}) + V({\bf r})V({\bf 0}) \right]_D \vspace{0.8cm} \\
   & = \frac{J^2}{8 \pi} \bar{\sigma} \left(\ln r - \cos^2 \theta \right).
   \end{split}
\end{equation}
 
Dieses Resultat wurde auch von Nelson \verweistext{nel83} bei der Berechnung der von ihm benutzten replizierten Hamiltonfunktion verwendet.
  
  \chapter{Berechnungen zur Polarisierbarkeit}
\section{Polarisierbarkeit in der Elektrodynamik}
Dieser Abschnitt soll zeigen, da"s die in Abschnitt 3.3 angewandte Methodik zur Bestimmung der Polarisierbarkeit genauso in der "ublichen Elektrodynamik benutzt werden kann und daraus ihre Berechtigung erf"ahrt.
 
Hierzu betrachten wir ein System von Punktladungen. Unordnung spiele keine Rolle. Die Wahrscheinlichkeit, an einem Paar von Pl"atzen im System mit Abstand $r$ einen Dipol zu finden, ist Boltzmann-verteilt $p(r) \propto \exp(-\beta E)$, E ist die Wechselwirkungsenergie der beiden Ladungen.
 
Legt man nun ein konstantes, homogenes {\bf E}-Feld an das System an, so ist die Wechselwirkung eines Dipols mit diesem Feld
\begin{equation}
   \label{c50}
   W = - \mu_i E_i.
\end{equation}
Hierbei ist ${\bf \mu} = {\bf r}$ das Dipolmoment eines Dipols mit Separation $r$ und Ladung $|q| = 1$.  
 
Analog zu \refformel{336a} l"a"st sich die {\it Polarisationsdichte} $ q_i(r)$ als
\begin{equation}
   \label{c51}
   q_i(r) = \frac{w_+ - w_-}{1 + w_+ + w_-} \mu_i
\end{equation}
mit
\begin{equation}
   \label{c52}
   w_{\pm} = \exp [-\beta (E \pm W)]
\end{equation}
bestimmen. Auch hier kann Tensor $A_{ij}$ eingef"uhrt werden
\begin{equation}
   \label{c53}
   A_{ij}(r) = \frac{1}{2} \int_0^{2 \pi} \frac{d \theta}{2 \pi}\, \frac{\partial q_i}{\partial E_j} \bigg\arrowvert_{E=0}
\end{equation}
Der Vorfaktor ber"ucksichtigt, da"s eine Orientierung in \refformel{c51} modulo $\pi$ ausgezeichnet ist. Ableiten gibt
\begin{equation}
\label{c54}
\begin{split}
     \frac{\partial q_i}{\partial E_j} \bigg\arrowvert_{E=0} & = \beta \mu_i \mu_j \frac{(w_+ + w_-)(1 + w_+ + w_-) - (w_+ - w_-)^2}{(1 + w_+ + w_-)^2}  \bigg\arrowvert_{E=0} \\
\\
      & = \beta \mu_i \mu_j \, 2 \, e^{-\beta E}.
\end{split}
\end{equation}
 
Damit kann der Tensor $A_{ij}$ berechnet werden
\begin{equation}
\label{c55}
\begin{split} 
    A_{ij}(r) & = \frac{1}{T}  \int_0^{2 \pi} \frac{d \theta}{2 \pi} x_i x_j p(r)\\
    & = \frac{r^2 \, p(r)}{2T} \delta_{ij}
\end{split}
\end{equation}
 
Da in der Elektrodynamik in einem isotropen Medium die induzierte Polarisation parallel zum anliegenden {\bf E}-Feld ist
\begin{equation}
   \label{c56}
   P_i = \chi_e \delta_{ij} E_j,
\end{equation}
nimmt die Polarisierbarkeitsdichte $A(r)$, f"ur die $\chi_e(r) = \int_0^r dr' A(r')$ gilt, folgende Form an
\begin{equation}
   \label{c57}
   A(r) = A_{xx}(r) = A_{yy}(r) = \frac{r^2 \, p(r)}{2 T}.
\end{equation}
 
Die {\it Polarisierbarkeit} $\alpha(r)$ wird damit zu
\begin{equation}
   \label{c58}
   \alpha(r) = \frac{A(r)}{p(r)} = \frac{r^2}{2T},
\end{equation}
dem aus der Elektrodynamik bekannten Ausdruck. In Abschnitt 3.3 wurde diese Methode zur Polarisierbarkeitsberechnung "ubertragen auf das zu untersuchende Problem eines Coulombgases mit Vektorladungen. 
  
\section{Winkelintegration des Dipoltensors}
In diesem Abschnitt soll das Integral
\begin{equation}
   \label{c01}
   \int_0^{2 \pi} \frac{d \theta}{2 \pi} \, Q_{ij} Q_{kl} \, e^{x \cos^2 \theta}
\end{equation}
gel"ost werden, welches bei der Berechnung der Polarisierbarkeitsdichte immer auftritt, lediglich der Wert von $x$ h"angt vom betrachteten Temperaturbereich ab. Im weiteren wird folgende Abk"urzung benutzt:
\begin{equation}
   \label{c02}
   <f(r,\theta)>_w := \int_0^{2 \pi} \frac{d \theta}{2 \pi} \, f(r,\theta) \, e^{x \cos^2 \theta}
\end{equation}
 
Da entsprechend Gleichung \refformel{334a} gilt
\begin{equation}
   \label{c03}
   Q_{ij} Q_{kl} = \frac{1}{4} \epsilon_{ir} \epsilon_{js} \epsilon_{kt} \epsilon_{lu} (b_r \epsilon_{sm} x_m + b_s \epsilon_{rm} x_m)(b_t \epsilon_{un} x_n + b_u \epsilon_{tn} x_n),
\end{equation}
beschr"anken wir uns zun"achst auf die Berechnung von
\begin{equation}
   \label{c04}
   <(b_r \epsilon_{sm} x_m + b_s \epsilon_{rm} x_m)(b_t \epsilon_{un} x_n + b_u \epsilon_{tn} x_n)>_w.
\end{equation}

Mittels elementarer Integration unter Beachtung der Additionstheoreme f"ur Cosinus und Sinus (vgl. auch Young \verweistext{you79}),  erh"alt man die folgenden Beziehung
\begin{equation}
   \label{c05}
   <x_i x_j>_w = \left< \frac{r^2}{2} \delta_{ij} + r^2 \left(b_i b_j - \frac{1}{2} \delta_{ij} \right) \cos 2 \theta \right>_w.
\end{equation}
 
Wendet man diese auf \refformel{c04} an so ergibt sich unter Verwendung von $\epsilon_{ij} \epsilon_{kl} = \delta_{ik} \delta_{ij} - \delta_{il} \delta_{jk}$: 
\begin{equation}
   \label{c06}
   \begin{split}
   & <(b_r \epsilon_{sm} x_m + b_s \epsilon_{rm} x_m)(b_t \epsilon_{un} x_n + b_u \epsilon_{tn} x_n)>_w \\
\\
   & = \, <(b_r b_t \epsilon_{sm} \epsilon_{un} x_m x_n + b_s b_u \epsilon_{rm} \epsilon_{tn} x_m x_n \\
   & \quad + b_r b_u \epsilon_{sm} \epsilon_{tn} x_m x_n + b_s b_t \epsilon_{rm} \epsilon_{un} x_m x_n)>_w \\
\\ 
   & = \, <[b_r b_t (\delta_{su} \delta_{mn} - \delta_{sn} \delta_{um}) + b_s b_u (\delta_{rt} \delta_{mn} - \delta_{rn} \delta_{tm}) \\
   & \quad + b_r b_u (\delta_{st} \delta_{mn} - \delta_{sn} \delta_{tm}) + b_s b_t (\delta_{ru} \delta_{mn} - \delta_{rn} \delta_{um})]\\
   & \quad \times r^2 \left(\delta_{mn}/2 + \left[b_m b_n - \delta_{mn}/2 \right] \cos 2 \theta \right)>_w.
\end{split}
\end{equation}
 
F"ur die Mittelung "uber alle Richtungen des Burgers-Vektors {\bf b} setzt man ein isotropes System an. Hier gilt
\begin{align}
\label{c07}
   &<b_m b_n>_{av} = \frac{1}{2} \delta_{mn} \\
\label{c08}
   &<b_k b_l b_m b_n>_{av} = \frac{1}{8} (\delta_{kl} \delta_{mn} + \delta_{km} \delta_{ln} + \delta_{kn} \delta_{lm}).
\end{align}
F"ur das Dreiecksgitter gelten beide Gleichungen wie f"ur das isotrope System exakt, im Fall des Quadratgitters kann \refformel{c08} als N"aherung benutzt werden. Wendet man diese Mittelung nun auf \refformel{c06} an, so resultiert
\begin{equation}
   \label{c09}
   \begin{split}
   & <(b_r \epsilon_{sm} x_m + b_s \epsilon_{rm} x_m)(b_t \epsilon_{un} x_n + b_u \epsilon_{tn} x_n)>_w \\
   & = \, \left< \frac{r^2}{2} (\delta_{rt} \delta_{su} + \delta_{ru} \delta_{ts}) - \frac{r^2}{2} \cos 2 \theta \delta_{rs} \delta_{tu} \right>_w.
\end{split}
\end{equation}

Mit der Umformung $\exp(x \, \cos^2 \theta) = \exp(\frac{x}{2})\, \exp(\frac{x}{2} \cos 2 \theta)$ und unter Benutzung der modifizierten Besselfunktionen $I_0$ und $I_1$, die wie folgt definiert sind 
\begin{equation}
\label{c10}
\begin{split}
   I_0(x) & := \int_0^{2 \pi} \frac{d \theta}{2 \pi} e^{x \cos 2 \theta}\\
\\
   I_1(x) & := \int_0^{2 \pi} \frac{d \theta}{2 \pi} \cos 2 \theta \, e^{x \cos 2 \theta},
\end{split}
\end{equation}
findet man schlie"slich f"ur das gesuchte Resultat
\begin{equation}
\label{c11}
\begin{split}
   & \int_0^{2 \pi} \frac{d \theta}{2 \pi} \, Q_{ij} Q_{kl} \, e^{x \cos^2 \theta}\\
\\
   & = \frac{1}{4} \epsilon_{ir} \epsilon_{js} \epsilon_{kt} \epsilon_{lu} \, e^{\frac{x}{2}} \, \left[Q_1(r) (\delta_{rt} \delta_{su} + \delta_{ru} \delta_{ts}) + Q_2(r)  \delta_{rs} \delta_{tu} \right],
\end{split}
\end{equation}
wobei
\begin{equation}
\label{c12}
\begin{split}  
   Q_1(r) & = \frac{1}{2} r^2 \, I_0\left(\frac{x}{2} \right)\\
   Q_2(r) & = -\frac{1}{2} r^2 \, I_1\left(\frac{x}{2} \right).
\end{split}
\end{equation}   
Diese Relation erlaubt die Berechnung der Polarisierbarkeiten und Polarisierbarkeitsdichten in Abschnitt 3.3.

  \chapter{Mittelung des Ordnungsparameters}
 
Unter Verwendung der Methoden von Nelson \verweistext{nel83} soll in diesem Abschnitt der in Kapitel 5 \refformel{513a} verwendete Ausdruck f"ur den Ordnungspar. $\left[ \langle \exp (i {\bf G} {\bf u(r)}) \rangle \right]_D$ hergeleitet werden. Hierbei steht $<...>$ f"ur eine thermische Mittelung und $[...]_D$ f"ur die Mittelung "uber die m"oglichen Unordnungskonfigurationen. Wir betrachten das System auf sehr gro"sen L"angenskalen, wo Versetzungen in der kristallinen Phase keine Rolle mehr spielen, da sie zu Dipolen mit kleinerer Separation gebunden sind. Folglich kann bei der thermischen Mittelung ein rein phononisches Verschiebungsfeld angesetzt werden, welches Gau"s-verteilt ist.

\section{Berechnung ohne Unordnung}
Zun"achst betrachten wir ein System, das keine Verunreinigungen aufweist; in der Hamiltonfunktion wird $\delta c$ null gesetzt. Mit einem beliebigen reziproken Gittervektor {\bf G} und dem Verschiebungsfeld {\bf u(r)} kann man den Ordnungsparameter entwickeln
\begin{equation}
   \label{d01}
   \langle \exp (i {\bf G} \, {\bf u(r)}) \rangle = \sum_{n=0}^{\infty} i^n \frac{\langle ({\bf G} \, {\bf u(r)})^n \rangle}{n!} = \sum_{n=0}^{\infty} (-1)^n \frac{\langle ({\bf G} \, {\bf u(r)})^{2n} \rangle}{(2n)!},
\end{equation}
die ungeraden Terme der Entwicklung fallen weg, da "uber eine gau"sische Wahr\-schein\-lich\-keits-Verteilung gemittelt wird. Das Wick-Theorem liefert nun 
\begin{equation}
   \label{d02}
   \langle ({\bf G} \, {\bf u})^{2n} \rangle = \frac{(2n)!}{2^n n!} \langle ({\bf G} \, {\bf u})^{2} \rangle^n.
\end{equation}
Somit ist 
\begin{equation}
   \label{d03}
   \langle \exp (i {\bf G} \, {\bf u(r)}) \rangle = \sum_n \left(- \frac{1}{2}\right)^n \frac{\langle ({\bf G} \, {\bf u})^{2} \rangle^n}{n!} = \exp \left(- \frac{1}{2} \langle ({\bf G} \, {\bf u})^{2} \rangle \right), 
\end{equation}
was es notwendig macht, den Ausdruck 
\begin{equation}
   \label{d04}
   \langle ({\bf G} \, {\bf u})^{2} \rangle = G_x^2 \langle u_x^2 \rangle + 2 G_x G_y  \langle u_x u_y  \rangle +  G_y^2 \langle u_y^2 \rangle
\end{equation}
zu berechnen. 
 
Um die Hamiltonfunktion
\begin{equation}
   \label{d05}
   H = \frac{1}{2} \int d^2r 2 \mu u_{ik}^2 ({\bf r}) + \lambda u_{jj}^2 ({\bf r})
\end{equation}
im Impulsraum darstellen zu k"onnen, f"uhren wir einen Basiswechsel von einer Darstellung $(u_x, u_y)$ in eine Darstellung $(u_l, u_t)$ durch. Hierbei ist ${\bf u_l}({\bf k})$ der longitudinale Anteil von ${\bf u}$ bez"uglich ${\bf k}$ und ${\bf u_t}({\bf k})$ der transversale Anteil
\begin{align}
  \label{d06}
  {\bf u_l} & = {\bf e_k} ({\bf e_k}{\bf u}) \\
  \label{d07}
  {\bf u_t} & = {\bf u} - {\bf u_l} = {\bf u} - {\bf e_k} ({\bf e_k}{\bf u}).
\end{align}
${\bf e_k}$ symbolisiert den Einheitsvektor in {\bf k}-Richtung. Die zugeh"orige Basiswechselmatrix ist 
\begin{equation}
  \label{d08}
  A = \left( \begin{array}{cc} \frac{k_x}{\sqrt{k^2}} & \frac{k_y}{\sqrt{k^2}} \\- \frac{k_y}{\sqrt{k^2}} & \frac{k_x}{\sqrt{k^2}} \end{array} \right),
\end{equation}
und die zugeh"orige inverse Matrix
\begin{equation}
  \label{d09}
  A^{-1} = \left( \begin{array}{cc} \frac{k_x}{\sqrt{k^2}} &  - \frac{k_y}{\sqrt{k^2}} \\\frac{k_y}{\sqrt{k^2}} & \frac{k_x}{\sqrt{k^2}} \end{array} \right).
\end{equation}  
Mit diesem Basiswechsel wird die fouriertransformierte Hamiltonfunktion zu (vgl.  auch \verweistext{chlu}):
\begin{equation}
   \label{d10}
   H = \frac{1}{2} \int \frac{d^2k}{2 \pi} (2 \mu + \lambda) k^2 u_l^2 + \mu k^2 u_t^2
\end{equation}
Durch Ableiten findet man
\begin{align}
   \label{d11}
   \frac{\partial H}{\partial u_l({\bf k})} & = (\lambda + 2 \mu) k^2 u_l({\bf k})\\
   \label{d12}
   \frac{\partial H}{\partial u_t({\bf k})} & = \mu k^2 u_t({\bf k}).
\end{align}
Benutzung des "Aquipartitionsprinzips 
\begin{equation}
   \label{d13}
   \langle u_i \, \frac{\partial H}{\partial u_j} \rangle = T \, \delta_{ij}
\end{equation}
liefert
\begin{equation}
   \label{d14}
   \langle u_l({\bf k}) u_l({\bf k}) \rangle = \frac{1}{(\lambda + 2 \mu) k^2} \langle u_l \, \frac{\partial H}{\partial u_l} \rangle = \frac{T}{(\lambda + 2 \mu) k^2},
\end{equation}
und
\begin{equation}
   \label{d15}
   \langle u_t({\bf k}) u_t({\bf k}) \rangle = \frac{T}{\mu k^2}.
\end{equation}
 
F"uhrt man jetzt die R"ucktransformation mit Hilfe der inversen Matrix $A^{-1}$ aus, so resultiert unter Beachtung von $\langle u_t({\bf k}) u_l({\bf k}) \rangle = 0$ (wegen \refformel{d13}):
\begin{equation}
\label{d16}
\begin{split}
   \langle u_i({\bf k}) u_j({\bf k}) \rangle & = \hat{k}_i \, \hat{k}_j \langle u_l u_l \rangle + ( \delta_{ij} - \hat{k}_i \, \hat{k}_j ) \langle u_t u_t \rangle \\
   & = \frac{T}{(\lambda + 2 \mu) k^2}  \hat{k}_i \, \hat{k}_j + \frac{T}{\mu k^2} ( \delta_{ij} - \hat{k}_i \, \hat{k}_j ).
\end{split}
\end{equation}
Hierbei stehen die $(i,j)$ f"ur Kombinationen der $(x,y)$, ferner ist $\hat{k}_i$ die $i$-te Komponente des Einheitsvektors ${\bf e_k}$. F"uhrt man nun eine Fouriertransformation durch, so erh"alt man wegen $\int_0^{2 \pi} \hat{k}_i \, \hat{k}_j d \theta = \pi \delta_{ij}$ folgenden Ausdruck
\begin{equation}
\label{d17}
\begin{split}
    \langle u_i({\bf r}) u_j({\bf r}) \rangle & = \frac{1}{(2 \pi)^2} \int d^2k \langle u_i({\bf k}) u_j({\bf k}) \rangle  \\
    & = \delta_{ij} \frac{1}{4 \pi} \int dk \, k \left( \frac{T}{(\lambda + 2 \mu) k^2} + \frac{T}{\mu k^2} \right) \\
    & = \delta_{ij} \frac{1}{4 \pi} \ln R \, \left( \frac{T}{(\lambda + 2 \mu} + \frac{T}{\mu} \right) \\ 
    & = T \, \ln R \, \frac{3 \mu + \lambda}{4 \pi \mu (\lambda + 2 \mu)}\\
    & =  \langle (u_x({\bf r}))^2 \rangle =  \langle (u_y({\bf r}))^2 \rangle.
\end{split}
\end{equation}
Setzt man dieses Resultat in \refformel{d04} ein, so findet man
\begin{equation}
   \label{d18}
   \langle ({\bf G} \, {\bf u})^{2} \rangle = G^2 \, T \, \ln R \, \frac{3 \mu + \lambda}{4 \pi \mu (\lambda + 2 \mu)}.
\end{equation}
 
Mit \refformel{d03} erh"alt man schlie"slich im Fall ohne Unordnung das bereits von Pokrovsky \verweistext{pok1} gefundene Ergebnis
\begin{equation}
  \label{d19}
   \langle \exp (i {\bf G} {\bf u(r)}) \rangle = R^{- \frac{T \, G^2 \, (3 \mu + \lambda)}{8 \pi \mu (\lambda + 2 \mu)}}.
\end{equation}

\section{Berechnung mit Unordnung}
Unter Ber"ucksichtigung der Unordnung nimmt die zu betrachtende Hamiltonfunktion die folgende Gestalt an
\begin{equation}
  \label{d20}
  H = H_0 - \int d^2r (\mu + \lambda) \Omega_0 \, \delta c({\bf r}) u_{kk}({\bf r}).
\end{equation}
Dabei steht $H_0$ f"ur die Hamiltonfunktion des Systems ohne Unordnung \refformel{d05}. Ebenso bezeichnen $<...>$ und $<...>_0$ im folgenden die thermische Mittelung "uber den vollen Hamiltonian, beziehungsweise die Mittelung mit $H_0$. Damit ist
\begin{equation}
   \label{d21}
   \left[ \langle \exp (i \, {\bf G} \, {\bf u(r)}) \rangle \right]_D = \left[ \frac{ \langle e^{i \, {\bf G} \, {\bf u(r)}} \, e^{\frac{(\mu + \lambda) \Omega_0}{T} \int d^2r \delta c \, {\bf \nabla u}} \rangle_0}{\langle e^{\frac{(\mu + \lambda) \Omega_0}{T} \int d^2r \delta c \, {\bf \nabla u}} \rangle_0} \right]_D.
\end{equation}
Setzt man nun
\begin{equation}
   \label{d22}
   x = i \, {\bf G \, u} + \frac{(\mu + \lambda) \Omega_0}{T} \int d^2r \delta c \, {\bf \nabla u},
\end{equation}
so gilt
\begin{equation}
   \label{d23}
   \langle e^x \rangle_0 = e^{\frac{1}{2} \langle x^2 \rangle_0},
\end{equation}
da die Mittelung "uber $H_0$ wiederum eine gau"sische Mittelung ist. Wir finden\begin{equation}
   \label{d24}
   \begin{split}
   \langle x^2 \rangle_0 = & - \langle ({\bf G \, u})^2 \rangle_0 + 2 \, i \, \frac{(\mu + \lambda) \Omega_0}{T} \int d^2r \delta c \, G_i \, \langle u_i(0) \, \nabla_j u_j ({\bf r}) \rangle_0 \\
& \\
& + \left(\frac{(\mu + \lambda) \Omega_0}{T} \right)^2 \langle \int d^2r \delta c \, {\bf \nabla u} \rangle_0^2.
\end{split}
\end{equation}
Setzt man dies in \refformel{d21} ein, so heben sich Nenner und der letzte Term des obigen Ausdrucks weg. Es bleibt:
\begin{equation}
  \label{d25}
  \begin{split}
   \left[ \langle \exp (i \, {\bf G} \, {\bf u(r)}) \rangle \right]_D & =  \left[ e^{- \frac{1}{2} \langle ({\bf G \, u})^2 \rangle_0} + e^{i \frac{(\mu + \lambda) \Omega_0}{T} \int d^2r \delta c \, G_i \nabla_j \langle u_i(0) \, u_j ({\bf r}) \rangle_0} \right]_D \\
   & =: \left[ e^{- \frac{1}{2} \langle ({\bf G \, u})^2 \rangle_0} + e^{i \, I(0)} \right]_D
\end{split}
\end{equation}
Die Unordnungsmittelung des ersten Terms ist unn"otig, da er nicht von der Unordnung abh"angt, er wurde in Abschnitt D.1 berechnet, das Resultat ist \refformel{d19}. Da die Unordnung gau"sisch verteilt ist, gilt f"ur die Unordnungsmittelung des zweiten Terms
\begin{equation}
   \label{d26}
   \left[ e^{i \, I(0)} \right]_D = e^{- \frac{1}{2} \left[I^2(0) \right]_D},
\end{equation}
mit $\left[ \delta c({\bf r}) \delta c({\bf r'}) \right]_D = \sigma \delta ({\bf r} - {\bf r'})$. Daraus folgt
\begin{equation}
   \label{d27}
    \left[I^2(0) \right]_D =  \left(\frac{(\mu + \lambda) \Omega_0}{T} \right)^2 \sigma G_r \, G_s \int d^2r \nabla_i \langle u_r(0) \, u_i({\bf r}) \rangle_0 \, \nabla_j \langle u_s(0) \, u_j({\bf r}) \rangle_0.
\end{equation}
Mit Hilfe der in D.1 beschriebenen Methodik erh"alt man hier folgende Darstellung 
\begin{equation}
\label{d28}
\begin{split}
   \left[I^2(0) \right]_D & =  \frac{(\mu + \lambda)^2 \Omega_0^2 \sigma}{(2 \mu + \lambda)^2} G_r \, G_s \int \frac{d^2k}{(2 \pi)^2} \frac{k_r \, k_s}{k^4} \\
\\
   & = \frac{(\mu + \lambda)^2 \Omega_0^2 \sigma}{4 \pi (2 \mu + \lambda)^2} (G_x^2 + G_y^2) \int dk \, k \frac{1}{k^2} \\
\\
   & = \ln R \frac{G^2 \, \sigma (\mu + \lambda)^2 \Omega_0^2 \sigma}{4 \pi (2 \mu + \lambda)^2}.
\end{split}
\end{equation}
Setzt man dies in \refformel{d25} ein, so ergibt sich die in Kapitel 5 benutzte Formel \refformel{513a}
\begin{equation}
   \label{d29}
    \left[ \langle \exp (i \, {\bf G} \, {\bf u(r)}) \rangle \right]_D = R^{- \frac{T \, G^2 \, (3 \mu + \lambda)}{8 \pi \mu (\lambda + 2 \mu)}} \, R^{- \frac{\sigma \, G^2 \, \Omega_0^2 (\mu + \lambda)^2}{8 \pi (2 \mu + \lambda)^2}},
\end{equation}
die in "ahnlicher Form bereits von Nelson \verweistext{nel83} hergeleitet worden ist.

\end{appendix}


\end{document}